\documentclass[10pt]{article}
\usepackage{theorem}

\newtheorem{theorem}{Theorem}[section]

\newtheorem{lemma}[theorem]{Lemma}

\def\indirizzo#1{{}}
\def\email#1{{}}
\def\qed{\raise1pt\hbox{\vrule height5pt width5pt depth0pt}}

\def\brm#1\erm{\vskip.8em{\bf\0Remark.}\ #1\vskip.8em}
\def\bpr{{\it\0Proof.}\ }
\def\epr{\quad\qed\vskip.8em}
\def\bac#1\eac{\vskip.8em{\bf\0Acknowledgements}\ #1\vskip.8em}

\usepackage{amsmath}
\usepackage{amssymb}
\usepackage{amsfonts}
\usepackage{graphicx}
\allowdisplaybreaks

\numberwithin{equation}{section}

\newcount\driver
\newcount\bozza

\def\a{\alpha}        \def\b{\beta}         \def\d{\delta}
\def\e{\varepsilon}   \def\z{\zeta}         \def\h{\eta}
\def\k{\kappa}        \def\l{\lambda}       \def\m{\mu}
\def\n{\nu}           \def\p{\pi}           \def\r{\rho}
\def\s{\sigma}        \def\t{\tau}          
\def\c{\chi}          \def\ps{\psi}         
    \def\g{\gamma}        
\def\D{\Delta}        \def\L{\Lambda}       
                  
\def\th{\vartheta}    \def\O{\Omega}        
\def\x{\xi}                  \def\o{\omega}
\def\Th{\Theta}       

\def\PP{{\mathcal P}}\def\EE{{\mathcal E}}\def\MM{{\mathcal M}}\def\VV{{\CMcal V}}
\def\CC{{\mathcal C}}\def\HH{{\mathcal H}}\def\WW{{\mathcal W}}
\def\TT{{\mathcal T}}\def\NN{{\mathcal N}}\def\BB{{\mathcal B}}
\def\RR{{\mathcal R}}\def\LL{{\mathcal L}}
\def\DD{{\mathcal D}}\def\AA{{\mathcal A}}\def\SS{{\mathcal S}}

\def\pp{{\bf p}}\def\qq{{\bf q}}\def\xx{{\bf x}}
\def\yy{{\bf y}}\def\kk{{\bf k}}\def\nn{{\bf n}}
\def\zz{{\bf z}}\def\uu{{\bf u}}\def\vv{{\bf v}}\def\ww{{\bf w}}

\def\bP{{\bf P}}\def\tt{{\bf t}}

\def\ha{{\widehat \a}}\def\hb{{\widehat \b}}
\def\hr{{\widehat \r}}
\def\hH{{\widehat  H}}
\def\hK{{\widehat  K}}
\def\hU{{\widehat  U}}\def\hp{{\widehat\ps}}\def\hF{{\widehat  F}}
\def\hJ{{\widehat  J}}\def\hg{{\widehat  g}}

\def\hG{{\widehat  G}}\def\hS{{\widehat S}}

\def\hh{{\hat \h}}    
    \def\hT{{\widehat  T}}

\def\uo{{\underline \o}}

\def\ux{{\underline\xx}}\def\uy{{\underline \yy}}
\def\uk{{\underline\kk}}\def\uq{{\underline\qq}}\def\up{{\underline \pp}}

\def\ue{{\underline \e}}\def\uz{{\underline\zz}}\def\us{{\underline  \s}}
                      \def\uw{{\underline\ww}}

\def\uTh{{\underline  \Th}}
\def\ual{{\underline \a}} \def\usp{{\underline s}} \def\uv{{\underline v}}
\def\uB{{\underline B}} \def\ub{{\underline b}}

\def\RRR{\mathbb{R}}

\def\ZZZ{\mathbb{Z}}

\def\VV{\mathcal{V}}

\let\dpr=\partial                  \let\bs=\backslash
\let\io=\infty                     \let\==\equiv

\def\lft{\left}                \def\rgt{\right}
\def\la{{\langle}}             \def\ra{{\rangle}}

       \def\tilde#1{{\widetilde #1}}

\def\Halmos{\hfill\vrule height10pt width4pt depth2pt \par\hbox to \hsize{}}

\def\wh#1{\widehat{#1}}
\def\hat#1{\wh{#1}}
\def\bar#1{\overline {#1}}

\def\erp{{\perp}}
\def\arp{{\parallel}}

\def\lb#1{\label{#1}}

\def\pref#1{(\ref{#1})}
\def\virg{\quad,\quad}
\def\*{\vspace{0.3cm}}
\let\0=\noindent

\def\be{\begin{equation}}    \def\ee{\end{equation}}
\def\brrm#1\errm{\vskip.8em{\bf\0Remarks.}\ \bd#1{\hfill\bf erm}\ed}
\def\bea{\begin{eqnarray}}   \def\eea{\end{eqnarray}}
\def\bean{\begin{eqnarray*}} \def\eean{\end{eqnarray*}}
\def\bfr{\begin{flushright}} \def\efr{\end{flushright}}
\def\bc{\begin{center}}      \def\ec{\end{center}}
\def\bal#1\eal{\begin{align}#1\end{align}}
\def\ba#1{\begin{array}{#1}} \def\ea{\end{array}}
\def\bd{\begin{enumerate}} \def\ed{\end{enumerate}}
\def\bv{\begin{verbatim}}    \def\ev{\end{verbatim}}
\def\bit{\begin{itemize}}    \def\eit{\end{itemize}}
\def\bsp#1\esp{\begin{split}#1\end{split}}
\def\nn{\nonumber}

\def\insertplot#1#2#3#4#5#6{%
\begin{figure}[h!]
\begin{center}
\includegraphics[
width=#1pt, height=#2pt,
clip=true, viewport=0 0 #1 #2]{#4.pdf}
\caption{\small#5}
\end{center}
\end{figure}
}

\begin{document}

\title{Universality of one-dimensional Fermi systems, II.
\\ The Luttinger liquid structure.}

%

\author{G. Benfatto$^1$ \and P. Falco$^2$ \and V. Mastropietro$^3$}

\maketitle

\footnotetext[1]{\normalsize Dipartimento di Matematica, Universit\`{a} di Roma
``Tor Vergata", 00133 Roma, Italy.}
\footnotetext[2]{\normalsize Department
of Mathematics, California State university, Northridge, CA 91330.}
\footnotetext[3]{\normalsize Dipartimento di Matematica F.Enriquez,
Universit\`{a} di Milano, Via Saldini 50, Milano, Italy.}

\begin{abstract}
We complete the proof started in \cite{BFM012_1} of the universal Luttinger
liquid relations for a general  model of spinning fermions on a lattice, by
making use of the Ward Identities due to asymptotically emerging symmetries.
This is done by introducing an effective model verifying extra symmetries and
by relating its critical exponents to those of the fermion lattice gas by
suitable fine tuning of the parameters.
\end{abstract}


\section{Main results}

We consider a standard (generically non solvable) model of a gas
of spinning fermions on a one dimensional lattice with a short
range repulsive interaction, with Hamiltonian
\be\lb{1.1} H=-{1\over 2}\sum_{x\in \CC\atop s=\pm}
(a^+_{x,s}a^-_{x+1,s}+a^+_{x,s}a^-_{x-1,s}) + \bar\m \sum_{x\in
\CC\atop s=\pm } a^+_{x,s}a^-_{x,s} +\l \sum_{x,y\in \CC\atop
s,s'=\pm } v_L(x-y)a^+_{x,s}a^-_{x,s} a^+_{y,s'}a^-_{y,s'} \ee
where $\CC=\{1,2,\ldots,L\}$ is a one dimensional lattice of $L$,
$a^\pm_{x,s}$ are fermion creation and annihilation operators at
site $x$ with spin $s$, and such that $a^\pm_{1,s}=a^\pm_{L+1,s}$
(periodic boundary conditions), $v_L(x)$ is a function on $\ZZZ$,
periodic of period $L$, such that $v_L(x)=v(x)$ for $-[L/2] \le x
\le [(L-1)/2]$, $v(x)$ being an even function on $\ZZZ$ satisfying
the short range condition $|v(x)|\le C e^{-\k |x|}$, and $-\bar\m
\in (-1,+1)$ is the {\it chemical potential}. In the special case
$\l v(x-y)=U \d_{x,y}$  the model is known as {\it Hubbard model},
which is exactly solvable by Bethe ansatz \cite{LW068}.

We define the operators $a_{\xx,s}^{\pm}=e^{x_0 H} a_x^{\pm}e^{-H
x_0}$, with $\xx=(x,x_0)$, $0\le x_0 < \b$ for some $\b>0$
($\b^{-1}$ is the temperature) and the densities $\r^\a_{\xx}$
with $\a=C,S_i,SC_i,TC_i$ denoting respectively the {\it Charge
density}, the {\it spin densities} and singlet and tripled {\it
Cooper densities}
\bea && \r^C_{\xx}= \sum_{s=\pm }a^+_{\xx,s}a^-_{\xx,s}\quad\quad
\r^{S_i}_{\xx}= \sum_{s,s'=\pm } a^+_{\xx,s}\s^{(i)}_{s,s'} a
^-_{\xx,s'}\label{ss111a}\nn\\
&& \r^{SC}_{\xx}= \frac12 \sum_{s=\pm \atop \e=\pm} s\,
a^\e_{\xx,s} a^\e_{\xx,-s}\quad\quad \r^{TC_i}_ \xx= \frac12
\sum_{s,s'=\pm \atop \e=\pm} a^\e_{\xx,s} \tilde \s^{(i)}_{s,s'}
a^\e_{\xx+{\bf e},s'}\virg {\bf e}=(1,0)\eea
where $\s^{(i)}$ are the Pauli matrices, while
$$\tilde\s^{(1)} = \begin{pmatrix} 1 &0\\ 0& 0 \end{pmatrix} \virg
\tilde\s^{(2)} = \begin{pmatrix} 0 &1\\ 1& 0 \end{pmatrix} \virg
\tilde\s^{(3)} = \begin{pmatrix} 0 &0\\ 0& 1 \end{pmatrix}$$

Similarly we will introduce the {\it paramagnetic} and {\it
diamagnetic} part of the current
\be J_\xx={1\over 2i}\sum_{s=\pm} [a^+_{\xx+{\bf e},s}
a^-_{\xx,s}- a_{\xx,s}^+ a^-_{\xx+{\bf e},s}] \virg
\t_\xx=-{1\over 2} \sum_{s=\pm}[a^+_{\xx,s} a^-_{\xx+{\bf
e},s}+a_{\xx+{\bf e},s}^+ a^-_{\xx,s}] \ee
Defining $\la \cdot\ra_{L,\b} := {{\rm Tr}[e^{-\b H}\cdot]\over
{\rm Tr}[e^{-\b H}]}$, the density and current {\it response
functions} are defined by the following truncated correlations:
\bea \O_{\a,\b,L}(\xx-\yy):= \la {\bf T}\r^\a_{\xx}
\r^\a_{\yy}\ra_{T;\b,L}:=\la {\bf T}\r^\a_{\xx}
\r^\a_{\yy}\ra_{\b,L}- \la \r^\a_{\xx}\ra_{\b,L}
\la\r^\a_{\yy}\ra_{\b,L}\lb{ss111}\nn\\
\O_{j,j,\b,L}(\xx-\yy):= \la {\bf T} J_{\xx}
J_{\yy}\ra_{T;\b,L}:=\la {\bf T} J_{\xx} J_{\yy}\ra_{\b,L}- \la
J_{\xx}\ra_{\b,L} \la J_{\yy}\ra_{\b,L} \eea
where, if $O_\xx$ is quadratic in the fermion operators, ${\bf
T} O_{\xx}O_{\yy}=O_{\xx} O_{\yy}$ if $x_0\ge y_0$ and
$O_{\yy}O_{\xx}$ if $x_0\le y_0$. If $\xx-\yy=(\x,\t)$, the
response functions are defined in $(-L,L)\times [-\b,\b]$ and are
$\b$-periodic in $\t$ and $L$-periodic in $\x$. If $F_{\b,L}$ is
any function of this type, we define its Fourier transform as
\be \hat F_{\b,L}(\pp)= \int_{-{\b\over 2}}^{\b\over 2}
dx_0\sum_{x\in\CC} e^{i\pp \xx}\;F_{\b,L}(\xx)\ee
where $\pp=(p, p_0)$, with  $p\in {2\pi \over L}n$, $-[L/2]\le
n\le [(L-1)/2]$ and $p_0\in {2 \pi\over \b}\ZZZ$.

In the following we will be interested in the zero temperature limit of some
truncated correlation functions, in particular the two-point function
$\d_{s,s'} S_2^{\b,L}(\xx-\yy) := \la {\bf T}\a^-_{\xx,s}
\a^+_{\yy,s'}\ra_{\b,L}$, the density and current response functions and
vertex functions (to be defined later), calculated in the thermodynamic limit.
We shall denote these functions by the same symbols, without the $\b$ and
$L$ labels; for example, we shall write: $\lim_{\b\to\io}\lim_{L\to\io}
\hat\O_{L,\b,\a}(\pp)\equiv \hat\O_{\a}(\pp)$. Note that the thermodynamic
limit $L\to\io$ is taken before the zero temperature limit $\b\to\io$; this
allows us to derive properties of the {\it thermal ground state}. To shorten
the notation, in the following we shall use the definition $\lim_{\b,L\to\io}
\= \lim_{\b\to\io} \lim_{L\to\io}$.

Several important thermodynamic quantities can be derived from the
knowledge of the response functions. In particular the {\it
susceptibility}, which is given by
\be\lb{kk} \k:=\lim_{p\to 0}\lim_{p_0\to 0} \hat\O_C(\pp)\;. \ee
and the {\it Drude weight}, related to the response of the system
to an e.m. field, is defined as
\be D=\lim_{p_0\to 0}\lim_{p\to 0}\hat D(\pp)\label{cc}\ee
with
\be \hat D_{\b,L}(\pp)= - \la \t_x\ra -\int_{-\b/2}^{\b/2}
dx_0\sum_{x\in\L} e^{i\pp\xx} \la J_\xx J_{\bf
0}\ra_{T;L,\b}\label{cc1}\ee
where the first term is a constant independent of $x$. If one
assumes analytic continuation in $p_0$ around $p_0=0$, one can
compute the conductivity in the linear response approximation by
the Kubo formula, that is $\s = \lim_{\o\to 0}\lim_{\d\to 0}{\hat
D(-i \o+\d,0)\over -i \o+\d}$. Therefore, a nonvanishing $D$
indicates infinite conductivity.

The conservation law
\be\lb{eqm} {\partial \r^C_\xx\over \partial x_0}= e^{H x_0}
[H,\r_x] e^{-H x_0} =-i\dpr^{(1)}_x J_{\xx} \= -i
[J_{x,x_0}-J_{x-1,x_0}]\nn\;, \ee
where $\dpr^{(1)}_x$ denotes the lattice derivative, implies exact
relations, called {\it Ward identities} (WI), among the
Schwinger functions, the density correlations and the {\it vertex
functions}, defined as $G^{2,1}_{\r,\b,L}(\xx,\yy,\zz) = \la {\bf
T}\r^{(C)}_\xx a^-_\yy a^+_\zz\ra_{T,\b,L}$ and
$G^{2,1}_{j,\b,L}(\xx,\yy,\zz) = \la {\bf T} J_\xx a^-_\yy
a^+_\zz\ra_{T,\b,L}$. Some Ward Identities, which will play an
important role in the following, are
\bal -ip_0\hat G^{2,1}_{\r,\b,L}(\kk,\kk+\pp)-i(1-e^{-ip}) \hat
G^{2,1}_{j,\b,L}(\kk,\kk+\pp)
&= \hat S_2^{\b,L}(\kk) - S_2^{\b,L}(\kk+\pp)\lb{ref1}\\
-i p_0  \hat\O_{C,\b,L}(\pp) -i(1-e^{-ip})
\hat\O_{j,\r,\b,L}(\pp) &=0\lb{ref2}\\
-i p_0  \hat\O_{\r,j,\b,L}(\pp) -i (1-e^{-ip}) \hat D_{\b,L}(\pp)
&=0\lb{ref3} \eal
where $\O_{\r,j,\b,L}(\xx,\yy)=\la \r^C_{\xx} J_{\yy}\ra_{T,\b,L}$
and $\O_{j,\r,\b,L}(\xx,\yy)=\la J_{\xx} \r^C_{\yy}\ra_{T,\b,L}$.

In the previous paper \cite{BFM012_1} we have analyzed the model
\pref{1.1} in the repulsive non half-filled band case and we have
proved, see Theorem 1.1 of \cite{BFM012_1}, that the zero temperature
response functions \pref{ss111} in the thermodynamic limit decay at large distances with {\it
critical exponents}, which are non trivial functions of the coupling and are
denoted by $X_C$, $X_{S_i}$, $X_{SC}, X_{SC}, X_{TCi}$, see (1.26)
of \cite{BFM012_1}; finally $\h$ is the critical exponent of the
interacting propagator, see (1.25) \cite{BFM012_1}.

Such exponents, together with $\k$ and $D$, depend on all
microscopic details of the model, for instance the form of the two
body interaction or the chemical potential. Nevertheless, according to the
{\it Luttinger liquid conjecture} proposed by Haldane \cite{Ha080}
(extending previous ideas by Kadanoff \cite{KW071}, and Luther and
Peschel \cite{LP075}) such quantities, through model dependent,
are believed to satisfy a set of model independent relations. Such
relations are true in a special solvable spinless models, the {\it
Luttinger model}, whose solvability relies on the absence of the
spin and on the linear dispersion relation of the fermions, see
\cite{ML065}. The content of the conjecture is that several of the
relations valid in the Luttinger model are generically valid in a
wide class of systems describing 1D fermions with with {\it non
linear} dispersion relation and in presence of {\it spin}. The
following Theorem proves for the first time the validity of the
universal Luttinger liquid relations in a wide class of models
(including the 1D Hubbard model) of spinning fermions on a one
dimensional lattice with a generic short range interaction satisfying a special positivity
condition, in the non half filled band case.

\begin{theorem}\lb{th1.2} Given the Hamiltonian \pref{1.1},
if $\bar\m \not=0$ and $\hat v(2\arccos(\bar\m))>0$, there exists
$\l_0>0$ such that, if $0\le \l \le \l_0$, it is possible to find
a continuous function $p_F\equiv
p_F(\bar\m,\l)=\arccos(\bar\m)+O(\l)$ verifying the conditions
\be\lb{ma} p_F \not=0,\p/2,\p \virg \hat v(2 p_F) > 0\ee
so that there exist continuous functions
\be\lb{g1} K\=K(\l)= 1-c\l + O(\l^{2}),\quad\quad\bar K\=\bar
K(\l)= 1-c\l + O(\l^2)\ee
with $c=2[\hat v(0)-\hat v(2p_F)/2](\pi \sin p_F)^{-1}$, such that
\bd \item  the critical exponents satisfy the extended scaling
formulas
\be\bsp\lb{1.8}
4\h = K + K^{-1}-2\;,\qquad& 2X_{C} = 2X_{S_i}=K+1\;,\\
2X_{TC_i} = 2X_{SC}=K^{-1}+1\;,\qquad& 2\tilde X_{SC} = K +
K^{-1}\;; \esp \ee
\item the small momentum asymptotic behavior of the response functions is
\be\lb{den} \bsp \hat \O_{C}(\pp) &= {\bar K\over \pi
v}{v^2p^2\over p_0^2+v^2
p^2}+A(\pp)\\
\hat D(\pp) &= {v\over \pi}\bar K {p_0^2\over p_0^2+v^2 p^2}
+B(\pp) \esp \ee
with $A(\pp)$, $B(\pp)$ continuous and vanishing at $\pp=0$,
$v=\sin p_F+O(\l)$; therefore the Drude weight $D$ and the
susceptibility $\k$ are $O(\l)$ close to their free values and
verify the Luttinger liquid relation \be\lb{hh1} v^2=D/\k \ee \ed
\end{theorem}
The relations \pref{1.8} were proposed in the spinless case by
Kadanoff \cite{Kad077} and Luther and Peschel \cite{LP075}, and
imply the exact determination of all the other exponents from the
knowledge of a single one of them. The relation \pref{hh1},
proposed by Haldane \cite{Ha080}, gives exact formulas relating
the susceptibility and the Drude weight (connected to the {\it
amplitudes} of the response functions) to the charge velocity $v$.
The importance of such relations, in addition to the fact that
they are among the very few cases in which the basic principle of
universality in statistical mechanics can be rigorously
established, rely in the fact that they allow for predictions which
could be possibly experimentally verified in real anisotropic
materials.

In order to prove such properties we introduce an effective model verifying
some extra symmetries (which are only asymptotic in the lattice model) like
the Lorentz symmetry and {\it chiral} local phase invariance implying many
Ward Identities. In the model \pref{1.1} such symmetries are not true, and
there is a much smaller number of Ward Identities, given by \pref{ref1},
\pref{ref2}, \pref{ref3}, which are not sufficient by themselves to derive
the universal relations. The critical exponents and the thermodynamic
quantities of the effective model are related to the ones of the model
\pref{1.1}, provided that a suitable fine tuning of the parameters is done;
on the other hand the Ward Identities valid for the effective model imply
relations from which at the end \pref{1.8} and \pref{hh1} follow. This method
is a way to implement the concept of {\it emerging symmetries} in a rigorous
mathematical setting. This fine tuning is possible thanks to the fact that
both model are analyzed by Renormalization Group methods, and have a two
dimensional manifold of fixed points. The strategy we followed can provide
several extra information on the Hubbard model; an example is provided by the
content of App. D, in which a detailed expression of the 2-point function of
the Hubbard model is given, which looks in agreement with the so called {\em
spin-charge separation}, that is the conjectured property that the spin and
charge waves have different velocities.

\section{Renormalization Group Analysis of the Effective Model} \lb{ss2.5b}

\subsection{Definition of the effective model} \lb{ss2.5c}
The effective model which we introduce in order to prove Theorem
1.1 is not Hamiltonian and is defined directly in terms of {\it
Grassmann variables}. Given $L>0$, we consider the set $\DD'_L$ of space-time
momenta $\kk=(k,k_0)$, with $k={2\pi\over L}(n+{1\over 2})$ and
$k_0={2\pi\over L}(n_0+{1\over 2})$; with each $\kk\in \DD'_L$ we
associate eight {\it Grassmann variables} (sometimes also called
fields) $\hat\psi^+_{\kk,\o,s},\hat\psi^-_{\kk,\o,s}$ with
$\o=\pm$ a {\it quasi-particle} index and $s=\pm$ a {\it spin}
index. We also define (only {\em formally} for the moment, since the number of Grassmann variables
is infinite) the {\it Grassmannian field} as
\be\lb{2.1a} \psi^\pm_{\xx,\o,s}={1\over L^2}\sum_{\kk\in \DD'_L} e^{\pm
i\kk\xx} \hat\psi^\pm_{\kk,\o,s}\ee
where $\xx=(x,x_0) \in \L_L$, $\L_L$ being a two dimensional square torus
of size $L^2$. To shorten notation we will also denote
$\int_{\L_L} d\xx$ by $\int d\xx$;
moreover, in general we shall not stress the dependence on $L$ of the various quantities we shall
consider.

Given two integers $l$ and $N$ independent of $L$, such that $l \ll 0 \ll N$,
the {\it Generating Function} of the effective model with {\em ultraviolet cutoff} $\g^N$
and {\em infrared cutoff} $\g^l$ is the following Grassmann integral:
\be\lb{vv1} \bsp e^{\WW_{l,N}(J,\h)} &= \int\!
P_Z^{[l,N]}(d\psi)\; \exp \left\{ -V(\sqrt{Z}\psi)
+ \sum_{\o,s} \int \!d\xx\; J_{\xx,\o,s}
\psi^{+}_{\xx,\o,s} \psi^{-}_{\xx,\o,t}\right.\\
& + \left. \sum_{\o,s} \int\! d\xx\; \lft[\psi^{+}_{\xx,\o,s}
\h^-_{\xx,\o,s} + \h^+_{\xx,\o,s}
\psi^{-}_{\xx,\o,s}\rgt]\right\}\;, \esp \ee
where $J,\h$ are (respectively commuting and anticommuting) {\it
external fields}, $P_Z^{[l, N]}(d\psi)$ is the {\it Grassmann-valued
Gaussian measure} with propagator
$\d_{\o,\o'}\d_{s,s'}g^{[l,N]}_{{\rm D},\o}(\xx-\yy)$, where
\be\lb{gth} g^{[l,N]}_{{\rm D},\o}(\xx-\yy)={1\over Z}{1\over
L^2}\sum_{\kk\in \DD'_L}e^{-i\kk(\xx-\yy)}{\chi^\e_{l,N}(|\tilde
\kk|)\over -ik_0+\o c k} \virg \tilde\kk = (ck, k_0)\ee
$Z, c>0$ and $\chi^\e_{l,N}(t)$ is a smooth cut-off function defined for $t\ge 0$,
depending on a small positive parameter $\e\ge 0$ with the following properties. If $\e>0$,
it is strictly positive for all $t>0$, which reduce, as $\e\to 0$, to a compact support function
$\chi_{l,N}(t)$ equal to $1$ for $\g^{l}\le t \le
\g^{N}$ and vanishing for $t \le \g^{l-1}$ or $t \ge \g^{N+1}$.
The model is not really dependent of $\e$, since we plan to perform the limit $\e\to 0$
in the expressions we get for the correlation functions and the kernels of the effective potential,
at fixed values of $L$, $N$ and $l$.
\insertplot{300}{100} {} {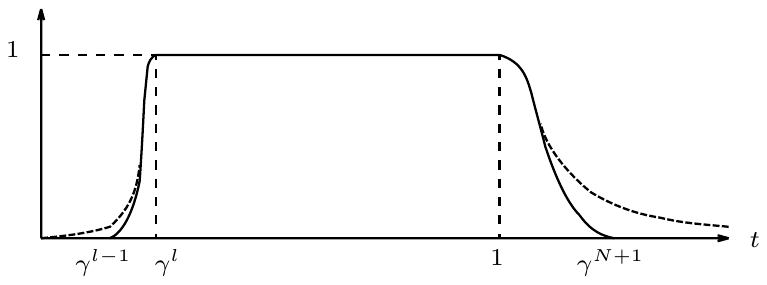} {The cut-off functions $\c^\e_{l,N}(|\tilde \kk|)$ (dashed line)
and $\c_{l,N}(|\tilde \kk|)$ (solid line)\lb{m21}}{0}

The interaction is
\be\lb{fs1} V(\ps)=g_{1,\perp} V_{1,\perp}(\ps)+ g_\arp
V_\arp(\ps) + g_\erp V_\erp(\ps) + g_4 V_4(\ps) \ee
with
\be\lb{five1} \bsp V_{1,\perp}(\ps) &= \frac12 \sum_{\o,s}\int
d\xx d\yy h_{L,K}(\xx-\yy)
\ps^+_{\xx,\o,s}\ps^-_{\xx,\o,-s} \ps^-_{\yy,-\o,s} \ps^+_{\yy,-\o,-s}\\
V_\arp(\ps) &= \frac12 \sum_{\o,s}\int d\xx d\yy h_{L,K}(\xx-\yy)
\ps^+_{\xx,\o,s}\ps^-_{\xx,\o,s}\ps^+_{\yy,-\o,s}\ps^-_{\yy,-\o,s}\\
V_\erp(\ps) &= \frac12 \sum_{\o,s}\int d\xx d\yy h_{L,K}(\xx-\yy)
\ps^+_{\xx,\o,s}\ps^-_{\xx,\o,s}\ps^+_{\yy,-\o,-s}\ps^-_{\yy,-\o,-s}\\
%
%
V_4(\ps) &= \frac12 \sum_{\o,s}\int d\xx d\yy h_{L,K}(\xx-\yy)
\ps^+_{\xx,\o,s}\ps^-_{\xx,\o,s}\ps^+_{\yy,\o,-s}\ps^-_{\yy,\o,-s}
\esp \ee
where $h_{L,K}(\xx)$ is defined in the following way. Let us fix an integer $K$
\footnote{it is possible but not advisable to choose $K=0$, because leaving
a generic $K$ will make clearer the
``power counting'' formulas: see for example Lemma \ref{t3.2}} and a smooth function $\hat h(\pp)$, defined
on $\RRR^2$ and rotational invariant, such that $|\hat h(\pp)|\le C e^{-\m |\pp|}$ for some
positive $C$ and $\m$, and $\hat h(0)=1$; moreover, let us call $\DD_L$ the set of space-time
momenta $\kk=(k,k_0)$, with $k={2\pi\over L}n$ and $k_0={2\pi\over L} n_0$. Then
\be\lb{defh} h_{L,K}(\xx) :={1\over L^2}\sum_{\pp\in \DD_L} \hat h(\g^{-K}\pp)
e^{i\pp\xx}\ee
Note that $h_{L,K}(\xx)$ is a smooth periodic function on $\L_L$, which converges, as
$L\to\io$, to a smooth function $h_K(\xx) = \g^{2K} h(\g^K\xx)$, $h(\xx)$ being
the Fourier transform of $\hat h(\pp)$; this function is fast decreasing on scale $\g^{-K}$.

Even if the properties of \pref{vv1} are largely independent of
the exact form of $\c^\e_{l,N}(t)$ and $\c_{l,N}(t)$, we find it convenient to choose them in the
following way. We introduce a function $\chi(t) \in C^\io(\RRR^+)$ such that $\chi(t)= 1$ if $t \le 1$
and $\chi(t)=0$ if $|t|\ge \g$;
then we define, for any integer $j$, $f_j(t)=\chi(\g^{-j} t)- \chi(\g^{-j+1}t)$ (hence the support of
$f_j(t)$ is contained in the interval $[\g^{j-1}, \g^{j+1}]$)  and we put
\be\lb{sumscale} \chi^\e_{l,N}(|\tilde \kk|)=\sum_{j=l}^N
f^\e_j(|\tilde \kk|)\ee
where $f_j^\e(t)=f_j(t)$, if $l+1 \le j \le N-1$, while $f^\e_l(t)$ and $f^\e_N(t)$ are obtained
by slightly modifying $f_l(t)$ and $f_N(t)$ in the following way, see Fig. \ref{m21}.
We put $f^\e_N(t) = f_N(t) + \e\D_N(\g^{-N}t)$, where $\D_N(t)$ is a Schwartz function with
support in $[1,+\io)$, such that $\D_N(t)>0$, if $t>1$.
Analogously, $f^\e_l(t) = f_l(t) + \e\D_l(\g^{-l}t)$,
where $\D_l(t)$ is a $C^\io$ function with support in $[0,1]$,
such that $\D_l(t)>0$, if $t\in(0,1)$.

In order to understand this definition, note that, if $\e=0$, the model is well defined,
since the family of Grassmann variables $\hat\psi^\pm_{\kk,\o,s}$, with
$\kk\in\tilde\DD'_L=\{\kk\in\DD'_L: \chi_{l,N}(|\tilde \kk|)>0\}$,
is finite, so that we can restrict the sum in \pref{gth} and \pref{2.1a} to the set $\tilde\DD'_L$
and we can write
\be\lb{2.8a} P_Z^{[l,N]}(d\psi) = \frac1{\NN} \exp\Big\{ -\frac{Z}{L^2}\sum_{\o,s\atop\kk\in\tilde\DD'_L}
\chi_{l,N}^{-1}(|\tilde \kk|) (-ik_0+\o c k) \hat\psi^+_{\kk,\o,s} \hat\psi^-_{\kk,\o,s} \Big\}
\prod_{\o,s\atop\kk\in\tilde\DD'_L} d\hat\psi^+_{\kk,\o,s} d\hat\psi^-_{\kk,\o,s}
\ee
where $\NN$ is a suitable
normalization constant. However, in the following we have to analyze the behavior of this measure
under the local gauge transformation $\psi^\pm_{\xx,\o,s}\to e^{\pm i\a_{\xx,\o,s}}
\psi^\pm_{\xx,\o,s}$ and this looks very difficult, since it is not
possible to give a useful representation of the measure \pref{2.8a} in terms of the
Grassmann field $\psi^\pm_{\xx,\o,s}$, even if it is now well defined, if we restrict in
\pref{2.1a} the sum to the set $\tilde\DD'_L$.
In order to solve this problem, we put $\e>0$ and, at the same time, in order to keep the number
of independent Grassmann variables finite, we introduce a {\em lattice cutoff}, by substituting the torus
$\L_L$ with a lattice of spacing $a$, such that $\g^{N+1} < \p a^{-1}$ and $L a^{-1}=2M$, $M$ integer.
We call $\L_L^a =\{\xx=(m a,m_0 a), m,m_0\in [-M,M-1]\}$ the lattice and we substitute everywhere
$\int d\xx$ with $\sum_{\xx\in\L_L^a} a^2$ and the set $\tilde\DD'_L$ with the set
$\tilde\DD'_{L,a} =\{\kk\in\DD'_L:|k|,|k_0|\le \p a^{-1}-\p L^{-1}\}$.

The outcome of this procedure is that the field $\{\psi^\pm_{\xx,\o,s}, \xx\in\L_L^a\}$ is related
to the field $\{\hat\psi^\pm_{\kk,\o,s}, \kk\in \tilde\DD'_L\}$, up to a constant, through the finite Fourier transform, so
that
\be\lb{FFT} d\psi := \prod_{\o,s,\kk\in\tilde\DD'_{L,a}} d\hat\psi^+_{\kk,\o,s} d\hat\psi^-_{\kk,\o,s}=
\frac1{\NN'}\prod_{\o,s,\xx\in\L_L^a} d\psi^+_{\xx,\o,s} d\psi^-_{\xx,\o,s}\ee
where $\NN'$ is a normalization constant; \pref{FFT} easily implies the
invariance of the {\em Grassmann-valued Lebesgue measure} $d\psi$ under the
local gauge transformation \pref{3.2}. Of course, after writing the Ward
identities following from the gauge invariance, we have to take the limit
$a\to 0$, followed by the limit $\e\to 0$, while keeping fixed $L$, $N$ and
$l$.

In agreement with these definitions, we shall define the Fourier transform of the external field $\h^\pm_{\xx,\o,s}$
through the analogous of \pref{2.1a}, while, for the external field $J_{\xx,\o,s}$, we shall use the definition
$J_{\xx,\o,s}=L^{-2}\sum_{\pp\in\tilde\DD_{L,a}} \hJ_{\pp,\o,s} e^{-i\pp\xx}$, where $\tilde\DD_{L,a}:=\{\pp\in\DD_L:
p,p_0\in [-\p/a, \p/a-2\p/L] \}$.
For the same reasons we modify the definition \pref{defh} of
the function $h_{L,K}(\xx)$, by restricting the sum over $\pp$ to the set $\tilde\DD_{L,a}$.

With this setup, as we shall prove, we can rigorously compute the correlations and the kernels of the effective
potentials at fixed values of $L$, $N$, $l$ and then perform the limit $a\to 0$, followed by the limit $\e\to 0$.

It is easy to see that the limit $a\to 0$ is essentially trivial; in fact, in the scale decomposition that we shall describe below,
the lattice spacing has a role only in two points:

\0a)  The bounds concerning the integration of the Grassmann variables with momenta larger then $\g^N$,
which has to be performed in a single step using the propagator $g_\o^{(N)}(\xx)$ defined as the r.h.s.
of \pref{gth} with $f_N^\e$ in place of $\chi^\e_{L,N}$, can not be uniform in $L$. The reason is that
we {\em do not} modify the free propagator, in order to make it periodic on the set $\tilde\DD'_{L,a}$, see \pref{2.8a};
it follows that boundary terms appear in the integration by part arguments which allow us to control the decreasing
properties in $\xx$ of $g_\o^{(N)}(\xx)$, see \pref{2.15} below. However, thanks to the
fast decreasing properties of the function $\D_N(t)$ introduced in the definition of $\tilde f^\e_N(|\tilde\kk|)$,
it is easy to see that  the boundary terms are negligible for $a$ small enough, so that the
dimensional bounds on $g_\o^{(N)}(\xx)$ are uniform
in $a\le \bar a_{L,N}$ and $\e\in [0,1]$, with $\bar a_{L,N}\to 0$ for $L,N\to \io$. This is not a source
of trouble, since we have to perform the limit $a\to 0$ at fixed values of $N$ and $L$.

\0b) The presence of the lattice introduces the so called {\em Umklapp terms}, when we write the kernels
of the effective potentials and the correlations in terms of Fourier transforms, a procedure that is
important in our analysis of the infrared scales. For example, our definitions
imply that $\sum_{\xx\in \L_L^a} J_{\xx,\o,s} \psi^+_{\xx,\o,s}
\psi^-_{\xx,\o,s} = L^{-2} \sum_{\pp\in \tilde\DD_{L,a}} \hJ_{\pp,\o,s} [\hr_\pp
+ \hr_{\pp+2\p a^{-1}} + \hr_{\pp-2\p a^{-1}}]$ with
$\hr_\pp = L^{-2} \sum_{\kk^+, \kk^-\in\tilde\DD'_{L,a}: \pp= \kk^+-\kk^-}
\hat\psi^+_{\kk^+,\o,s} \hat\psi^-_{\kk^-,\o,s}$. It is easy to see that the contribution of the
terms proportional to $\hr_{\pp+2\p a^{-1}}$ and $\hr_{\pp-2\p a^{-1}}$, as well of all similar Umklapp
terms does not qualitatively modifies the structure of our multiscale expansion for $a\le \bar a_{L,N}$ and
that all quantities of interest are well defined in the limit $a\to 0$, at fixed values of $L$ and $N$ and $\e\le 1$.

Hence, in the following we shall write the results of our calculations directly in the limit $a\to 0$; in
particular we shall use the symbols
$\DD'_L$ and $\DD_L$ in place of $\tilde\DD'_{L,a}$ and $\tilde\DD_{L,a}$, respectively.

As concerns the functions $\D_N(\g^{-N}|\tilde \kk|)$ and $\D_l(\g^{-l}|\tilde \kk|)$, their strict positivity has a role
only when we discuss the Ward Identities following from the local gauge invariance. This calculation
involves the expression in the exponent of \pref{2.8a}, which is very singular for $\e\to 0$, but, as we
shall see, the Ward Identities for the correlations and the kernels of the effective
potentials have a simple well defined limit, which contains very important terms,
which are at the origin of the anomalous critical exponents.

\*

Let us now give a look at the interaction.
The coupling $g_{1,\erp}$ has a special role:  if $g_{1,\perp}=0$
the model is invariant under the global phase transformation
\be\lb{2.7} \psi^\pm_{\xx,\o,s} \to e^{\pm i\a_{\o,s}}
\psi^\pm_{\xx,\o,s} \ee with the constant phase $\a_{\o,s}$ which
can depend both on $\o$ and $s$. Otherwise,  if
$g_{1,\perp}\not=0$, the spin invariance is broken; the invariance
is under the transformation \be\lb{2.8} \psi^\pm_{\xx,\o,s} \to
e^{\pm i\a_{\o}} \psi^\pm_{\xx,\o,s} \ee with the phase
independent of $s$.

For finite values of $N,l,L$ it is a consequence of the {\it
Brydges-Battle-Federbush} formula and of the Gram-Hadamard
inequality, see (2.43) and (2.47) of the companion paper
\cite{BFM012_1}, that the the functional integral \pref{vv1} is well
defined and analytic in the couplings $\vec g=(g_{1,\perp},
g_\arp, g_\erp, g_4)$ in a disk in the complex plane. In order to
get results valid in the limit of removed cut-off we need a
multiscale analysis.

It will turn out that the limit $N\to\io$ at $l$ or $L$ finite
(that is the solution of the {\it ultraviolet problem}) can
be controlled by only assuming that $\vec g$ is
small enough. The analysis of the ultraviolet problem is somewhat
similar to the one in \S 2.2, 2.3 of the companion paper
\cite{BFM012_1} for the ultraviolet problem of the model \pref{1.1};
in particular a tree expansion and a multiscale analysis are
necessary. However, in the case of the model \pref{1.1} the
lattice plays the role of an ultraviolet cut-off for spatial
momenta and this makes the problem much simpler. In the case of the
continuum model \pref{vv1}, on the contrary, there is no such
cut-off and the propagator \pref{gth} in the limit $N\to\io$ has a
low $O(\kk^{-1})$ decay for large \kk; power counting arguments
suggest that ultraviolet divergences, similar to those
appearing in $d=1+1$ quantum Field Theory models, could appear.
They are however avoided, as we will see, thanks to cancelations
and the {\it non-locality} of the interaction \pref{1.1}, see
lemma 2.2 below.

While the removal of the ultraviolet cut-off (at finite infrared
cut-off) only requires that the couplings are small, the removal
of the infrared cut-off is much more subtle and depends
critically on the values of the couplings
$g_{\erp},g_{1,\perp},g_\perp,g_4$; this reflects the fact that
different long distance decay properties of the correlations are
expected to depend on the nature of the
interaction. We will show that the infrared cut-off can be safely
removed in the following two cases:
\begin{enumerate}
\item the case $g_{1,\perp}=0$ \item the case
$g_\arp=g_{\erp}-g_{1,\perp}$ and $g_{1,\perp}>0$
\end{enumerate}

\subsection{The ultraviolet integration}\lb{ss2.2}

For simplicity, we shall put in \pref{vv1} $\h=0$ and, for notational convenience we write \pref{fs1} as
\be \lb{interac} V(\ps)= \frac12 \sum_{\Th,\Th'}\int\! d\xx
d\yy\; \ps^+_{\xx,\o,s}\ps^-_{\xx,\o,t}
h^{L,K}_{\Th,\Th'}(\xx-\yy)\ps^+_{\yy,\o',s'}\ps^-_{\yy,\o',t'} \ee
with $\Th=(\o,s, t)$,  $\Th'=(\o',s', t')$ and
$$
h^{L,K}_{\Th,\Th'}(\xx-\yy) =
\begin{cases}
-g_{1,\erp} h_{L,K}(\xx-\yy)  &  \text{for $\o'=-\o$ and $s=t'=-s'=-t$}\\
g_\arp h_{L,K}(\xx-\yy)  &  \text{for $\o'=-\o$ and $s=t=s'=t'$}\\
g_\erp h_{L,K}(\xx-\yy)  &  \text{for $\o'=-\o$ and $s=t=-s'=-t'$}\\
g_4   h_{L,K}(\xx-\yy)   &  \text{for $\o'=\o$ and $s=t=-s'=-t'$}\\
0 & \text{otherwise}
\end{cases}
$$
To exploit certain identities, we have to put in \pref{vv1} a more general
source term; hence, we substitute the term proportional to the $J$ field with
$\sum_\Th \int\! d\xx\; J_{\xx,\Th} \psi^{+}_{\xx,\o,s} \psi^{-}_{\xx,\o,t}$,
where again $\Th=(\o,s,t)$, and we define:
\be\lb{ep0} \VV(\psi,J)= \sum_{\Th} \int\! d\xx\; J_{\xx,\Th} \psi^{+}_{\xx,\o,s} \psi^{-}_{\xx,\o,t}
-V(\sqrt{Z}\psi)
\ee
 The graphical representation of the interaction is given  in Fig. \ref{z0q}.
\insertplot{450}{50} {} {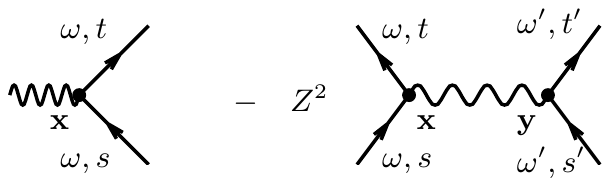}
{Graphical representation of \pref{ep0}\lb{z0q}}{0}

By using \pref{sumscale}, we can write
\be g^{[l,N]}_{D,\o}(\xx)=\sum_{j=l}^N g^{(j)}_{\o}(\xx) \ee
where $g^{(j)}_{\o}(\xx)$ is defined as $g^{[l,N]}_{D,\o}(\xx)$ with
$\chi^\e_{l,N}(|\tilde \kk|)$ replaced by $f^\e_j(|\tilde \kk|)$.
Therefore, for a positive constant $c_0$,
\be\lb{2.15}\bsp
\|g^{(k)}\|_{L_\io}:=\sup_{\xx,\o} |g^{(k)}_\o &(\xx)|\le c_0 \g^{k}
\virg \|g^{(k)}\|_{L_1} :=\max_{\o}\int\!d\xx\;|g^{(k)}_\o(\xx)|\le c_0 \g^{-k}\\
&\|g^{(k)}\|_{\tilde L_1}
:=\max_{\o}\int\!d\xx\; \|\xx\| \,|g^{(k)}_\o(\xx)| \le c_0 \g^{-2k}
\esp\ee
where $\|\xx-\yy\|$ is the distance on the torus, and, from \pref{defh}
\bal
\|h_{L,K}\|_{L_\io}:=\sup_{\xx}|h_{L,K}(\xx)|\le c_0  \g^{2K}&\virg
\|\dpr h_{L,K}\|_{L_1}:=\sup_{j=0,1}\int\!d\xx \;|\dpr_j h_{L,K}(\xx)| \le
c_0 \g^{K}
\notag\\
\|h_{L,K}\|_{L_1}&:=\int\!d\xx \; |h_{L,K}(\xx)|\le c_0 \eal
The scales $N,N-1,\ldots,K$ are called {\em ultraviolet scales}, while
the remaining ones are called {\em infrared scales}; if $\psi^{(j)}$, $j\ge
K$ is the field with propagator $g^{(j)}_{\o}(\xx)$ and we call $P_Z(d\psi^{(j)})$
the corresponding Grassmann measure, the
integration of the fields $\psi^{(N)}, \ldots, \psi^{(K)}$ is done
iteratively in the usual way, by using the decomposition
$P_Z(d\psi^{[l,N]}) \prod_{j=l}^K P_Z(d\psi^{(j)})$, where we have modified the previous
notation by writing $P_Z(d\psi^{[l,N]})$ in place of $P^{[l,N]}_Z(d\psi)$. After integrating the fields
$\psi^{(N)}, \psi^{(N-1)},...,\psi^{(h+1)}$, $h\ge K$, we get an expression
of the form:
\be \int
P_Z(d\psi^{[l,N]})e^{\VV(\psi^{[l,N]},J)}=e^{-L^2
E_h+S_h(J)}\int
P_Z(d\psi^{[l,h]})e^{\VV^{(h)}(\psi^{[l,h]},J)}\label{ssssaa}
\ee
where
\be\lb{2.17b}
\VV^{(h)}(\psi,J)=\sum_{m=1}^\io\sum_{n=0}^\io\sum_{\uTh,\uTh'} \int d\uz d\ux d\uy
W^{(n;2m)(h)}_{\uTh,\uTh'}(\uz;\ux,\uy) [\prod_{j=1}^n J_{\zz_j, \Th_j
}][\prod_{j=1}^m \ps^+_{\xx_j, \o'_j,s'_j} \ps^-_{\yy_j, \o'_j,t'
_j}] \ee
and $\uTh=(\Th_1, \ldots, \Th_n)$, $\uTh'=(\Th'_1, \ldots,
\Th'_m)$, $\Th'_j=(\o'_j,s'_j,t'_j)$.

\0{\bf Remark} - The compact support properties of the single scale propagators $g_\o^{(j)}$ imply that\\
$\VV^{(h)}(\psi^{[l,h]},J)$ depends on $N$ and $L$, but is independent of the IR cutoff $l$, if $l<K$.

The integration of the field $\psi^{(j)}$
is done after resumming the marginal terms, which is done by rewriting the r.h.s. of \pref{ssssaa} as
\be e^{-L^2 E_h+S_h(J)}\int P_Z(d\psi^{[l,h-1]}) \int P_Z(d\psi^{(h)}) e^{\LL
V^{(h)}(\psi^{[l,h]},J)+\RR V^{(h)}(\psi^{[l,h]},J)}\label{ssss2} \ee
where $\RR=1-\LL$ and $\LL$ is a linear operation acting on the kernels of
$\VV^{(h)}$ so that
\be\lb{lcz} \LL W^{(n;2m)(h)}(\uz;\ux,\uy):=
\begin{cases}
\displaystyle
W^{(n;2m)(h)}(\uz;\ux,\uy)&\text{ if $n+m\le 2$}\\
0 &\text{ otherwise}
\end{cases}
\ee
%
%
%
%
Therefore, if we define, as in \S2.3 of \cite{BFM012_1}, $e_h=E_h-E_{h+1}$ and $s_h(J)=\SS_h(J)-\SS_{h+1}(J)$,
then $-L^2 e_h+s_h(J) + \VV^{(h)}(\psi,J)$ can be expressed as a tree expansion very similar
to the one used in Lemma 2.2 of the companion paper \cite{BFM012_1}
that here we briefly recall.

\insertplot{450}{150}{}%
{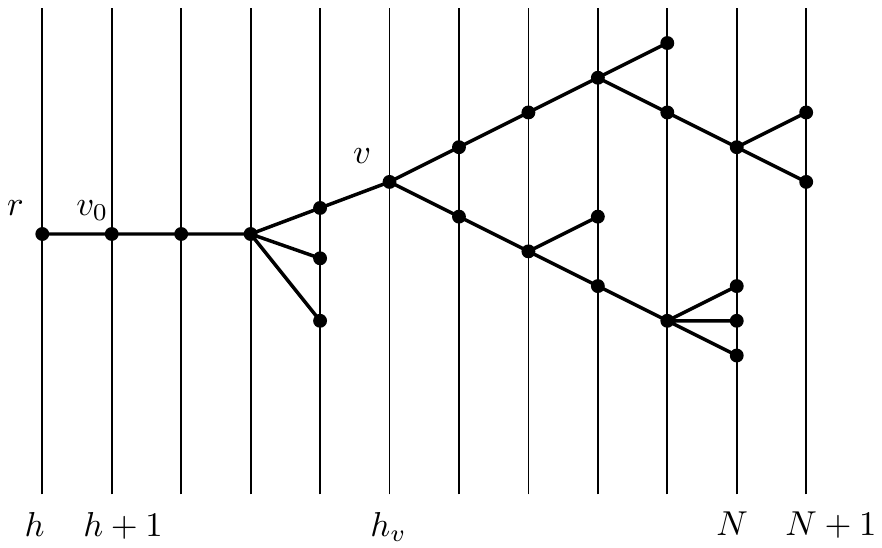}{A renormalized tree appearing in the graphic
representation of $\VV^{(k)}$\lb{h2}}{0}

Let us consider the family of all trees which can be constructed
by joining a point $r$, the {\it root}, with an ordered set of
$\bar n\ge 1$ points, the {\it endpoints} of the {\it unlabeled
tree}, so that $r$ is not a branching point. $\bar n$ will be
called the {\it order} of the unlabeled tree and the branching
points will be called the {\it non trivial vertices}. The
unlabeled trees are partially ordered from the root to the
endpoints in the natural way; we shall use the symbol $<$ to
denote the partial order. Two unlabeled trees are identified if
they can be superposed by a suitable continuous deformation, so
that the endpoints with the same index coincide. It is then easy
to see that the number of unlabeled trees with $\bar n$ end-points
is bounded by $4^{\bar n}$. We shall also consider the set
$\TT_{N,h,n_g,n_J}$ of the {\it labeled trees} with $n_g+n_J$ endpoints
(to be called simply trees in the following); they are defined by
associating some labels with the unlabeled trees, as explained in
the following items.

\0 1) We associate a label $J$ or $g$ to each endpoint, so that
there are $n_g$ endpoints with label $g$, to be called {\em normal
endpoints}, and $n_J$ endpoints with label $J$ to be called {\em
special  endpoints}.

\0 2) We associate a label $h\le N$ with the root. Moreover, we
introduce a family of vertical lines, labeled by an integer taking
values in $[h,N+1]$, and we represent any tree
$\t\in\TT_{N,h,n_g,n_J}$ so that, if $v$ is an endpoint or a non
trivial vertex, it is contained in a vertical line with index
$h_v>h$, to be called the {\it scale} of $v$, while the root $r$
is on the line with index $h$. In general, the tree will intersect
the vertical lines in set of points different from the root, the
endpoints and the branching points; these points will be called
{\it trivial vertices}. The set of the {\it vertices} will be the
union of the endpoints, of the trivial vertices and of the non
trivial vertices; note that the root is not a vertex. Every vertex
$v$ of a tree will be associated to its scale label $h_v$,
defined, as above, as the label of the vertical line whom $v$
belongs to. Note that, if $v_1$ and $v_2$ are two vertices and
$v_1<v_2$, then $h_{v_1}<h_{v_2}$.

\0 3) There is only one vertex immediately following the root,
which will be denoted $v_0$; its scale is $h+1$. If $v_0$ is an
endpoint, the tree is called the {\em trivial tree}; this can
happen only if $n_g+n_J=1$.

\0 4) Given a vertex $v$ of $\t\in\TT_{N,h,n_g,n_J}$ that is not an
endpoint, we can consider the subtrees of $\t$ with root $v$,
which correspond to the connected components of the restriction of
$\t$ to the vertices $w\ge v$. If a subtree with root $v$ contains
only $v$ and one endpoint on scale $h_v+1$, it will be called a
{\it trivial subtree}.

\0 5) Given an end-point, the vertex $v$ preceding it is surely a
non trivial vertex, if $n_g+n_J>1$.

\vspace{.3cm}

Our expansion is build by associating a value to any tree $\t\in
\TT_{N,h,n,m}$ in the following way. First of all, given an
endpoint $v\in\t$ with $h_v=N+1$, we associate to it one of the
terms contributing to the potential in \pref{ep0}, while, if
$h_v\le N$, we associate to it one of the terms appearing in

\bal  &\sum_{\Th,\o'} \int d\xx d\xx d\yy W^{(1;2)(h_v)}_{\Th,(\o',s,t)}(\zz;\xx,\yy)
J_{\zz, \Th} \ps^{+(<h_v)}_{\xx, \o',s}\ps^{-(<h_v)}_{\yy,\o',t}+
\sum_{\o,s}\int d\xx d\yy
W^{(0;2)(h_v)}(\xx,\yy) \ps^{+(h_v)}_{\xx, \o,s}\ps^{-(h_v)}_{\yy,\o,s}\nn\\
&\hspace{1cm}+\sum_{\underline\o,\underline s} \int d\xx_1 d\xx_2 d\xx_3
d\xx_4 W^{(0;4)(h_v)} \ps^{+(h_v)}_{\xx_1,
\o_1,s_1}\ps^{-(h_v)}_{\xx_2, \o_2,s_2} \ps^{+(h_v)}_{\xx_3,
\o_3,s_3}\ps^{-(h_v)}_{\xx_4, \o_4,s_4} \lb{ref}
\eal
All these possible choices will be distinguished by a
label $\a$ in a set $A_\t$, depending on $\t$. Finally, the operator $\RR$
is associated with each non-trivial vertex $v$.

The previous definitions imply that the following iterative equations are satisfied:
\be \VV^{(h)}( \psi^{(\le h)},J) + s_h(J)-L^2 e_h=
\sum_{n=1}^\io \sum_{\t\in\TT_{N,h,n_g,n_J}\atop \a\in A_\t}  {\cal
V}^{(h)}(\t,\a, \psi^{(\le h)},J)\;,\lb{B.12}\ee
where, if $v_0$ is the first vertex of $\t$ and
$\t_1,\ldots,\t_s$, $s\ge 1$, are the subtrees with root in $v_0$,
\be\lb{B.13} \bsp &{\cal V}^{(h)}(\t,\a, \psi^{(\le h)},J)=\\
{1\over s!} \EE^T_{h+1} &\big[\bar{\cal V}^{(h+1)}(\t_1,\a_1, \psi^{(\le h+1)},J);
\ldots;\bar{\cal V}^{(h+1)} (\t_{s},\a_s,
 \psi^{(\le
h+1)},J)\big]\esp\ee
where $\bar{\cal V}^{(h+1)}(\t_i,\a_i,\psi^{(\le h+1)},J)$ is equal to
$\RR {\cal V}^{(h+1)}(\t_i,\a_i,\psi^{(\le h+1)},J)$ if the subtree
$\t_i$ contains more than one end-point (that is $v$ is a non
trivial vertex), otherwise it is given by one of the terms in \pref{ep0}, if $h_v=N+1$,
or one of the terms in \pref{ref}, if $h_v\le N$.

Note that, by the definition
\pref{lcz}, the tree value can be different from $0$ only if, for any non trivial vertex $v\not=v_0$,
$|P_v|/2+n^J_v>2$, where $P_v$ is the number of $\psi$ variables not yet contracted in the vertex $v$
(see App. \ref{appA} for a precise definition) and $n^J_v$ is the number of $J$-endpoints of the subtree with root $v$.
Note also that $|P_v|$ has to be positive for any $v\not= v_0$ and that $P_{v_0}=0$ for the trees contributing
to $e_h$ and $s_h(J)$; in the following, we shall denote by $W^{(0;0)(h)}$ and $W^{(n;0)(h)}_{\uTh}(\uz)$
the corresponding kernels, while $\tilde W^{(0;0)(h)}= \sum_{j=h}^N W^{(0;0)(j)}$ and
$\tilde W^{(n;0)(h)}_{\uTh}(\uz)= \sum_{j=h}^N W^{(n;0)(j)}_{\uTh}(\uz)$ will be
the kernels of $E_h$ and $\SS_h(J)$

By \pref{B.12} and \pref{B.13} it is straightforward to verify,
see App. A, that the kernels $W^{(n;2m)(h)}(\uz;\ux,\uy)$ in
\pref{2.17b}, $h>K$ are represented by integrals of power series expansions in
the running coupling functions $W^{(0;2)(k)}
,W^{(1;2)(k)},W^{(0;4)(k)}$ with $k>h$. Consider now the $L^1$
norm
\be\lb{norm} \|W^{(n;2m)(k)}\| :=\max_{\uTh,\uTh'} \frac{1}{L^2}
\int\! d\uz d\ux d\uy\; \lft|W^{(n;2m)(k)}_{\uTh;\uTh'}
(\uz;\ux,\uy)\rgt| \ee
Of course, since the kernels may contain delta functions, we
extend, as usual, the definition of the $L^1$ norm by treating the
delta as positive functions. We also define
\bal
&w^{(1;2)(h)}_{\Th,\Th'}(\zz;\xx,\yy):= \d_{\Th,\Th'}\d(\zz-\xx)
\d(\zz-\yy)\lb{2.23}\\
&w^{(0;4)(h)}_{\Th,\Th'}(\xx_1,\xx_2, \yy_1,\yy_2)
:= Z^2 h^{L,K}_{\Th,\Th'}(\xx_1-\xx_2) \d(\xx_1-\yy_1) \d(\xx_2-\yy_2) \eal
that are equal to $W^{(1;2)(h)}_{\Th,\Th'}$ and
$W^{(0;4)(h)}_{\Th,\Th'}$ when $h=N$.
%
\begin{theorem}\lb{t2.1} Let $K\le k\le N$ and assume that
\be \sup_{h>k} \lft[ \g^{-h}\|W^{(0;2)(h)}\|+\|W^{(0;4)(h)}\| \rgt] \le
\e_0\quad\quad \sup_{h>k}\|W^{(1;2)(h)}\|\le 2\label{ind}\ee
with $\e_0$ independent of $k,L,N$; there exists a
constant $\bar\e$ such that, if $\e_0\le \bar \e$ then, for a
suitable constant $C$,
\be\lb{pc1} \|W^{(n;2m)(k)}\| \le \e_0^{d_{n,m}} C^{n+m}
\g^{k(2-n-m)} \ee
where $d_{n,m}=\max(m-1,0)$ if $n>0$, and $d_{n,m}=\max(m-1,1)$ if
$n=0$. Moreover the limits $\lim_{N\to\io} W^{(n;2m)(k)}_{\Th,\Th'}$ do exist
and are reached uniformly in $N$.
\end{theorem}

The proof of  this theorem is very similar to the proof of
Lemma 2.2 of the companion paper \cite{BFM012_1}, and is in appendix
\ref{appA}. Note however a crucial difference; in the present case the
scaling dimension, which can be read from the r.h.s. of
\pref{pc1}, is $n+m-2$; this explains why, according to
\pref{lcz}, we renormalize the kernels in $\VV^h$ such that $n+m\le 2$,
that is those which are the only ones with $m>0$ and scaling dimension $\le 0$.

Note that Theorem \ref{t2.1} implies a bound also on the $L^1$ norm of the kernels
$\tilde W_\uTh^{(n;0)(k)}(\uz)= \sum_{j=k}^N W_\uTh^{(n;0)(j)}(\uz)$.
This bound is finite uniformly in $N$ for $n\ge 3$,
while it is divergent for $N\to\io$, if $n\le 2$. The cases $n=0$ and $n=1$
do not give any trouble; in fact $W^{(1,0)(h)}=0$, because of the oddness of the free propagator, while
the divergence of the free energy is not a problem in a QFT model. As concerns $\|\tilde W^{(2;0)(k)}\|$,
the logarithmic divergence for $N\to\io$ of its bound is a problem in the following analysis.
However, by using the symmetry $g_\o^{[k,N]}(x,x_0)=-i\o g_\o^{[k,N]}(-x_0,x)$, one can see
that $\int d\zz d\zz' \tilde W_{\Th,\Th'}^{(2;0)(k)}(\zz,\zz')=0$;
this identity (or better its validity for the term of order $0$ in $\vec g$)
will be a crucial ingredient in the proof of Lemma \ref{t3.2} below, together with
the non locality of the interaction, which allows us to improve some bounds
by substituting the $L_\io$ norm of the propagator with its $L_1$ norm. In this
way we will be able to verify the assumptions \pref{ind} on the
running coupling functions by improving their dimensional bound.

\begin{lemma}\lb{t3.2}
There exist positive constants $C_1,C_2,C_3$, such that, if $\bar
g=\max\{|g_{1,\erp}|,|g_\arp|, |g_\erp|, |g_4|\}$ is small enough and if
$K\le h\le N$, then
\bal \lb{hb1}  \|W^{(0;2)(h)}\| &\le
C_1\bar g\g^{h}\g^{-2(h-K)}&\\
\lb{hb2} \|W^{(1;2)(h)}- w^{(1;2)(h)}\|
&\le C_2\bar g \g^{-(h-K)}&\\
\lb{hb3} \|W^{(0;4)(h)}- w^{(0;4)(h)}\| &\le C_3 \bar g \e_0
\g^{-(h-K)}&\eal
\end{lemma}
By the above lemma we see that we can choose $\vec g$ so small that Theorem
\ref{t2.1} holds and the bound \pref{pc1} is true.
Note that the crux of the above bounds is the
fact that $C_1$, $C_2$ and $C_3$ are independent of the scale $h$. In other
words,  we have standard ``power counting'' bounds even though we do not
study beta functions of marginal or relevant terms: that is obtained by the
cancelations that we will provide in the proof below.  For $K\le h\le N$, the
factor(s) $\g^{-(h-K)}$ do not play any decisive role, and in applications
will be bounded by 1. However, note that they are unbounded if, instead,
$h<K$: this explains why this Lemma is stated only for $K\le h\le N$.
\vskip.1cm

{\bf\0Proof of Lemma 2.2} The proof
is by induction. We assume that the bounds
\pref{hb1}-\pref{hb3} hold for $k+1\le h\le N$ (for $h=N$ they are true with
$C_1=C_2=C_3=0$); then we can use
Theorem \ref{t2.1} to have \pref{pc1} on scale $k$, assuming that
$\bar g\le \bar\e/\bar C$ where $\bar C$ is the minimum among
$C_1,C_2,C_3$; finally, we prove that \pref{hb1}-\pref{hb3} holds
for $h=k$.
Note that the definition of the kernels (see remarks after \pref{B.13}) imply that, if $m>0$ and $n\ge 0$,
\be\lb{2.29} \bsp &W^{(n;2m)(h)}_{\uTh;\uTh'} (\uz;\ux,\uy)
:=\left.\prod_{j=1}^n\frac{\dpr}{\dpr J_{\zz_j, \Th_j }}\times
\prod_{j=1}^m\frac{\dpr^2} {\dpr \ps^+_{\xx_j, \o'_j,s'_j}\dpr
\ps^-_{\yy_j, \o'_j,t'_j}} \VV^{(h)}(\sqrt{Z}\ps^{[l,h]}, J)\right|_{\psi=J=0}\\
&\tilde W^{(n+1;0)(h)}_{\uTh} (\uz)
:=\left.\prod_{j=1}^{n+1}\frac{\dpr}{\dpr J_{\zz_j, \Th_j }} S_h( J)\right|_{J=0}
\virg \tilde W^{(0.0)(h)}=E_h
\esp \ee
Let us now put $\bar V^{(h)}(\psi,J) = \VV^{(h)}(\psi,J) - L^2 E_h + \SS_h(J)$.
We will make repeated use of the following two identities. The first one,
graphically represented in Fig. \ref{z8q}, says that, for any $\Th=(\o,s,t_0)$,
\bal &{\dpr \VV^{(k)}\over \dpr \ps^+_{\xx,\o,s}}(\ps,J)
=\sum_{t_0}J_{\xx,\Th}\ps_{\xx,\o,t_0}^-
+\sum_{t_0}J_{\xx,\Th}\int\! d\uu\
 g^{[k+1,N]}_\o(\xx-\uu){\dpr \VV^{(k)}\over \dpr \ps_{\uu,\o,t_0}^+}(\ps,J)
\lb{prc1}
\notag\\
&+\sum_{t_0,\Th_1}\int\!d\ww d\uu\
h^{L,K}_{\Th,\Th_1}(\xx-\ww)g^{[k+1,N]}_\o(\xx-\uu)\lft[
{\dpr^2\VV^{(k)}\over\dpr J_{\ww,\Th_1} \dpr \ps^+_{\uu,\o,t_0}}
+{\dpr \VV^{(k)}\over\dpr J_{\ww,\Th_1}} {\dpr \VV^{(k)}\over\dpr
\ps^+_{\uu,\o,t_0}}\rgt](\ps,J)
\notag\\
&+\sum_{t_0,\Th_1}\ps_{\xx,\o,t_0}^-\int\!d\ww\
h^{L,K}_{\Th,\Th_1}(\xx-\ww)
 {\dpr \bar V^{(k)}\over\dpr
J_{\ww,\Th_1}}(\ps,J)\;, \eal
\insertplot{400}{140}{}%
{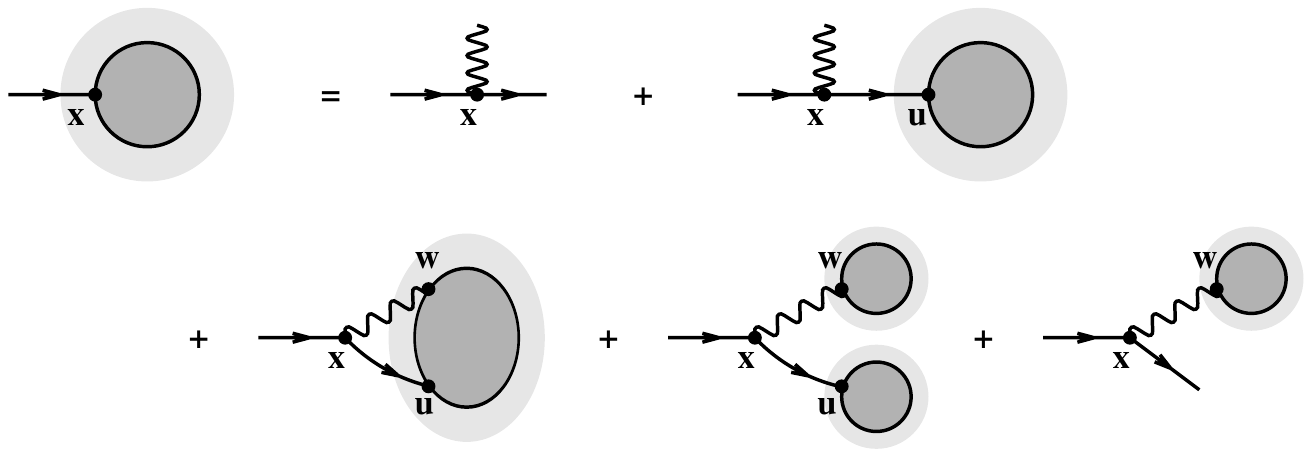}{Graphical representation of \pref{prc1}. The wiggly
line and the solid arrow between two
points represent $h^{L,K}_{\Th, \Th'}$ and $g_{\o}^{[k+1,N]}$, respectively;  the external wiggly
line and the external arrow represent $J_{\xx,\Th}$ and
$\ps^{[l,k]+}_{\xx,\o,s}$, respectively; the blobs represent the derivatives of
$\bar V^{(k)}$ with respect to their external fields, with a 'halo' that reminds the possibility of
taking other derivatives w.r.t. $J$ or $\ps$ fields.\lb{z8q}}{0}

The second identity, graphically represented in Fig. \ref{z8a}, says that
\bal &{\dpr \VV^{(k)}\over \dpr J_{\xx,\Th}}(\ps,J)
=\ps^+_{\xx,\o,s}\ps^-_{\xx,\o,t_0}+\int\!d\uu\
g^{[k+1,N]}_\o(\xx-\uu)\lft[ \ps^+_{\xx,\o,s}{\dpr \VV^{(k)}\over
\dpr \ps^+_{\uu,\o,t_0}}- {\dpr \VV^{(k)}\over \dpr
\ps^-_{\uu,\o,s}}\ps^-_{\xx,\o,t_0} \rgt](\ps,J) \cr\cr
&+\int\!d\uu d\uu'\
g^{[k+1,N]}_\o(\xx-\uu)g^{[k+1,N]}_\o(\uu'-\xx) \lft[ {\dpr^2
\VV^{(k)}\over \dpr \ps^+_{\uu,\o,s}\dpr \ps^-_{\uu',\o,t_0}}+
{\dpr\VV^{(k)}\over \dpr \ps^+_{\uu,\o,s}} {\dpr\VV^{(k)}\over\dpr
\ps^-_{\uu',\o,t_0}}\rgt](\ps,J)\;. \lb{prc2} \eal
\insertplot{430}{140}
{}%
{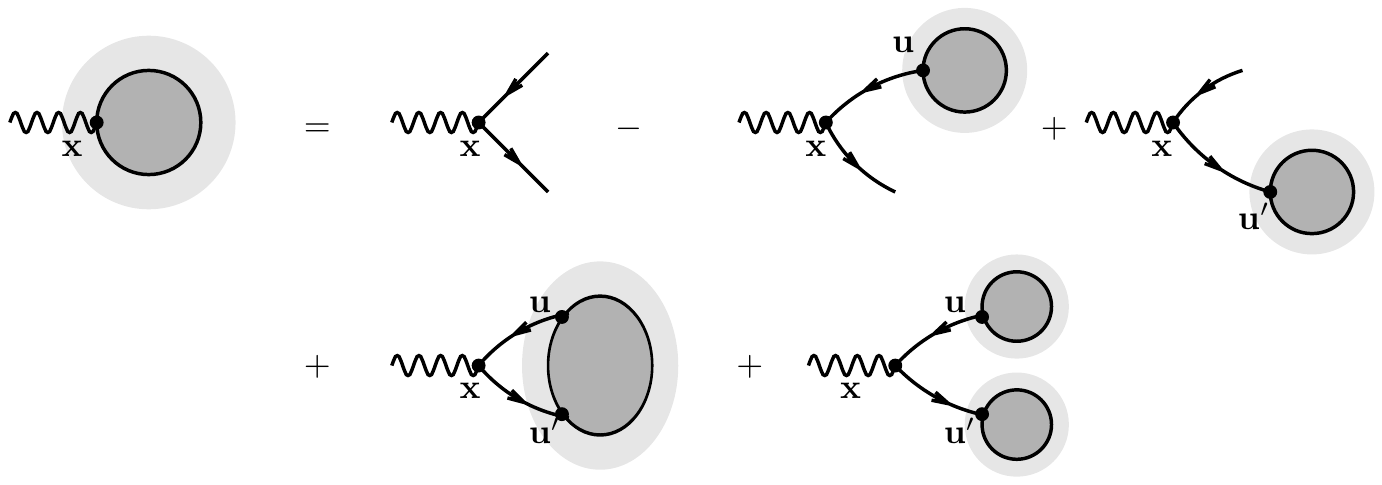}{Graphical representation of \pref{prc2}. The notation is
the same as in Fig. \ref{z8q}\lb{z8a}}{0}

A proof of these identities is given in appendix \ref{appB}. Let us show
that \pref{pc1} can be improved to \pref{hb1}-\pref{hb3} case by
case.\*

{\it\0a) Bound for $W^{(0;2)(k)}_{\Th}$.}
If we take in \pref{prc1} one derivative w.r.t. $\psi^-_{\yy,\o,t}$ and, after that,
we put $J=\psi=0$,  we obtain an identity for $W^{(0;2)(k)}_{\Th}(\xx,\yy)$, $\Th=(\o,s,t)$,
which, since $W^{(1;0)}_\Th({\bf 0})=0$ by  the oddness of the fermion
propagator, reads
\be \lb{111b} W^{(0;2)(k)}_{\Th}(\xx-\yy) =\sqrt{Z}
\sum_{t_0,\Th'}\int\!d\ww d\ww'\; h^{L,K}_{(\o,s,t_0),\Th'}(\xx-\ww)
g_\o^{[k+1,N]}(\xx-\ww')W^{(1;2)(k)}_{\Th';(\o,t_0,t)}(\ww;\ww',\yy)\;,
\ee
where $\Th=(\o,s,t)$.
\insertplot{450}{55}{}%
{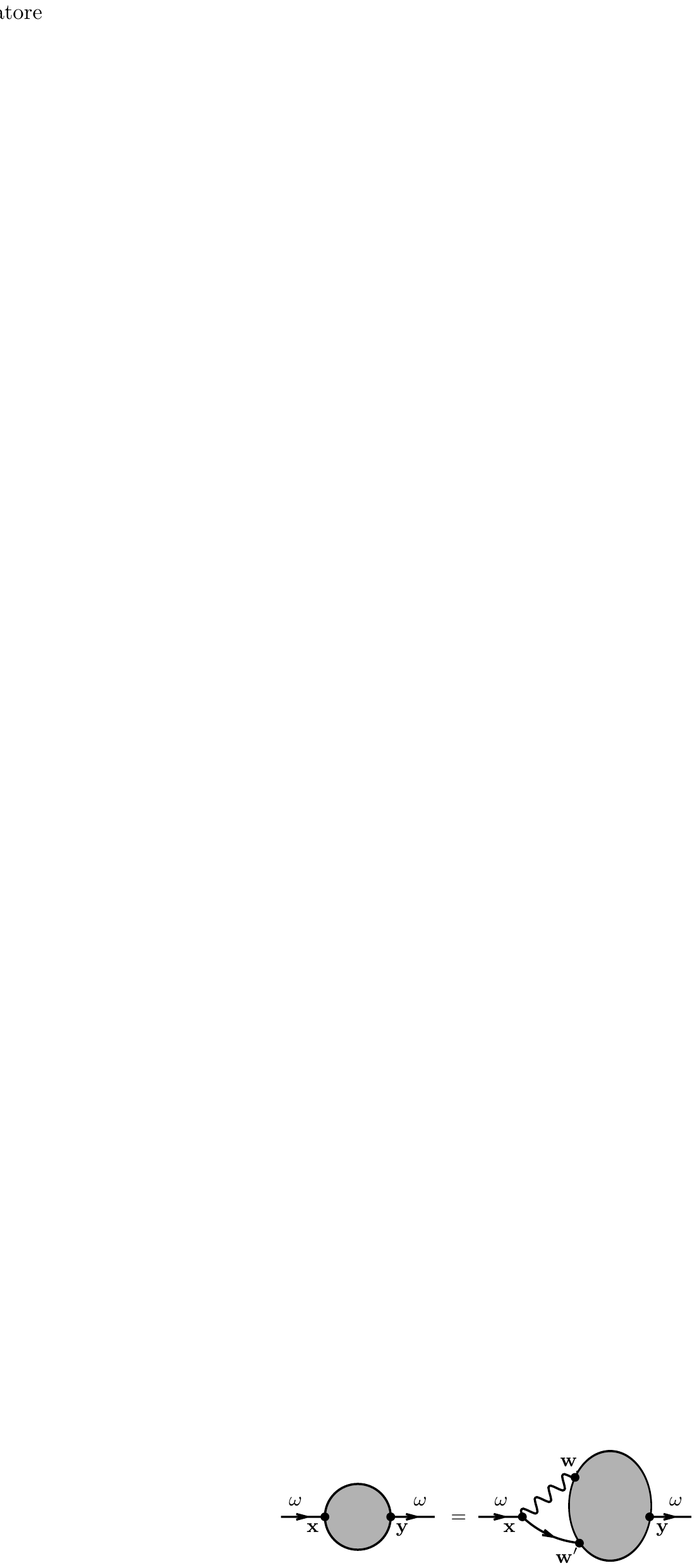}{Graphical representation of \pref{111b}; the gray
blobs represent $W^{(0;2)(k)}_{\Th}$ and
$W^{(1;2)(k)}_{\Th';(\o,t_o,t)}$; the solid lines are fermion propagators
$g_\o^{[k+1,N]}$ or external lines; the wiggly line is
$h^{L,K}_{\Th,\Th'}$.\lb{z1q}}{0}
We now bound $h_K$ by its $L_\io$-norm, the fermion propagator by its $L_1$-norm
and $W^{(1;2)(k)}$ by \pref{pc1},
$\|W^{(1;2)(k)}\| \le C^2$. We get
\bal && \|W^{(0;2)(k)}_\Th\| \le 16\bar g \|h_{L,K}\|_{L_\io}
\|W^{(1;2)(k)}\| \sum_{j\ge k+1} \|g^{(j)}\|_{L_1}\le \bar g C_1
\g^{k} \g^{-2(k-K)}\label{233}\;, \eal
This proves \pref{hb1}.\*

{\it\0b) Bound for $W^{(1;2)(k)}$.}
If we take in \pref{prc1} two derivatives w.r.t. $\psi^-_{\yy,\o,t}$ and
$J_{\zz,\Th'}$ and we put $J=\psi=0$, we can decompose
$W^{(1;2)(k)}_{\Th';\Th}-w^{(1;2)(k)}_{\Th';\Th}$ into the four
terms in Fig.\ref{z2q}.
\insertplot{410}{150}{}%
{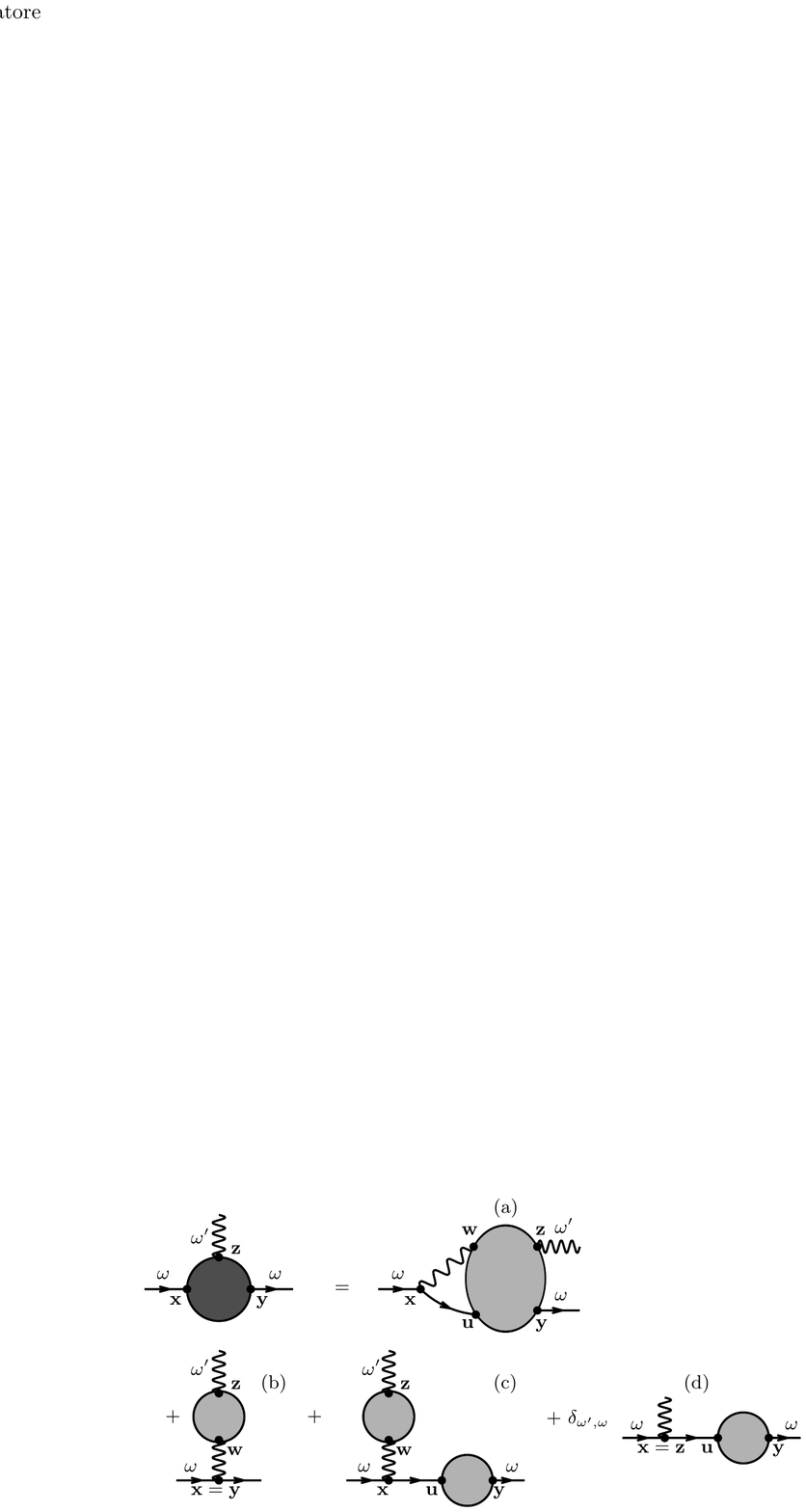}{\lb{z2q}: Graphical representation of the decomposition of
$W^{(1;2)(k)}_{\Th';\Th}$; the darker bubble represents the difference
$W^{(1;2)(k)}_{\Th';\Th}-w^{(1;2)(k)}_{\Th';\Th}$; the other bubbles
represent the kernels in \pref{2.29}; the internal vertices are integrated.}{0}

Consider first the graph $(a)$. As in the previous case, take the
$L_1$-norm of the fermion propagator and the $L_\io$-norm of the
function $h^{L,K}_{\Th,\Th'}$; then use for  $W^{(2;2)(k)}$ the bound \pref{pc1}, that is
$\|W^{(2;2)(k)}\| \le C^3 \e_0 \g^{-k}$. We get:
\bal \|W^{(1;2)(k)}_{(a)}\| \le 16\bar g \|h_{L,K}\|_{L_\io}
\|W^{(2;2)(k)}\| \sum_{j\ge k+1} \|g^{(j)}\|_{L_1} \le {C_2\over
4} \bar g \g^{-2(k-K)}\;. \eal
Analogously, by using \pref{233}, we get for the graph (d) the bound
\bal \|W^{(1;2)(k)}_{(d)}\| \le 2
\|W^{(0;2)(k)}\| \sum_{j\ge k+1} \|g^{(j)}\|_{L_1} \le {C_2\over
4} \bar g \g^{-2(k-K)}\;. \eal

Let us now consider the terms (b) and (c). In these cases the previous procedure,
based on the non locality of the interaction, which allows us to improve the bound
by substituting the $L_\io$ norm of the propagator with its $L_1$ norm, can not be directly applied
and we have to face the problem that the trivial bound is proportional to $\|\tilde W^{(2;0)(k)}\|$,
which is divergent for $N\to\io$ (see the remarks after Theorem \ref{t2.1}).
We shall bypass this difficulty, by exploiting the fact
that $\int d\zz d\zz' \tilde W_{\Th,\Th'}^{(2;0)(k)}(\zz,\zz')=0$ in the following way.
Let us consider first the term (b) and note that
$$W^{(1;2)(k)}_{(b)\Th';\Th}(\zz;\xx,\yy)=\d(\xx-\yy)Z
\sum_{\Th''} \int d\ww \tilde W_{\Th',\Th''}^{(2;0)(k)}(\zz,\ww) h^{L,K}_{\Th'',\Th}(\ww-\xx)$$

If we take one derivative w.r.t. $J_{\zz,\Th'}$ in \pref{prc2}
and then we put $\psi=J=0$, we get a representation of $\tilde W_{\Th',\Th''}^{(2;0)(k)}(\zz,\ww)$,
that we insert in the previous expression; the result can be graphically
represented as in the left side of the first line in Fig. \ref{z3q}. We further expand this expression
by substituting the kernel $W^{(1;2)(k)}_{(b)\Th';\Th''}(\zz;\uu',\uu)$ with its expansion in
Fig. \ref{z2q}; the result is represented in Fig.\ref{z3q}.
\insertplot{400}{170} {} {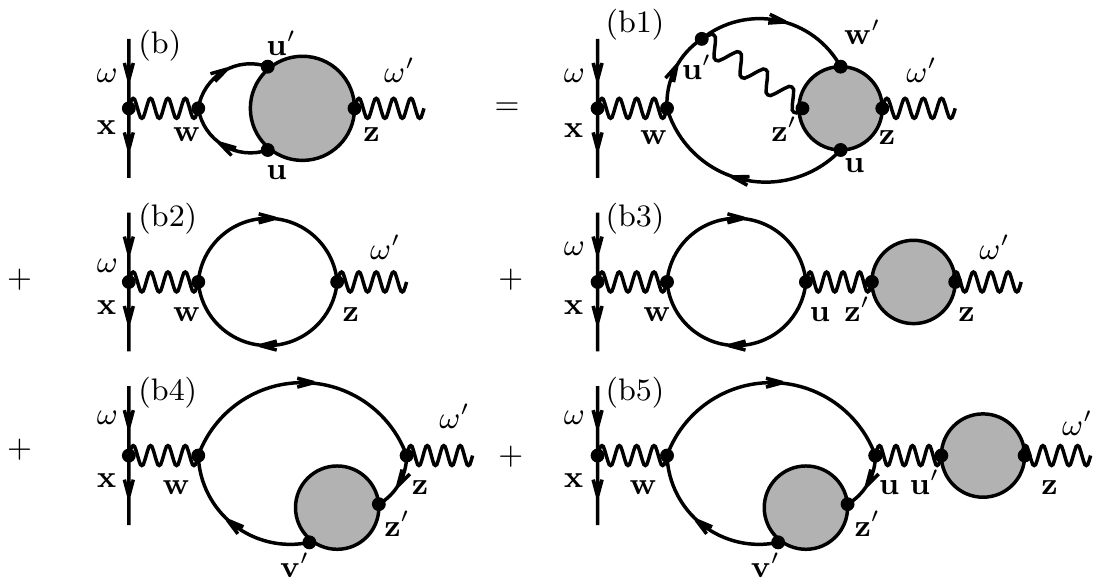}{Graphical representation
of the term $(b)$ in the decomposition in Fig.\ref{z2q}\lb{z3q}}{0}

Consider the term $(b1)$. The new estimation procedure is now
slightly more elaborated; decompose the three propagators
$g^{[k+1,N]}_{\o''}(\ww-\uu')$, $g^{[k+1,N]}_{\o''}(\uu-\ww)$ and
$g^{[k+1,N]}_{\o''}(\uu'-\ww')$ into scales and  take the
$L_\io$-norm of the lowest scale propagator, while the two others
are used to control the integration over the inner space variables
through their $L_1$-norm; then, for $W^{(2;2)(k)}$ use again the
bound \pref{pc1}, that is $\|W^{(2;2)(k)}\| \le C^3 \e_0 \g^{-k}$. We get:
\be\lb{27}\bsp & \|W^{(1;2)(k)}_{(b1)}\|\le 16\bar g^2
\|h_{L,K}\|_{L_1}\|h_{L,K}\|_{L_\io} \|W^{(2;2)(k)}\| \cdot\\
&\cdot 3!\sum_{i\ge j \ge i' \ge k+1} \|g^{(i)}\|_{L_1}
\|g^{(j)}\|_{L_1} \|g^{(i')}\|_{L_\io} \le C_{2,1} \bar g^2
\g^{-2(k-K)}\esp \ee
In a similar way, by using the bound \pref{233}, we see that $\|W^{(1;2)(k)}_{(b4)}\|\le
C_{2,4} \bar g \g^{-2(k-K)}$

The bound of $(b2)$ and $(b3)$ looks as problematic as the bound
of $(b)$, but we can now use in a simple way the cancelations related to the
propagator symmetry $g_\o^{[k,N]}(x,x_0)=-i\o g_\o^{[k,N]}(-x_0,x)$, which implies
that $\int d\zz d\zz' \tilde W_{\Th,\Th'}^{(2;0)(k)}(\zz,\zz')=0$, at any order in $\vec g$,
in particular at order $0$, that is
\be\lb{mas3} \int\!  d\uu\ \lft[g^{[k+1,N]}_{\o'}(\uu)\rgt]^2=0\;.
\ee
Let us consider first $(b2)$; we get
\be W^{(1;2)(k)}_{(b2)\Th';\Th}(\zz;\xx,\yy) =
\d(\xx-\yy)\sum_{\Th''}\int\!d\ww\ h^{L,K}_{\Th,\Th''}(\xx-\ww)
\lft[g^{[k+1,N]}_{\o''}(\ww-\zz)\rgt]^2 \ee
By using the identities \pref{mas3} and
\be \lb{idb} h^{L,K}_{\Th,\Th''}(\xx-\ww)= h^{L,K}_{\Th,\Th''}(\xx-\zz)+
\sum_{j=0,1} (z_j-w_j) \int_0^1\!\!ds\ \big(\dpr_j
h^{L,K}_{\Th,\Th''}\big)\big(\xx-\zz+s(\zz-\ww)\big)\ee
we get:
\be\bsp & W^{(1;2)(k)}_{(b2)\Th';\Th}(\zz;\xx,\yy) = \d(\xx-\yy)
\cdot\\
\cdot \sum_{\Th''\atop j=0,1} &\int_0^1\!\!ds\ \int\! d\ww\
\big(\dpr_j h^{L,K}_{\Th,\Th'}\big) \big(\xx-\zz+s(\zz-\ww)\big)
(z_j-w_j) \lft[g^{[k+1,N]}_{\o'}(\ww-\zz)\rgt]^2 \esp\ee
Hence,
\bea && \|W^{(1;2)(k)}_{(b2)}\| \le 8\bar g \|\dpr  h_{L,K}\|_{L_1}
\sum_{i\ge j\ge k+1} \|g^{(i)}\|_{\tilde L_1}\;
\|g^{(j)}\|_{L_\io}\; \le C_{2,2} \bar g \g^{-(k-K)}\;.
\eea
Following the same strategy, we obtain a bound for $(b3)$ which is obtained from the
bound for $(b2)$ by multiplying it by $\|W^{(1;2)(k)}_{(b)}\|$; we get:
\be \|W^{(1;2)(k)}_{(b3)}\| \le C_{2,2} \bar g
\g^{-(k-K)}\|W^{(1;2)(k)}_{(b)}\|\ee
Analogously, the bound for $(b5)$ is obtained from the
bound for $(b4)$ by multiplying it by $\|W^{(1;2)(k)}_{(b)}\|$; hence
$\|W^{(1;2)(k)}_{(b5)}\|\le C_{2,4} \bar g \g^{-2(k-K)} \|W^{(1;2)(k)}_{(b)}\|$.

Therefore, summing all the bounds for $(b1)$ to $(b5)$, we obtain
$$
\|W^{(1;2)(k)}_{(b)}\|\le {C_2\over 2}\bar g \g^{-(k-K)} +{\bar g}
C_5\|W^{(1;2)(k)}_{(b)}\|
$$
with $C_2=2(C_{2,1}+C_{2,2}+C_{2,4})$ and $C_5=C_{2,2}+C_{2,4}$.
Hence, if $\bar g$ is small enough,
\be\lb{b2b}
\|W^{(1;2)(k)}_{(b)}\|\le C_2 \bar g\g^{-(k-K)} \ee
It is now easy to complete the proof of  \pref{hb2}.

\*

{\it\0c) Bound for $W^{(0;4)(k)}$.}
If we take three derivatives w.r.t. $\psi$ in \pref{prc2}
and then we put $\psi=J=0$, we get the decomposition of $W^{(0;4)(k)}$
depicted in Fig.\ref{z4q}.
\insertplot{350}{190}{}%
{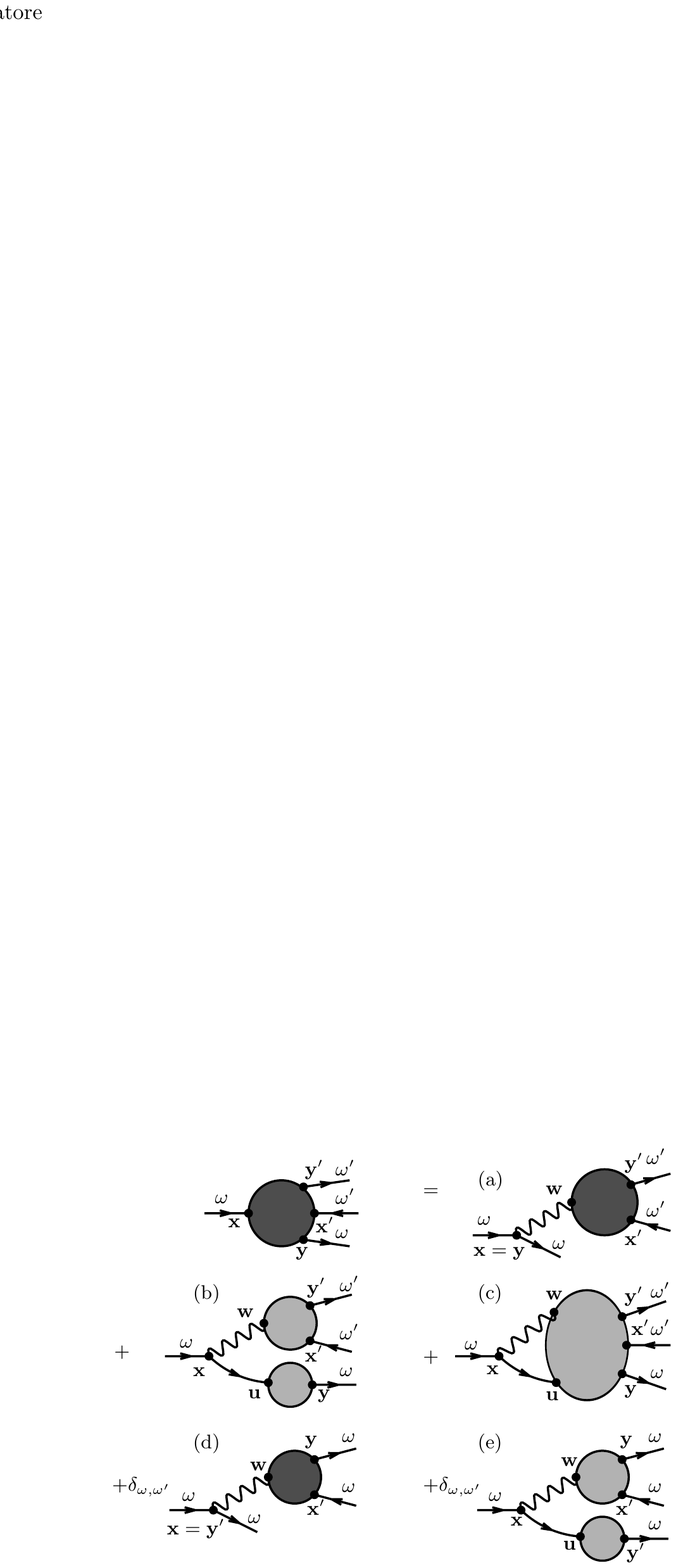}{\lb{z4q}Decomposition of the class of graphs
$W^{(0;4)(k)}_{\a,\a'}$. The darker bubbles represents
$W^{(0;4)(k)}_{\a,\a'}-w^{(0;4)}_{\a,\a'}$ or
$W^{(1;2)(k)}_{\a;\a'}-w^{(1;2)}_{\a;\a'}$. }{0}

Let us consider first the graphs $(a)$ and $(d)$, which differ only for the
interchange of $\yy$ with $\yy'$; by using \pref{hb2}, we get:
$$\|W^{(0;4)(k)}_{(a)}\| + \|W^{(0;4)(k)}_{(d)}\| \le 16 \|h_{L,K}\|_{L_1} C_2 \bar g \g^{-(h-K)}
\le C_{3,1} {\bar g}^2 \g^{-(h-K)} $$
By using \pref{hb2} and the bound in the right side of \pref{ind}, we get
$$\|W^{(0;4)(k)}_{(b)}\| + \|W^{(0;4)(k)}_{(e)}\| \le 16 \|h_{L,K}\|_{L_1}
\|W^{(1;2)(k)}\| \|W^{(0;2)(k)}\| \sum_{j=k}^N \|g^{(j)}\|_{L_1}
\le C_{3,2} {\bar g}^2 \g^{-2(h-K)} $$
Finally, by using \pref{ind} to bound $\|W^{(1;4)(k)}\|$, we get
$$ \|W^{(0;4)(k)}_{(c)}\| \le 8 \|h_{L,K}\|_{L_\io}
\|W^{(1;4)(k)}\| \sum_{j\ge k} \|g^{(j)}\|_{L_1} \le C_{3,3} \e_0 \bar g \g^{-2(k-K)}$$
Then, if $\bar g\le \e_0$, by summing the three previous bounds, we get \pref{hb3},
with $C_3=C_{3,1}+ C_{3,2} +C_{3,3}$.
The proof of Theorem \ref{t3.2} is complete. \qed

\subsection{The infrared integration}\lb{ss2.3}

The parameter $K$ has only the role of clarifying the dimensional content of the bounds in
\S\ref{ss2.2}; hence, from now on, we shall put $K=0$. For analogous reasons we shall also put $Z=1$.

Note that Theorem \ref{t2.1} implies that the effective potential on scale $0$ of the model \pref{vv1}
is equivalent, from the RG point of view, to the effective potential on scale $0$ of the extended Hubbard model,
which has been studied in detail in the companion paper \cite{BFM012_1}.
The reason is in the following remarks:

\0a) The free propagator $g_{D,\o}^{[l,0]}(\xx)$ has the same asymptotic behavior of the Hubbard free
propagator, see eq. (2.100) in \cite{BFM012_1}, at the infrared scales; the fact that, in \cite{BFM012_1},
there is a different definition of the finite space region and of the infrared cutoff has negligible effects,
as it is easy to see by looking at the description of the infrared integration in \S2.4 of \cite{BFM012_1}.

\0b) The local parts of the effective potential in the two models
have the same structure. As a consequence, the running coupling constants (r.c.c.)
of the two models can be defined in a similar way, with two differences: in the model \pref{vv1} there is one more
quartic marginal term, but, thanks to the oddness of the free propagator, the r.c.c. $\n_h$ (that multiplying the
relevant monomial $\psi^+_{\xx,\o,s} \psi^-_{\xx,\o,s}$) vanishes exactly, so that the singularity of the interacting propagator
stays fixed at $\kk=(0,0)$. As concerns the renormalization constants (ren.c.),
in the model \pref{vv1} we put a restrict set of source terms, which are related to the charge and spin
ren.c.'s of the Hubbard model; hence, we have to consider
three ren.c.'s, that is $Z_h$, the {\em free measure ren.c.}, $Z_h^{(1)}$ and $Z_h^{(2)}$,
related to the source terms with $s=t$ and $s=-t$, respectively.

Hence we can repeat word by word the analysis of \S2.4 of \cite{BFM012_1}, with the simplifications
that the free propagator scales exactly and $\n_h=0$. In particular, we can define in a similar way
the functionals $\tilde\SS_j(J,\h)$, $\hat\VV^{(j)}(\psi)$, $\VV^{(j)}(\psi)$, $\hat\BB^{(j)}(\psi,J,\h)$,
$\BB^{(j)}(\psi,J,\h)$ and the renormalized single scale free measure $P_{Z_{j-1},\tilde f_j^{-1}}(d\psi)$,
with propagator
\be\lb{prop_ir}\hg^{(j)}_\o(\kk) =
\begin{cases}\frac1{Z_{j-1}} \frac{\tilde f_j(|\tilde \kk|)}{D_\o(\kk)} & \text{$l<j\le 0$}\\
\frac1{\tilde Z_{l-1}(\kk)} \frac{f_l(|\tilde \kk|)}{D_\o(\kk)} & \text{$j=l$}
\end{cases}\ee
where $\tilde f_j(t)$ has the same support ans smoothness properties of $f_j(t)$ and
$\tilde Z_{l-1}(\kk)=1+(Z_{l-1}-1)f_l(|\tilde \kk|)$; for details see also \S 3 of \cite{BM002}.

\0{\bf Remark} - Even if $\tilde Z_{l-1}(\kk)$ takes values between $1$ and $Z_{l-1}$,
$|\hg^{(j)}_\o(\kk)|$ is bounded by $c\g^{-j}Z_{j-1}^{-1}$ for all $j\in[l,0]$. However, in the following,
in particular in the prove of the bounds \pref{61}, \pref{61a} and \pref{C.6a}, the strong dependence on $\kk$ of
$\tilde Z_{l-1}(\kk)$ will be relevant.

The r.c.c. are defined as the coefficients of the local part of $\hat\VV^{(j)}(\psi)$, which is given by
\be \LL \hat\VV^{(j)}(\psi) = \d_j F_\d(\psi) + g_{1,\perp,j} F_{1,\perp}(\ps)+ g_{\arp,j|} F_\arp(\psi)
+g_{\perp,j} F_{\perp}(\ps)+ g_{4,j}F_4(\psi)\lb{blue}\ee
where $F_\d(\psi)=\sum_{\o,s} \int d\xx \psi^+_{\xx,\o,s} (i\o\dpr_x) \psi^-_{\xx,\o,s}$, while the other
functions are defined as the potentials \pref{five1} with $\d(\xx-\yy)$ in place of $h_{L,K}(\xx-\yy)$.
Moreover,
\be\lb{LB} \LL \hat\BB^{(j)}(\psi,J,\h) = \sum_{\o,s} \int d\xx \lft[ J_{\xx,\o,s,s} \frac{Z_j^{(1)}}{Z_{j-1}} \psi^+_{\xx,\o,s}
\psi^-_{\xx,\o,s} +J_{\xx,\o,s,-s} \frac{Z_j^{(2)}}{Z_{j-1}} \psi^+_{\xx,\o,s}
\psi^-_{\xx,\o,-s}\right]\ee
We introduce a label $\a\in\{1,2\}$ to distinguish, in the tree expansion, the endpoints of type $J$ associated with the
two terms of \pref{LB}.

\*

\0{\bf Remark} - As we have remarked in \S\ref{ss2.2} for the UV scales, the compact support properties of the single scale
propagators $\hat g_\o^{(j)}(\kk)$ imply that the effective potential on scale $j$ depends on $N$ and $L$,
but is independent of the IR cutoff $l$, if $j>l$. The same remark is of course true for the r.c.c. and the ren.c.;
however, since the r.c.c. are defined only up to scale $l+1$, they are all independent of $L$.

\*

Let us now call $W^{(h)}_{\uo,\ual,\ue,\usp, m_\psi,m_J,m_\h}(\ux)$ the
kernels of the various terms contributing to $\VV^{(j)}(\psi)$, $\BB^{(j)}(\psi,J,\h)$ and
$\tilde\SS_j(J,\h)$, where $m_\psi$, $m_J$, $m_\h$ are the numbers of external fields of type $\psi$,
$J$, $\h$, respectively; $\uo$, $\ue$ and $\usp$ are the sets of $\o$, $\e$ and $s$ indices (possibly void)
of the $\psi$ and $\h$ fields, while $\ual$ is the the set of $\a$ indices of the $J$ fields.
Let us define
\be\lb{epsh} \e_h=\max_{h\le j \le 0} \max\{|\d_j|, |g_{1,\perp,j}|, |g_{\arp,j}|,
|g_{\perp,j}|, |g_{4,j}|\}\ee
Lemma 2.3 of \cite{BFM012_1} and the analysis of the ren.c. flow in \S2.6 of the same paper
imply the following Theorem.

\begin{theorem}\lb{th2.2}
There is $\bar\e$, $c_1$, $C$, independent of $N,l,L$, such that, if $\e_h\le \bar\e$, then
\be\lb{ass} \sup_{j> h} Z_j/Z_{j-1} \le e^{c_1 \e_h^2} \virg  \sup_{j> h\atop i=1,2} Z^{(i)}_j/Z_{j-1}\le e^{c_1
\e_h}\ee
\be\lb{IRbound} \int d\ux \big|
W^{(h)}_{\uo,\ual,\ue,\usp, m_\psi,m_J,m_\h}(\ux)\big| \le \b L C^{m+m_S}
\e_h^{k_{m,m_S}} \g^{-h D_{m_\psi,m_J,m_\h}} \ee
where $2m=m_\psi+m_\h$, $m_S=m_J+m_\h$,
$D_{m_\psi,m_J,m_\h}=-2+m+m_J (1+ c_1\e_0) + m_\h (1+ \frac12 c_1\e_0^2)$,
$k_{m,m_s}=\max\{1,m-1\}$, if $m_S=0$, otherwise
$k_{m,m_s}=\max\{0,m-1\}$.
\end{theorem}

This Theorem is sufficient to perform the limit $N\to\io$, followed by the limit $L\to\io$, at fixed $l$,
but to prove that $\bar\e$ is independent of $l$ is not trivial. It can be proved only under one of the
conditions: 1) $g_{1,\perp}= 0$; 2) $g_\arp=g_{\erp}-g_{1,\perp}$ and $g_{1,\perp}>0$. Moreover,
we have to use the strong cancelations
related to the local gauge invariance of the interaction \pref{ep0} in the case $g_{1,\perp}=0$.
In the following sections, we shall explain how to use this symmetry in order to prove iteratively
that $\bar\e$ stays away from $0$ for $l\to-\io$ and, at the same time, the important bound
\be\lb{2.48}
\left| \frac{Z_{j}^{(1)}}{Z_j}-1 \right|\le C |\e_j^2|
\ee

\*

The properties of the tree expansion allow us also to prove that all Schwinger functions and the correlation
functions, at {\em fixed distinct values of the space variables}, are well defined in the limit $L\to\io$, when all the cutoffs
are removed (the condition that there are not coinciding space points is to avoid the
singularities related with the UV divergence of the free propagator).
The same is true for their Fourier transforms, if one puts the following condition on the
corresponding momenta $\{\qq_i, i=1,\ldots,m\}$ of the $m$ external fields of type $J$ or $\h$ :
\be\lb{nonex} \sum_{i=1}^m \s_i\qq_i \not=0 \virg \forall \s_i=\pm1\ee
If this condition is satisfied, we shall say that {\em the momenta are not exceptional}; it is imposed to
avoid the singularities related to the possible divergence of the $L^1$ norm for $L,-l\to\io$,
in agreement with the bound \pref{IRbound}; these singularities are of course related to the divergence
of the $L^1$ norm of  the free propagator.
The proof of these properties depends only on the structure of the tree expansion and on the possibility
of controlling the r.c.c.'s flow, while the details of the model are not important. Hence, we refer for their
proof to paper \cite{BFM007}, where they are precisely discussed, especially the first one, with a different localization
procedure, which involves also the terms with one $\h$ field and no $J$ field.

In \S\ref{ss5.1} we shall also need a bound on some particular correlation functions in the limit
$L,N\to\io$ at fixed $l$, when the external momenta are of order $\g^l$.
A similar property, whose proof is based on a technique introduced in \S3 of \cite{BM002},
has been used in \cite{BM005}, Theorem 1, in the case of a model with spin $0$, local interaction and a fixed UV cutoff.
To extend this theorem to the actual model is essentially trivial; hence we shall write the result without proof. Let us define
$<\cdot>_T$ the truncated expectation of the model \pref{vv1} and $\r_{\xx,\o,s}:=\ps^+_{\xx,\o,s}
\ps^-_{\xx,\o,s}$.

\begin{theorem}\lb{th3.3}
In the hypotheses of Theorem \ref{th2.2}, if $|\tilde\kk|=\g^l$, then, in the limit $L\to\io$,
followed by the limit $N\to\io$, at fixed $l$,
\be\lb{corr_h}\bsp
\la \hp^-_{\kk,\o,s}\hp^+_{\kk,\o,s}\ra &= \frac{1}{Z_l D_\o(\kk)}\big[
1+O(\e_l)\big]\\
\la \hr_{2\kk,\o,s}; \hp^-_{\kk,\o,s} \hp^+_{-\kk,\o,s}\ra_T &= -\frac{Z_l^{(1)}}{Z_l^2 D_\o(\kk)^2}\big[
1+O(\e_l)\big]\\
\la\hp^+_{\kk,\o,s}; \hp^-_{-\kk,\o,\m s}; \hp^+_{-\kk,\o',s'}; \hp^-_{\kk,\o',\m s'}\ra_T &=
-\frac{1}{Z_l^2 |\tilde\kk|^4} \Big\{ -g_{1,\erp,l} \d_{\m,-1} \d_{\o,-\o'}\d_{s,-s'}\\ + \d_{\m,1}\big[
g_{\arp,l} \d_{\o,-\o'}\d_{s,s'} &+ g_{\erp,l} \d_{\o,-\o'}\d_{s,-s'}+
g_{4,l} \d_{\o,\o'}\d_{s,-s'} +O(\e_l^2)\big] \Big\}
\esp\ee

\end{theorem}


\section{Ward Identities}

\subsection{Ward Identities in the $g_{1,\perp}=0$ case in presence of cut-offs}\lb{ss3.1}

In this section we want to analyze the Ward Identities (WI) related
to the local gauge invariance of the interaction \pref{fs1} in the case $g_{1,\perp}=0$.
As explained in \S\ref{ss2.5c}, this can only be done
by putting $\e>0$ in \pref{gth} and by introducing a lattice spacing $a$ in $\L_L$. At the end,
we have to perform the limit $a\to 0$, followed by the limit $\e\to 0$. In \S\ref{ss2.5c} we have also
explained why the limit $a\to 0$ is essentially trivial; hence, in the following, we shall write the WI
directly in the continuum limit $a=0$. As concerns the limit $\e\to 0$, it is trivial too, in the
contributions coming from the effective potentials, that we shall then suppose evaluated at $\e=0$,
while it needs an accurate analysis for the terms coming from the free measure.

If $g_{1,\perp}=0$, the model \pref{vv1} is invariant under the global gauge transformation \pref{2.7}
and the interaction $\VV(\psi,J)$ (see \pref{ep0}, where $\int d\xx$ has to be understood
as $\sum_{\xx\in \L_L^a} a^2$) is invariant even under the local gauge transformation
\be\lb{3.2} \psi^\pm_{\xx,\o,s}\to e^{\pm i\a_{\xx,\o,s}}
\psi^\pm_{\xx,\o,s} \virg \xx\in\L^a_L\ee
for any lattice periodic function $\a_{\xx,\o,s}$. Hence, if we make in the r.h.s. of \pref{vv1}
this transformation, we get in the limit $a\to 0$, followed by the limit $\e\to 0$
(see \S2 of \cite{BM002} for details in a similar problem), the
following {\em functional Ward Identity (WI)}, where $\pp\in \DD_L$ and $\hJ_{\pp,\o,s}$ is defined so that
$J_{\zz,\o,s}=L^{-2}\sum_{\pp\in \DD_L} e^{-i\pp\zz} \hJ_{\pp,\o,s}$.
\be\lb{12}
 D_\o(\pp) \frac{\dpr\WW_{l,N}(J,\h)}{\partial \hJ_{\pp,\o,s}} =
B_{\pp,\o,s}(J,\h)+ R_{\pp,\o,s}(\pp;J,\h) \virg D_\o(\pp)= -ip_0 +c\o p\ee
where (since $Z=1$)
\be\lb{3.4a} B_{\pp,\o,s}(J,\h):= {1\over L^2}\sum_{\kk\in \DD'_L}
\lft[\hh^+_{\kk+\pp,\o,s} {\partial \WW_{l,N}(J,\h)\over \partial \hh^+_{\kk,\o,s}}-
{\partial \WW_{l,N}(J,\h)\over \partial \hh^-_{\kk+\pp,\o,s}} \hh^-_{\kk,\o,s}\rgt]
\ee
and
\be
 R_{\pp,\o,s}(J,\h)= e^{-\WW_{l,N}(J,\h)}\lim_{\e\to 0}
\frac1{L^2} \sum_{\qq\in \DD'_L} C_\o(\qq+\pp,\qq)\int P_Z^{[l,N]}(d\ps)\
\hp^+_{\qq+\pp,\o,s}\hp^-_{\qq,\o,s}e^{\VV(\ps,J,\h)} \ee
with
\be\lb{ccc} C_\o(\kk^+,\kk^-) := \lft[{1\over \c_{l,N}^{\e}(|\tilde\kk^-|)}-1 \rgt]
D_\o(\kk^-) -\lft[{1\over \c_{l,N}^{\e}(|\tilde\kk^+|)}-1\rgt] D_\o(\kk^+)\ee
Note that, here and in the following, the derivatives with respect to
the momenta have to be understood as the ordinary derivatives times $L^2$, so that
they formally become the usual functional derivatives when $L\to\io$. Moreover, we are defining
the derivatives w.r.t. the Grassman variables so that, if $\{\chi^{\e_i}_i, i=1,\ldots,n\}$ is a set of
distinct $\psi^\e_{\kk,\o,s}$ variables and $n>1$, then
$$\frac{\dpr}{\dpr\chi^{\e_j}_j} \prod_{i=1}^n \chi^{\e_i}_i=
\e_j(-1)^{j-1} \prod_{i\not=j} \chi^{\e_i}_i$$
The last term $R_{\pp,\o,s}(J,\h)$ formally vanishes if
we replace the cut-off function $\chi^\e_{l,N}$ in \pref{gth}
(necessary to make the functional integral \pref{vv1} well
defined) with $1$. This term can be therefore considered a {\it
correction} to the {\em formal} functional WI (the one in which
$R_{\o,s}(\pp;J,\h)$ is neglected); the origin of this correction is in the fact that
the momentum cut-offs break the (formal) {\it local} gauge invariance. We will
show that such corrections are not vanishing even in the limit of
removed cut-offs, a phenomenon known as {\it quantum anomaly}.

A crucial role is played by the properties of the function
$C_\o(\kk^+,\kk^-)$. This function looks very singular; however, since it appears
in the expansions always multiplied by two propagators, what it is really important are the
properties of the function
\be \lb{defU} \hU^{(i,j)}_{l,N,\o}(\qq+\pp,\qq) :=
C_{\o}(\qq+\pp,\qq) \hg^{(i)}_\o(\qq+\pp) \hg^{(j)}_\o(\qq) \ee
with $\hg^{(j)}_\o(\kk)$ defined as in \pref{prop_ir}, for all $j$, if we put
$\tilde f_j(|\tilde\kk|)=f_j(|\tilde\kk|)$ and $Z_{j-1}^{-1}=1$, if $j>0$.
The main property, which we will extensively use in the following, is that, since $\c_{l,N}^{\e}(\kk)=1$
if $\hg^{(j)}_\o(\kk)\not =0$ and $l<j<N$, then
\be\lb{proU} \hU^{(i,j)}_{l,N,\o}(\qq+\pp,\qq)=0\quad\quad l<i,j<N \ee
In the following we shall use the function $\hU^{(i,j)}_{l,N,\o}(\qq+\pp,\qq)$ only under the condition
that $|\tilde\pp|\le 2\g^{N+1}$; hence we can multiply it by $\tilde\chi_N(\pp) = \tilde\chi_0(2^{-1}\g^{-N-1}|\tilde\pp|)$,
where $\tilde\chi_0(t)$ is a smooth positive function of support in $[0,2]$ and equal to $1$ for $t\le 1$.
In App. \ref{appC} we show that it possible to define two functions
$\hS_{l,N,\o',\o}^{(i,j)}(\kk^+,\kk^-)$, such that (recall that $Z_j=1$ for $j\ge 0$)
\be\lb{du} \tilde\chi_N(\pp) \lim_{\e\to 0}\, \hU_{l,N,\o}^{(i,j)}(\qq+\pp,\qq)
=\frac1{Z_{i-1} Z_{j-1}} \sum_{\o'=\pm\o}D_{\o'}(\pp) \hS_{l,N,\o',\o}^{(i,j)}(\qq+\pp,\qq)
\ee
and, if we define the Fourier transform of $\hS_{l,N,\o',\o}^{(i,j)}(\qq+\pp,\qq)$ as
\be\nn S_{l,N,\o',\o}^{(i,j)}(\zz;\xx,\yy) = \frac1{L^4} \sum_{\kk^+,\kk^-\in\DD'_L}
e^{-i\kk^+(\xx-\zz)} e^{i\kk^-(\yy-\zz)} \hS_{l,N,\o',\o}^{(i,j)}(\kk^+,\kk^-) \ee
then, if $j\ge l$, for any positive integers $r$ (which in the following will be fixed large enough,
e.g. $r=4$),
\be\lb{61} |S_{l,N\o',\o}^{(N,j)}(\zz;\xx,\yy)| \le (1+\d_{j,l}Z_{l-1}) b_{r,N}(\xx-\zz) b_{r,j}(\yy-\zz)
\virg  b_{r,k}(\xx):= C_{p}{\g^{k}\over 1+[\g^k||\yy-\zz||]^r} \ee
where $||\xx-\yy||$ is the distance on the torus $\L_L$.  The bound \pref{61} has an essential
role in the integration of the UV scales; in the integration of the IR scales, we shall need a bound
on the Fourier transform of the functions (independent of $N$)
$$\tilde S_{l,\o'\o}^{(j,l)}(\qq+\pp,\qq) := \frac12 [1-\tilde\chi_l(\pp)] \lim_{\e\to 0}
\frac{\hU_{l,N,\o}^{(j,l)}(\qq+\pp,\qq)}{D_{\o'}(\pp)} \virg j\ge l$$
where $\tilde\chi_l(\pp) := \tilde\chi_0[2^{-1}\g^{-l-1}|\tilde\pp|]$, with $\tilde\chi_0(t)$
is the same function used in the definition of $\tilde\chi_N(\pp)$. These functions allows to write
an identity similar to \pref{du} for $\hU_{l,N,\o}^{(j,l)}$.

In App. \ref{appC} we also show that, for any positive integers $r$,
\be\lb{61a}
\bsp |\tilde S_{l,\o'\o}^{(j,l)}(\zz;\xx,\yy)| &\le \frac{(1-\d_{j,l})}{Z_{j-1}} \tilde b_{r,j}(\xx-\zz)
\tilde u_{r,l}(\yy-\zz)
\\  \tilde b_{r,j}(\xx):= C_{p}{1\over 1+[\g^k||\yy-\zz||]^r}
&\virg \tilde u_{r,l}(\xx):= C_{p}{\g^{2l}\over 1+[\g^k||\yy-\zz||]^r}\esp\ee
Finally, in App. \ref{appC} we show that, if we define (recall that we have put $Z=1$ and $K=0$)
\be \t^\pm_N := \sum_{i,j=K+1}^N  {1\over L^2}\sum_{\qq\in\DD'_L}
\hS_{\pm\o,\o}^{(i,j)}(\qq,\qq)= \sum_{i,j=K+1}^N
\int d\uu S^{(i,j)}_{\pm\o,\o}(\zz;\uu,\uu) \label{3.14a}\ee
then
\be |\t^+_N|\le C \frac{\g^{-N}}L \virg |\t^-_N- \t| \le C \frac{\g^{-N}}L \virg
\t={1\over 4\pi c}\lb{3.14} \ee

Let us now observe that, for any choice of the functions $\n_s^\o(\pp)$,
we can write $R_{\pp,\o,s}(\pp;J,\h)$ in the following way:
\be\lb{13}\bsp &R_{\pp,\o,s}(\pp;J,\h) = -\t^-_N D_{-\o}(\pp)
\hJ_{-\pp,\o,s} +\\
&+ D_{-\o}(\pp)\sum_{\o',s'}\n_{ss'}^{\o\o'}(\pp)
{\partial \WW_{l,N}(J,\h)\over\partial \hJ_{\pp,-\o',s'}}
+ \lim_{\e\to 0} {\dpr \HH_{l,N}\over \dpr\ha_{\pp,\o,s}}(0,J,\h)\esp\ee
where, if $\VV(\ps, J, \h)$ is the expression inside the braces
in the r.h.s. side of \pref{vv1} and $\a(\zz,\o,s)$ is a periodic function, whose Fourier
transform is defined (analogously to $J(\zz,\o,s)$) so that $\a(\zz,\o,s)=
L^{-2}\sum_{\pp\in\DD_L} e^{-i\pp\zz}\ha_(\pp,\o,s)$, then
\be\lb{acca} e^{\HH_{l,N}(\a,J,\h)}:=e^{{1\over L^2}
\sum_{\pp\in\DD_L} \ha_{\pp,\o,s} \t^-_N D_{-\o}(\pp) \hJ_{-\pp,\o,s}}
\int\! P_Z^{[l,N]}(d\ps)\;
e^{\VV(\ps,J,\h)+\AA_0(\ps,\a)-\AA_{-}(\ps,\a)} \ee
and
\bal \AA_0 (\a,\ps) &= \sum_{\o,s}{1\over L^2}\sum_{\pp\in\DD_L} {1\over
L^2}\sum_{\qq\in\DD'_L}
C_{\o}(\qq+\pp,\qq)\ha_{\pp,\o,s}\hp^+_{\qq+\pp,\o,s}\hp^-_{\qq,\o,s}\nn\\
\AA_-(\a,\ps) &=\sum_{\o,\o'\atop s,s'}{1\over
L^2}\sum_{\pp\in\DD_L} {1\over L^2}\sum_{\qq\in\DD'_L}
D_{-\o}(\pp)\n_{ss'}^{\o\o'}(\pp)
\ha_{\pp,\o,s}\hp^+_{\qq+\pp,-\o',s'} \hp^-_{\qq,-\o',s'} \nn\eal

Let us call {\em removed cutoffs limit} the limit $L\to\io$, followed by the limit $N\to\io$ and, finally,
from the limit $l\to-\io$.
In the following sections we shall prove that, in the removed cutoffs limit, the last term in the r.h.s. of \pref{13}
vanishes, if the functions $\n_{s}^{\o}(\pp)$ are chosen as
\be\lb{4.2} \n^{\o}_{s}(\pp) = \t^-_N\hat h_{L,K}(\pp) \lft[\d_{\o,1}\lft(\d_{s,-1}g_\erp + \d_{s,1}g_\arp\rgt)+
\d_{\o,-1}\d_{s,-1}g_4\rgt]\ee

Note that the first two terms in the r.h.s. of \pref{13} are, in this case, the corrections to the formal
functional WI we are looking for. Hence, we shall rewrite the functional WI \pref{12}, by using
\pref{13}, in the more convenient form
\be\lb{grez} \bsp & D_{\m}(\pp){\partial \WW_{l,N}(J,\h)\over \partial
\hJ_{\pp,\m,s}} -D_{-\m}(\pp)\sum_{\s,r}\n^{\m\s}_{sr}(\pp)
{\partial \WW_{l,N}(J,\h)\over\partial
\hJ_{\pp,-\s,r}}= \\
&- \t^-_N D_{-\o}(\pp) \hJ_{-\pp,\o,s} +
B_{\pp,\m,s}(J,\h)+ \lim_{\e\to 0} {\dpr \HH_{l,N}\over
\dpr\ha_{\pp,\m,s}}(0,J,\h) \esp \ee
Summing \pref{grez} over $s$ we obtain the {\it charge Ward
identity}:
\be\lb{char}\bsp \Big[D_{\m}(\pp) &-\n_4(\pp) D_{-\m}(\pp)\Big]
\sum_s{\partial \WW_{l,N}(J,\h)\over
\partial\hJ_{\pp,\m,s}} -2\n_\r(\pp) D_{-\m}(\pp) \sum_s{\partial \WW_{l,N}(J,\h)\over
\partial\hJ_{\pp,-\m,s}}=\\
&\sum_s\lft[ -\t^-_N D_{-\m}(\pp) \hJ_{-\pp,\m,s}
+B_{\pp,\m,s}(J,\h)+ \lim_{\e\to 0} {\dpr \HH_{l,N}\over
\dpr\ha_{\pp,\m,s}}(0,J,\h) \rgt]\esp \ee
with
\be\n_4(\pp)= \t^-_N g_4 \hat h_{L,K}(\pp) \virg 2\n_\r(\pp)=
\t^-_N (g_\arp + g_\erp) \hat h_{L,K}(\pp)\ee
Multiplying \pref{grez} by $s$ and summing over $s$ we obtain the
{\it spin Ward identity}:
\be\lb{spin} \bsp \Big[D_{\m}(\pp) &+\n_4(\pp) D_{-\m}(\pp)\Big]
\sum_ss{\partial \WW_{l,N}(J,\h)\over
\partial\hJ_{\pp,\m,s}} -2\n_\s(\pp) D_{-\m}(\pp) \sum_ss{\partial \WW_{l,N}(J,\h)\over
\partial\hJ_{\pp,-\m,s}}\\
&=\sum_s s\lft[ - \t^-_N D_{-\m}(\pp) \hJ_{-\pp,\m,s}
+B_{\pp,\m,s}(J,\h)+ \lim_{\e\to 0} {\dpr \HH_{l,N}\over
\dpr\ha_{\pp,\m,s}}(0,J,\h)\rgt] \esp \ee with
\be 2\n_\s(\pp)= \t^-_N (g_\arp - g_\erp) \hat h_{L,K}(\pp)\ee
By doing suitable functional derivatives of the charge WI
\pref{char}, we get many WI involving  the correlation functions.
For example, if we take two derivatives w.r.t.
$\hh^+_{\pp+\kk,\o,s}$ and $\hh^-_{\kk,\o,s}$ in both sides of
\pref{char}, we sum over $\m$ and we put $\h=J=0$, we get the
following {\em charge vertex Ward identity}:
\be\lb{chara} \bsp \hspace{-.3cm} -ip_0
\Big[1-2\n_\r(\pp)-\n_4(\pp)\Big] &G_{\r;\o,s}(\pp;\pp+\kk)+ cp
\Big[1-2\n_\r(\pp)+\n_4(\pp)\Big] G_{j;\o,s}(\pp;\pp+\kk)\\
&=G_{2;\o,s}(\kk)-G_{2;\o,s}(\pp+\kk) +R_{\o,s}(\pp;\pp+\kk)
\esp \ee
where
\be G_{\r;\o,s}(\pp;\pp+\kk) = \sum_{\m,t} {\dpr^3 \WW\over
\dpr\hJ_{\pp,\m,t} \dpr\hh^-_{\kk,\o,s} \dpr\hh^+_{\pp+\kk,\o,s}}(0,0)\lb{vertr}\ee

\be G_{j;\o,s}(\pp;\pp+\kk) = \sum_{\m,t} \m\, {\dpr^3 \WW\over
\dpr\hJ_{\pp,\m,t} \dpr\hh^-_{\kk,\o,s} \dpr\hh^+_{\pp+\kk,\o,s}}\lb{vertj}(0,0)\ee

\be G_{2;\o,s}  ={\dpr^2
\WW_{l,N}\over \dpr\hh^-_{\kk,\o,s} \dpr\hh^+_{\kk,\o,s}}(0,0)\virg
R_{\o,s}(\pp;\pp+\kk)=\lim_{\e\to 0}  \sum_{\m,t} {\dpr^3 \HH_{l,N}\over
\dpr\ha_{\pp,\m,t} \dpr\hh^-_{\kk,\o,s} \dpr\hh^+_{\pp+\kk,\o,s}}(0,0,0)
 \ee
Similarly, if we define $\r^{(c)}_{\xx,\o}:=\sum_s \psi^{[l,N]+}_{\xx,\o,s} \psi^{[l,N]-}_{\xx,\o,s}$,
so that $\la\hr^{(c)}_{\pp,\o} \hr^{(c)}_{-\pp,\o'}\ra_T=\sum_{s,s'}{\dpr^2 \WW\over
\dpr\hJ_{\pp,\o,s} \dpr\hJ_{-\pp,\o',s'}}(0,0)$, we get from \pref{char}:
\be\bsp [D_\o(\pp)-\n_4(\pp) D_{-\o}(\pp)] &\la\hr^{(c)}_{\pp,\o}
\hr^{(c)}_{-\pp,\o'}\ra_T -2 \n_\r(\pp) D_{-\o}(\pp)
\la\hr^{(c)}_{\pp,-\o} \r^{(c)}_{-\pp,\o'}\ra_T\\
&= -\d_{\o,\o'}  \t^-_N D_{-\o}(\pp) +R^{(c)}_{\o,\o'}(\pp)\esp\ee
with $R^{(c)}_{\o,\o'}(\pp)= \lim_{\e\to 0} \sum_{s,s'} {\dpr^2 \HH_{l,N}\over
\dpr\ha_{\pp,\o,s}\hJ_{-\pp,\o',s'}}(0,0,0)$.

Finally, from \pref{char} and \pref{spin} and some algebra we get
the following identity, which will be useful in the following:
\be\lb{wis}\bsp {\dpr \WW_{l,N}\over \dpr\hJ_{\pp,\m',s'}}(J,\h) &=\sum_{\m,s}
{M^\r_{\m',\m}(\pp)+s's M^\s_{\m',\m}(\pp)\over 2}
\Big[ - \t^-_N D_{-\m}(\pp) \hJ_{-\pp,\m,s}+\\
&+B_{\pp,\m,s}(J,\h)+ \lim_{\e\to 0} {\dpr \HH_{l,N}\over
\dpr\ha_{\pp,\m,s}}(0,J,\h)\Big] \virg\qquad \pp\not=0\esp\ee
where, if $\g=\r,\s$, and setting $\n_{4,\r}=-\n_{4,\s}=\n_4$,
\be\lb{MM} \bsp M^\g_{\m',\m}(\pp)
&={\Big[D_{-\m}(\pp)-\n_{4,\g}(\pp)D_\m(\pp)\Big]\d_{\m',\m}+
\Big[2\n_{\g}(\pp) D_{\m}(\pp)\Big]\d_{\m',-\m}\over
\Big[D_+(\pp)-\n_{4,\g}D_-(\pp)\Big]\Big[D_-(\pp)-\n_{4,\g}(\pp)D_+(\pp)\Big]
-4\n_{\g}^2(\pp) D_+(\pp)D_-(\pp)}\\
&={u_{\g,+}(\pp)\d_{\m',\m}+w_{\g,+}(\pp)\d_{\m',-\m} \over
-iv_{\g,+}(\pp)\big(p_0+i\m v_{\g}(\pp)c p_1\big)} +
{u_{\g,-}(\pp)\d_{\m',\m}+w_{\g,-}(\pp)\d_{\m',-\m} \over
-iv_{\g,+}(\pp)\big(p_0-i\m v_{\g}(\pp)c p_1\big)} \esp \ee
for
\be\lb{MM1} \bsp u_{\g,\m}(\pp) &={1\over
2}\lft[{1-\n_{4,\g}(\pp)\over v_{\g,+}(\pp)}
+\m{1+\n_{4,\g}(\pp)\over v_{\g,-}(\pp)}\rgt]\\
w_{\g,\m}(\pp) &=\n_\g(\pp)\lft[{1\over v_{\g,+}(\pp)} -\m{1\over
v_{\g,-}(\pp)}\rgt] \esp \ee
\be\lb{MM2} v^2_{\g,\m}(\pp)=\Big(1-\m\n_{4,\g}(\pp)\Big)^2
-4\n_\g^2(\pp)\virg v_\g(\pp) =v_{\g,-}(\pp)/v_{\g,+}(\pp) \ee

\subsection{Analysis of the correction term \pref{acca}. Integration of the UV scales.}
\lb{ss3.2}
The functional $\HH_{l,N}(\a,J,\h)$ defined in \pref{acca} has a form close to the functional integral
$\WW_{l,N}(J,\h)$ of \pref{vv1}; hence, the integration of the UV scales can
be studied by a Renormalization Group analysis very similar to the
one in \S 2. In particular, the external field $\a$ in $\AA_0(\a,\psi)$ and $\AA_-(\a,\psi)$ plays the
same role of the field $J$ in \pref{vv1}; the main difference
is due to the presence, in the definition of $\AA_0$, of the singular function \pref{ccc},
whose peculiar properties will play a crucial role.

After the integration of the fields $\psi^{(N)},...\psi^{(K+1)}$, where $K$ is
the fixed integer defined in \S\ref{ss2.5c}, we get
\be\lb{ep1}\; e^{\HH_{l,N}(\a,J,\h)}= \int\!P_Z^{[l,K]}(d\ps)\
e^{\VV^{(K)}(\ps, J, \h) +\AA^{(K)}(\a,J,\h,\ps)} \ee
where $\VV^{(K)}(\psi,J,\h)$ is defined as in \pref{ep2} and $\AA^{(K)}(\a,J,\h,\ps)$
satisfies the identity
\be\lb{ep1a}e^{\AA^{(K)}(\a,J,\h,\ps)} = e^{{1\over L^2}
\sum_{\pp\in\DD_L} \ha_{\pp,\o,s}  \t^-_N D_{-\o}(\pp) \hJ_{-\pp,\o,s}} \int P_Z^{[K+1,N]} (d\z)
e^{\VV(\psi+\z,J,\h)+\AA_0(\psi+\z,\a)- \AA_-(\psi+\z,\a)}\ee

As in \S\ref{ss2.2}, we shall consider, for simplicity, only the contributions to $\AA^{(K)}(\a,J,\h,\ps)$
with $\h=0$; the result can be easily extended to the general case. Hence, since we have to evaluate only the
terms linear in $\a$, we shall study in detail only the terms linear in $\a$
of $\AA^{(K)}(\a,J,0,\ps)$, whose kernels are given, for $\O$ a generic multi-index, such as
$(\uo',\us';\uo,\us)$, by
\be\lb{1n2m} H^{(n+1;2m)(K)}_{\o'', s'';\O}(\zz;\uw;\ux,\uy) :=
\prod_{i=1}^{n}{\dpr \over \dpr J_{\ww_i,\o'_i, s'_i}} \times
\prod_{i=1}^{m}{\dpr \over \dpr \ps^+_{\xx_i,\o_i,s_i}} {\dpr
\over \dpr \ps^-_{\yy_i,\o_i,t_i}} {\dpr \AA^{(K)}\over
\dpr\a_{\zz,\o'',s''}}(0,0,0,0)\;. \ee
By using the definitions of $\AA_0$ and $\AA_-$, we can get an explicit formula for the
Fourier transforms of the kernels $H^{(n+1;2m)(h)}_{\o'', s'';\O}$. In fact, \pref{ep1a}
implies that
\be\lb{prch}\bsp &{\dpr \AA^{(K)}\over \dpr\ha_{\pp,\o,s}}(0,J,0,\ps) =
\t^-_N D_{-\o}(\pp) \hJ_{-\pp,\o,s}+
\\
+e^{-\VV^{(K)}(\psi,J,0}) &\Bigg[
\frac1{L^2}\sum_{\qq\in\DD'_L} C_{\o}(\qq+\pp,\qq) {\dpr^2
e^{\VV^{(K)}}\over \dpr\hh^+_{\qq,\o,s} \dpr\hh^-_{\pp+\qq,\o,s}}(\psi,J,0)-\\
&-\sum_{\o'',s''}D_{-\o}(\pp)\n^{\o\o''}_{ss''}(\pp) {\dpr
e^{\VV^{(K)}}\over \dpr\hJ_{\pp,-\o'',s''}}(\psi,J,0)\Bigg] \esp\ee
In a similar way, see \pref{wt2} in App. \ref{appB}, one can prove that:
\be\lb{gpc} {\dpr e^{\VV^{(k)}}\over\dpr
\h^\e_{\xx,\o,s}}(\ps,J,\h) =\ps_{\xx,\o,s}^{-\e}e^{\VV^{(k)}(\ps,J,\h)}
+\e\int\! d\uu\ g^{[k+1,N]}_\o(\xx-\uu) {\dpr e^{\VV^{(k)}}\over
\dpr \ps_{\uu,\o,s}^\e}(\ps,J,\h) \ee
By using this identity, we can expand the r.h.s. of \pref{prch} and we get, by some
simple algebra, that
\be\nn\bsp {\dpr \AA^{(K)}\over \dpr\ha_{\pp,\o,s}}(0,J,0,\ps) &=
\t^-_N D_{-\o}(\pp) \hJ_{-\pp,\o,s}+
{1\over L^2}\sum_{\qq\in\DD'_L} C_{\o}(\qq+\pp,\qq)
\hp^+_{\qq+\pp,\o,s}\hp^-_{\qq,\o,s} +\\
&+ W_{\AA,1}(J,\ps) + W_{\AA,2}(J,\ps)\esp\ee
where $W_{\AA,1}(J,\ps)$ and $W_{\AA,2}(J,\ps)$ are graphically represented in
Fig. \ref{z5c} and Fig. \ref{z5cc}, respectively; the filled triangle together with the
wiggly line represents $\dpr \AA_0/ \dpr\ha_{\pp,\o,s}$, while the wiggly line enclosed between
two filled points represents $\sum_{\o'',s''} D_{-\o}(\pp)\n^{\o\o''}_{ss''}(\pp)$.
Note that in the terms contributing
to $W_{\AA,1}(J,\ps)$ both $\ps$ fields of the interaction $\AA_0$ or $V(\psi)$ are
contracted; in the case of $W_{\AA,2}(J,\ps)$, only one of the $\ps$ fields
of the interaction $\AA_0$ is contracted.
\insertplot{350}{70} {} {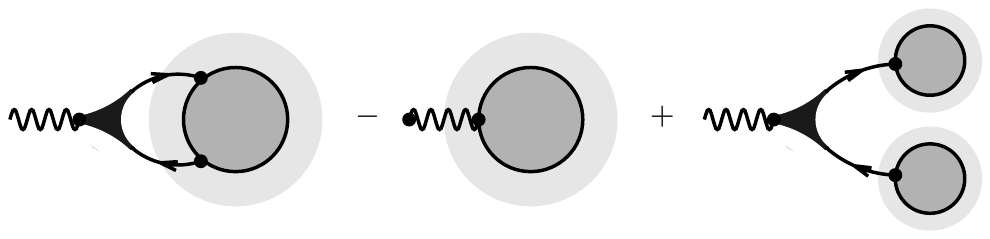}{\lb{z5c}: Graphical representation
of $W_{\AA,1}(J,\ps)$.
The corresponding kernels $H^{(n+1;2m)(K)}$ are obtained by substituting
the halos with $2m$ fermion lines and $n$ $J$ lines.
The full triangle represents the $C_\o$ operator;
the wiggly line represents, in the first and the third graph, the $\a$ external field;
the wiggly line between two points represents the Fourier transform of $D_{-\o}(\pp)\n^{\o\o''}_{ss''}(\pp)$.}{0}
\insertplot{300}{75} {} {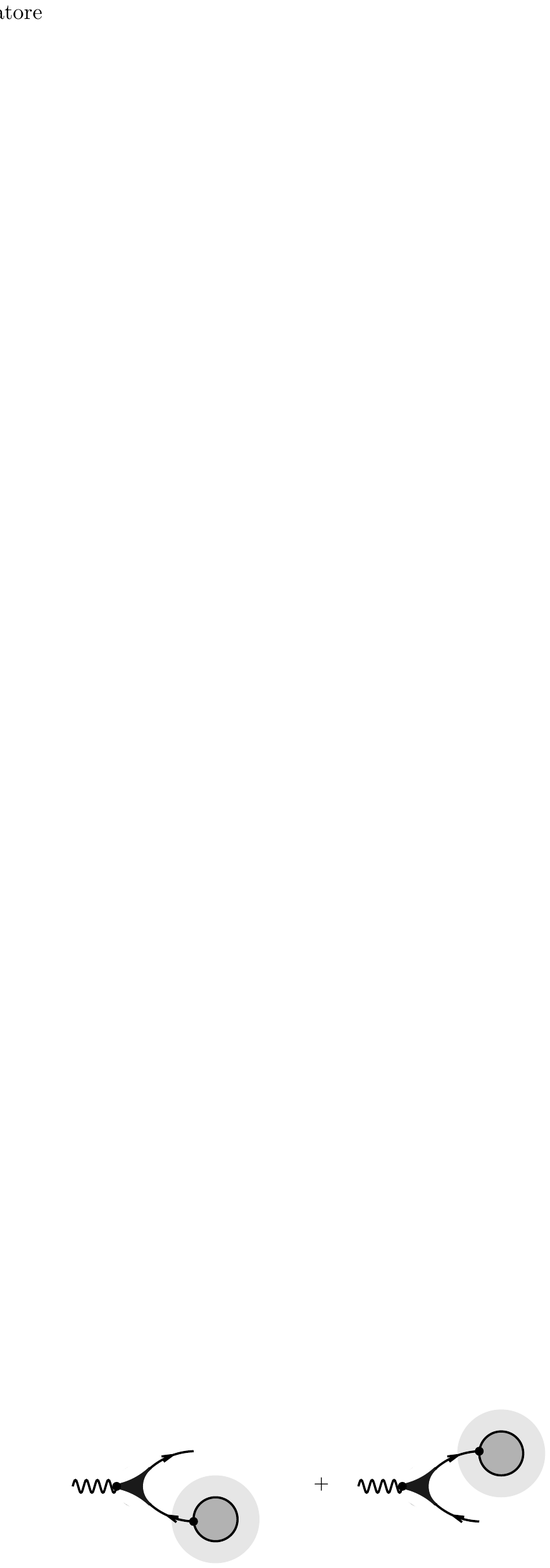}{Graphical representation
of $W_{\AA,2}(J,\ps)$. The corresponding kernels $H^{(n+1;2m)(K)}$ are obtained by substituting
the halos with $2m-1$ fermion lines and $n$ wiggly lines.\lb{z5cc}}{0}
%
%
%

For the analysis of the corresponding kernels,
it is convenient to extract from the $\AA_0$ vertex
the factors $D_{\o'}(\pp)$ of \pref{du} and to separate the first two terms in Fig. \ref{z5c}
from the third one. Hence, we make the following decomposition of the kernels Fourier
transforms (without the momentum conservation delta):
\be\lb{102m} \bsp &\hH^{(n+1;2m)(K)}_{\o,s;\O}(\pp;\up';\uk^+,\uk^-) =
\sum_{\s=\pm}D_{\s\o}(\pp)\hK^{(n+1;2m)(K)}_{\s;\o,s;\O}(\pp;\up';\uk^+,\uk^-)+\\
&+ \sum_{\s=\pm}D_{\s\o}(\pp)\hH^{(n+1;2m)(K)}_{\s;\o,s;\O}(\pp;\up';\uk^+,\uk^-)
+\hH^{(n+1;2m)(K)}_{\#;\o,s;\O}(\pp;\up';\uk^+,\uk^-)\esp\ee
where $\hK^{(n+1;2m)(K)}_{\s;\o,s;\O}$ is related to the sum of the first two
graphs in Fig. \ref{z5c}, $\hH^{(n+1;2m)(K)}_{\s;\o,s;\O}$ is related to the third
one and $\hH^{(n+1;2m)(K)}_{\#;\o,s;\O}$ is related to the
sum of the graphs in Fig. \ref{z5cc}. Our notation implies that $\up'$ are the momenta
of the $n$ $J$ fields, while $\uk^\pm$ are the momenta of the $m$ external $\psi^\pm$ fields
and $\pp=-\sum_{i=1}^n\pp'_i + \sum_{j=1}^m (\kk^+_j - \kk^-_j)$.
Note that $\hH^{(n+1;0)(K)}_{\s;\o,s;\O}=0$, for any $n$.

Let us consider first the kernels corresponding to the functional $W_{\AA,1}(J,\ps)$.
The following lemma shows that, if $\n^\o_s(\pp)$ is chosen as in \pref{4.2} and $m\ge 1$,
the kernels $\hK^{(n;2m)(K)}_{\s;\o,s;\O}$ and $\hH^{(n;2m)(K)}_{\s;\o,s;\O}$
are {\it vanishing} in the limit $N\to\io$ and that the same is true for
$\hK^{(2;0)(K)}_{\s;\o',s'}(\pp)$.

\begin{lemma}\lb{l3.3}
If $\n^\o_s(\pp)$ is chosen as in \pref{4.2}, there exist two
constants $C_0>0$ and $\th\in (0,1)$, such that, if
$\bar g$ is small enough, $n\ge 1$ and $m\ge 1$, then, for $\e\ge 0$,
\bal\lb{hbh2} &|\hK^{(n;2m)(K)}_{\s;\o,s;\O}(\pp;\up';\uk^+,\uk^-)|\le
C_0^{n+m}\bar g^{d_{1,n,m}}\g^{(2-m-n)K}\g^{-\th(N-K)}
\\\lb{hbh3}
&|\hH^{(n;2m)(K)}_{\s;\o,s;\O}(\pp;\up';\uk^+,\uk^-)|\le C_0^{n+m}\bar
g^{d_{2,n,m}}\g^{(2-m-n)K}\g^{-\th(N-K)} \eal
with $d_{1,n,m}\ge 1$ and $d_{2,n,m}\ge 2$.
Moreover,
\be\lb{hbh4} |\hK^{(2;0)(K)}_{\s;\o,s;\o',s'}(\pp)| \le C_0 \bar g \g^{-\th(N-K)}\ee
\end{lemma}
\bpr
The bound \pref{61} implies that $|S_{\o',\o}^{(N,j)}(\zz;\xx,\yy)|$ has the same bound of
$|g^{(N)}_\o(\xx-\zz)|\;|g^{(j)}_\o(\yy-\zz)|$.
Therefore, if we bound $|\hK^{(n;2m)(K)}_{\s;\o,s;\O}|$ and
$|\hH^{(n;2m)(K)}_{\s;\o,s;\O}|$ with the $L_1$ norm of $K^{(n;2m)(K)}_{\s;\o,s;\O}$ and
$H^{(n;2m)(K)}_{\s;\o,s;\O}$, respectively, we can proceed as in the proof of Lemma \ref{t3.2}.

An important role in the proof will have the property \pref{proU} of $\hU^{(i,j)}_\o(\qq+\pp,\qq)$,
which implies that in the multiscale expansion of $K^{(n;2m)(k)}_{\s;\o,s;\O}$ and $H^{(n;2m)(k)}_{\s;\o,s;\O}$
at least one of the two $\ps$-fields in $\AA_0$ has to be integrated at scale $N$.
%
%
This implies, in particular, that we can bound $\hH^{(n;2m)(K)}_{\s;\o,s;\O}$ by
$$C_2 \|b_N\|_{L_1}\sum_{j=K+1}^N \|b_j\|_{L_1}
\sum_{n_1+n_2=n-1\atop
 m_2+m_2=m+1}\|W^{(n_1;2m_1)(K)}\|\;\|W^{(n_2;2m_2)(K)}\|\;;$$
Hence, if we use \pref{pc1} with $k=K$, we easily get \pref{hbh3} with
$\th=1$.

In order to prove the bounds \pref{hbh2} and \pref{hbh4}, instead, we have to take
advantage of  partial cancelations.  We can write
\be\lb{71ter} \bsp &K^{(n;2m)(K)}_{\s;\o,s;\O}(\zz;\ux;\uy)
=\sum_{i,j=K}^N\int\! d\uu d\ww\;Z^2
S^{(i,j)}_{\s\o,\o}(\zz;\uu,\ww)W^{(n-1;2m+2)(K)}_{\O,\o,s}(\ux;\uy,\uu,\ww)\\
&-\d_{\s,-1}\sum_{\o'', s''}\int\! d\ww\;
 \n_{ss''}^{\o\o''}(\zz-\ww)W^{(n;2m)(K)}_{-\o'',s'',\O}(\ww;\ux,\uy)
+\d_{\s,-1} \d_{m,0} \d_{n,2} \d_{(\o,s),\O} \frac{\t^-_N}{Z^2}
\esp\ee
where $\n_{ss''}^{\o\o''}(\zz)$ is the Fourier transform of
$\n_{ss''}^{\o\o''}(\pp)$.

\*

{\it\0a)  Bound for $K^{(n;2m)(K)}_{-;\o,s;\O}$ for  $n,m\ge 1$.}
If we take in \pref{prc1} $2m+1$ derivatives with respect to the $\psi$ field and
$n-1$ derivatives with respect to the $J$ field, calculated at $J=\psi=0$, we
get an expansion for $W^{(n;2m+2)(K)}_{\O,\o,s}$, which has the structure
of the r.h.s. of Fig. \ref{z8q}, without the first term (since $m\ge 1$). If we
now insert this expansion in the first term of the r.h.s. of \pref{71ter}, we get an
expansion for $K^{(n;2m)(K)}_{-;\o,s;\O}$, which is graphically represented in Fig.\ref{z5d}.
\insertplot{600}{310} {} {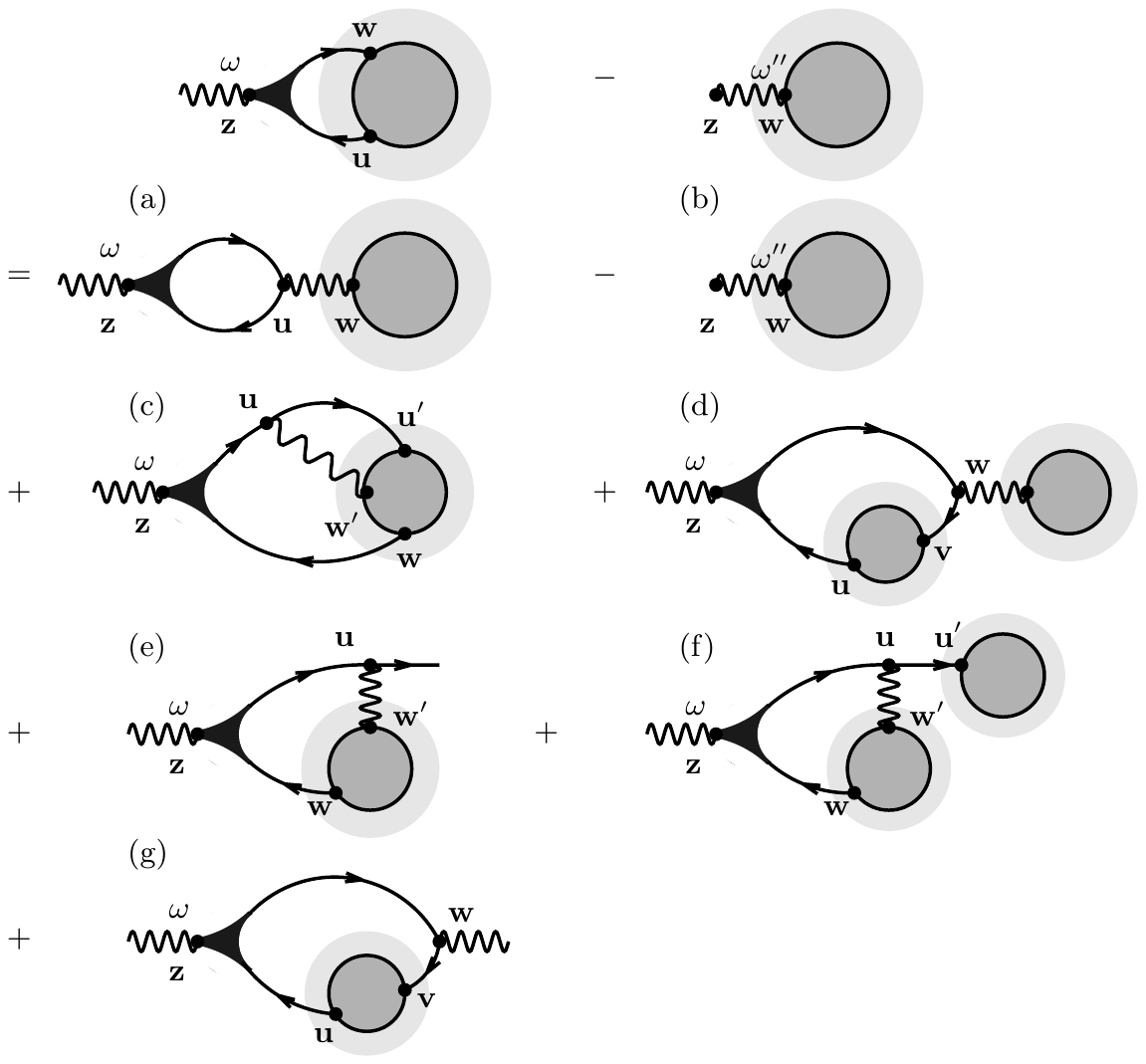}{\lb{z5d}: Graphical representation
of $K^{(n;2m)(k)}_{-;\o,s;\O}$. The identity is valid if $m\ge1$.
The internal wiggly lines represent $h_{\Th,\Th'}^{L,K}$; the others represent the
$\a$ or $J$ fields or, if they are closed by two full points, the Fourier transform of
$\n^{\o\o''}_{ss''}(\pp)$; the full triangle represents here the operator $C_\o(\qq+\pp,\qq) D_{-\o}(\pp)$.
The graph (g) is present only if $n\ge 2$ (one wiggly line being already attached to $\ww$).}{0}

We shall bound the terms corresponding to the graphs in the r.h.s.
of Fig. \ref{z5d}, by the same procedure used in \S\ref{ss2.5b};
though this time we also want to find the  exponential
small factor $\g^{-\th(N-K)}$.
A crucial role plays a cancelation between the terms corresponding to the
graphs $(a)$ and $(b)$, whose sum can be represented in the following way:
\be\lb{75} \sum_{\o''}\int\! d\uu\; \lft[\frac{Z^2}{\t^-_N} \sum_{i,j=K+1}^N
S^{(i,j)}_{-\o,\o}(\zz;\uu,\uu) -\d(\zz-\uu)\rgt]
\int\!d\ww\;
\n^{\o\o''}_{ss''}(\uu-\ww)W^{(n;2m)(K)}_{-\o'',s'';\O}(\ww;\xx,\yy)
\ee
Using the identity \pref{idb} and the definition \pref{3.14a} of $\t^-_N$, we have
\be\lb{125b} \bsp &\sum_{\o''}\sum_{i,j=K+1}^N\int\! d\uu d\ww \;
\frac{Z^2}{\t^-_N} S^{(i,j)}_{-\o,\o}(\zz;\uu,\uu)\n_{ss'}^{\o\o''}(\uu-\ww)
W^{(n;2m)(K)}_{-\o'',s'';\O}(\ww;\xx,\yy)\\
&=\sum_{\o''}\int\! d\ww\;  \n_{ss'}^{\o\o''}(\zz-\ww)
W^{(n;2m)(K)}_{-\o'',s'';\O}(\ww;\xx,\yy)
+ \sum_{p=0,1}\sum_{\o''}\sum_{i,j=K+1}^N \int\! d\uu \;
S^{(i,j)}_{-\o,\o}(\zz;\uu,\uu)\cdot\\
&\cdot \int_0^1\!dt\; \int\! d\ww\;(u_p-z_p)  (\dpr_p\n_{ss'}^{\o\o''})(\zz-\ww+t(\uu-\zz))
W^{(n;2m)(K)}_{-\o'',s'';\O}(\ww;\xx,\yy) \esp\ee
The first term in the r.h.s. of \pref{125b}, whose dimensional bound is divergent for $N\to\io$,
is indeed finite and is exactly canceled by the second term in \pref{75},
which then acts as a (finite) counterterm. Note that
the interpolation \pref{idb} was used also to control the graphs $(b2)$
and $(b3)$ of Fig. \ref{z3q}; however,
in that case the cancelation was due only to the symmetries of the propagator.

We are then left with the second term in the r.h.s. of \pref{125b}, which has the
estimate we wanted; in fact, by using that one between $i$ and $j$ is equal to
$N$, the bound \pref{pc1} on $\|W^{(n;2m)(K)}\|$, the bound \pref{61} on
$b_j(\xx)$ and the bound $\|\dpr h_{L,K}\|_{L_1}\le c_0 \g^K$, we can bound its norm by
\be\hspace{-.115cm} C_3\bar g \|W^{(n;2m)(K)}\| \|\dpr h_K\|_{L_1} \|{|\xx|}b_N\|_{L_1}
\sum_{j=K+1}^N \|b_j\|_{L_\io}  \le C_4
C^{n+m}(C_1\bar g)^m\g^{-(N-K)}\g^{(2-m-n)K} \ee

Let us now consider the graph $(c)$. Since either $i$ or $j$ of the
kernel $S^{(i,j)}_{-\o,\o}$ has to be equal to $N$, we obtain the bound
\be\lb{3.42} C_5 \bar g \|h_{L,K}\|_{L_\io}
\|W^{(n;2+2m)(K)}\| {\sum_{i,j,k=K}^N}^{\raisebox{-12pt}{\hspace{-6pt}*\ }}
\sup_{\ww,\uu'}\int\!d\zz d\uu \; b_i(\zz-\ww) b_j(\zz-\uu) |g^{(k)}_\o(\uu-\uu')| \ee
where $*$ is the constraint that at least one between $i$ and $j$
has to be $N$. Since the $L_\io$ and $L_1$ norm of $b_j(\xx)$ and $g_\o^{(j)}(\xx)$
are dimensionally equivalent, the best bound for the integral in \pref{3.42}
(as in the analogous bound of the graph $(b1)$ of  Fig.\ref{z3q}) is obtained
by taking the $L_\io$ norm of the function with the smallest scale index and the $L_1$ norm of the
others; it is then easy to see that
\be\lb{cases}E_{N,K} :=  {\sum_{i,j,k=K}^N}^{\raisebox{-12pt}{\hspace{-6pt}*\ }}
\sup_{\ww,\uu'}\int\!d\zz d\uu \; b_i(\zz-\ww) b_j(\zz-\uu) |g^{(k)}_\o(\uu-\uu')|\le
C \g^{-N} (N-K)\ee
By using also the bound \pref{pc1} for $\|W^{(n;2+2m)(K)}\|$ and the
bound $\|h_{L,K}\|_{L_\io} \le c_0\g^{2K}$, we see that \pref{3.42} can be bounded by
\be\lb{222} C_6 C^{n+m+1}(C_1\bar g)^{m+1} \g^{-\th(N-K)}
\g^{(2-m-n)K}\ee
for any $0<\th <1$ ($C_6$ is divergent for $\th\to 1$).

For graph $(d)$ a convenient estimate is given by
\be\lb{3.45} E_{N,K} \sum_{n_1+n_2=n, n_1\ge 1\atop m_1+m_2=m} F_{n_1,2m_1}
\|W^{(n_2;2+2m_2)(K)}\| \ee
where $F_{n_1,2m_1}$ is equal to either $\|h_K\|_{L_1}\|W^{(n_1;2m_1)(K)}\|$, if $(n_1,2m_1)\neq (2;0)$,
or $\|W^{(1;2)(K)}_{(b),\O}\|$, if $(n_1,2m_1)=(2;0)$ (see Fig.
\ref{z3q} and bound \pref{b2b}).
Therefore, by using \pref{pc1}, \pref{b2b} and \pref{cases}, we can bound
\pref{3.45} by
\be  C_7 C^{n+m+1} (C_1\bar g)^{\tilde d_{n,m}} \g^{-\th(N-K)} \g^{(2-m-n)K}
\virg \tilde d_{n,m} \ge 2\ee

For the graphs $(e)$ and $(f)$ the argument is similar; for $(f)$, for
example, the bound is
$$ \|h_K\|_{L_\io} \|b_N\|_{L_1}\sum_{j=K+1}^{N}\|b_j\|_{L_1}
\|g^{[K+1,N]}\|_{L_1} \sum_{n_1+n_2=n-1\atop
m_1+m_2=m+1, m_i\ge 1}\|W^{(n_1;2m_1)(K)}\| \|W^{(n_2;2m_2)(K)}\|]\;,
$$
which is less than $C_8 C^{n+m+1} (C_1\bar g)^{\tilde d_{n,m}} \g^{-(N-K)}
\g^{(2-m-n)K}$, with $\tilde d_{n,m} \ge 2$. Finally, the graph $(g)$ has to
be considered only in the case $n\ge 2$; it is easy to see that
the wanted bound can be obtained with the same procedure used for
the graph $(d)$.

\*

{\it\0b) Bound for  $K^{(n;2m)(K)}_{+,\O}$ for  $n,m\ge1$.} The
graph expansion of $K^{(n;2m)(K)}_{+,\O}$ is given again by
Fig.\ref{z5d};  the only  differences is that  the graph $(b)$ is
missing (because of $\d_{\s,-1}$ in \pref{71ter}).  Hence a bound
can be obtained as before, with only one
important difference: the contribution that in the previous
analysis was compensated by $(b)$ now is of order $\g^{-N}/L$ by the first
bound in \pref{3.14}. Hence, the result is the same.

\*

{\it\0c) Bound for  $K^{(2;0)(K)}_{\s;\o,s;\O}$.}
In this case the last term in the r.h.s of \pref{71ter} gives a contribution
different from $0$, for $\s=-1$, and the kernel $W^{(1;2)(K)}_{\O,\o,s}$ contains a term of
order $0$ is $\bar g$, corresponding to the kernel $w^{(1;2)}_{\Th,\Th'}$, defined in \pref{2.23}.
Hence, if we expand $W^{(1;2)(K)}_{\O,\o,s}$, $\O=(\o',s')$, as in Fig.\ref{z2q}, we obtain the
graphical representation of Fig.\ref{z6q}. In particular, the graph
$(e)$ comes from the kernel $w^{(1;2)}_{\Th,\Th'}$ that is included
in the darker bubble of Fig.\ref{z2q}; moreover, the graphs $(d)$ and $(f)$ are
obtained by also using the identity in Fig.\ref{z1q} to extract a further
wiggly line.
%
\insertplot{420}{260} {} {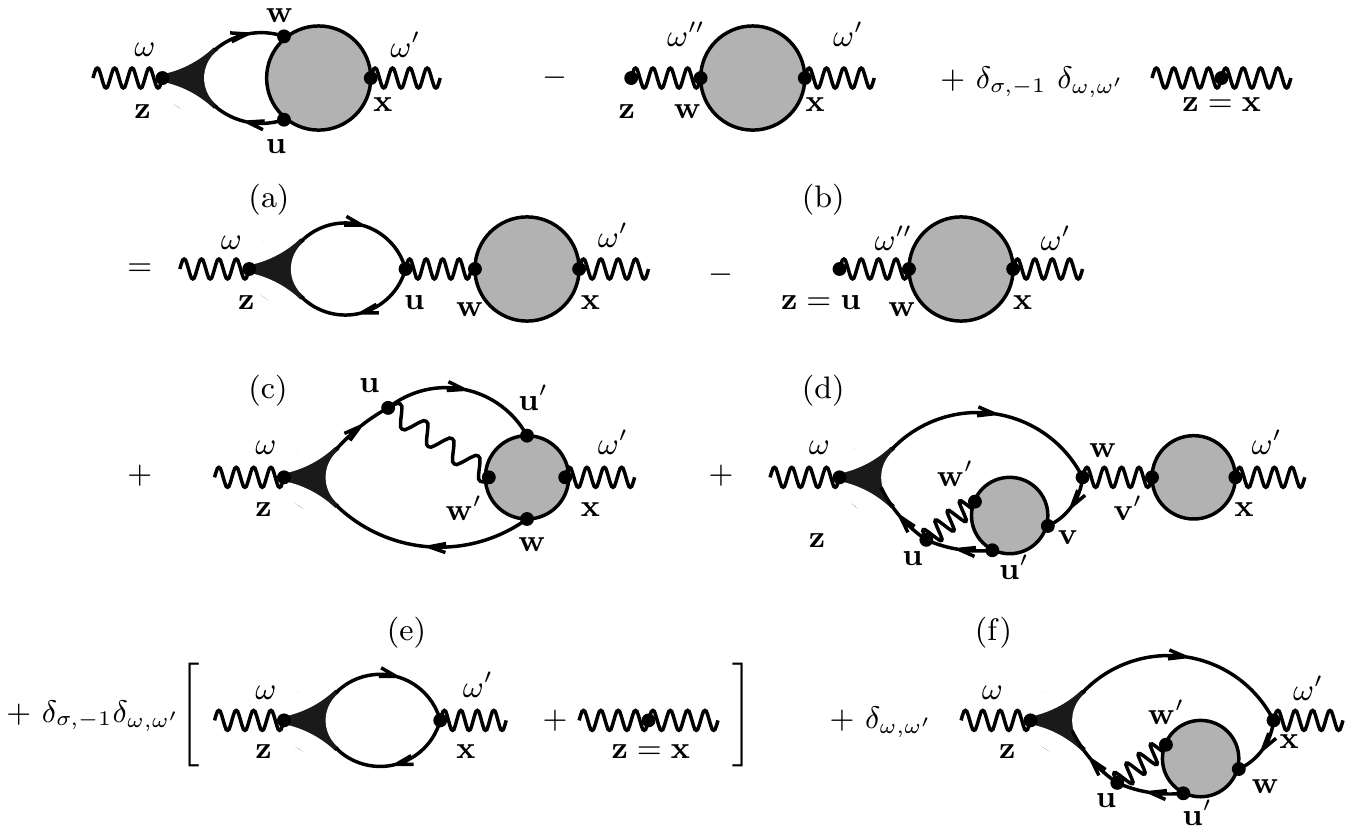}{Graphical representation of
$K^{(2;0)(K)}_{\s;\o,s;\O}$. All graphical objects are defined as in the previous pictures.\lb{z6q} }{0}
It is evident that the terms corresponding to graphs $(a)$, $(b)$, $(c)$, $(d)$ and $(f)$
can be bounded as those related to the graphs $(a)$, $(b)$, $(c)$, $(d)$ and $(g)$ in
Fig.\pref{z5d}, respectively; the only difference is in the number
and nature of the possible external lines, but that does not
change the topology of the graph, and therefore does not change
the approach to the bound. The only term that is not bounded by
the exponential small factor is that corresponding to graph $(e)$, which is
present only if $\s=-1$; however it is easy to see that this terms is a constant equal
to $-\t^-_N$, hence it is exactly canceled by the constant represented by the
graph with a single point between two wiggly lines.
%
Therefore also \pref{hbh4} is proved.
\epr

\*

Let us now analyze the kernels $\hH^{(n+1;2m)(K)}_{\#;\o,s;\O}(\pp;\up';\uk^+,\uk^-)$,
defined in \pref{102m}; recall that they are the kernels of the functional $W_{\AA,2}(J,\ps)$, graphically
represented in Fig. \ref{z5cc}. Since the two graphs in this picture have the same structure,
it is sufficient to analyze the contribution of the first one, which has the form
\be\lb{3.47} C_\o(\pp+\qq,\qq)\  \hg_\o^{[K+1,N]}(\qq) \psi^{+(\le K)}(\pp+\qq)
W^{(n;2m)(K)}_{\o,s;\O}(\up';({\uk'}^+,\qq),\uk^-)\ee
where  $\uk^+=({\uk'}^+,\pp+\qq)$.
By using \pref{ccc} and the definitions of \S\ref{ss2.5c}, it is easy to see that, if we
define $u_N(\kk)=1-\chi_{[-\io,N]}(|\kk|)$ (see also App. \ref{appC}), the expression
at left of $W^{(n;2m)(K)}_{\o,s;\O}$ in \pref{3.47} can be written as
\be\lb{3.48} u_N(\tilde\qq) \psi^{+(\le K)}(\pp+\qq) -  \hg_\o^{[K+1,N]}(\qq)
\lft[{1\over \c_{l,N}^{\e}(|\tilde\pp+\tilde\qq|)}-1\rgt] D_\o(\pp+\qq) \psi^{+(\le K)}(\pp+\qq)\ee
In the following sections, when we shall analyze also the integration of the infrared scales, the following
remarks will play an important role.

\*

\0a) The first term in \pref{3.48} gives no contribution, if $\pp$ is fixed and $N$ is large enough (how large
depending on $\pp$). In fact, the support properties of the field $\psi$ implies that $|\tilde\pp+\tilde\qq|\le
\g^{K+1}$, while $u_N(\tilde\qq)=0$ for $|\tilde\qq|\le \g^N$.

\*

\0b) The second term can give a non zero contribution only if the field $\psi^{+(\le K)}(\pp+\qq)$ is
contracted on scale $l$, that is if $|\tilde\pp+\tilde\qq| \le \g^{l+1}$.
However, at finite volume, $|\tilde\pp+\tilde\qq|\ge \sqrt{2(\p/L)^2}$, so that this condition can not be satisfied
if $\g^{l+1} \le \sqrt{2(\p/L)^2}$. It follows that the second term gives no contribution, if $L$ is fixed
and $l$ is small enough.

\subsection{Analysis of the correction term \pref{acca}. Integration of the IR scales and removed cutoffs limit.}
\lb{ss3.3}

The integration of the IR scales can be done as in \S\ref{ss2.3}, with $K=0$ for simplicity, by starting
the localization procedure at scale $0$. Since we need to analyze the correction term only for external
momenta of order one, we shall impose the condition that
\be\lb{smallp} 0<|\tilde\pp| \le 1 \ee
Moreover, we have to take into account that at scale $0$ the effective potential can be expanded in terms
of trees with one and only one $\a$ endpoint, which can be divided into three classes:

\0a) Those graphically represented in Fig. \ref{z5d}, that is those containing either one $\a$ endpoint with
interaction $\AA_-(\a,\psi)$ or one $\a$ endpoint with interaction $\AA_0(\a,\psi)$ (see \pref{acca}, whose
$\psi$ fields are both contracted on the UV scales; their kernels, by Lemma \ref{l3.3},
have the property that their Fourier transforms are bounded by a dimensional factor times a $\g^{-\th N}$ factor.

\0b) Those corresponding to the first term in \pref{3.48}, that is those with only one $\psi$ field of the
$\a$ endpoint with $\AA_0$ interaction contracted on the UV scales and a smooth kernel proportional to
$u_N(\tilde\qq)$; these terms vanish for $N\ge 2$, under the condition \pref{smallp}, so that we can
neglect them, since we want to send $N$ to $\io$.

\0c) Those with at most one $\psi$ field of the $\a$ endpoint with $\AA_0$ interaction contracted on the UV scales and singular kernel, that is $L^{-2}\sum_{\qq\in\DD'_L}
C_{\o}(\qq+\pp,\qq) \hp^+_{\qq+\pp,\o,s}\hp^-_{\qq,\o,s}$ and those  corresponding to the second term in \pref{3.48}.

Let us now suppose that the condition $\e_h\le \bar\e$ of Theorem \ref{th2.2} is satisfied, so that we can control the
tree expansion by localizing, besides the terms involved in the tree expansion of $W_{l,N}(J,\h)$, the only new marginal terms, that is those with one $\a$-vertex and two external $\psi$ field. If we do not localize the trees having a subtree
at scale $0$ of class c) above (whose kernel is singular), the localization procedure gives rise to a local term of the form
$$ \sum_{\o,s} \int d\xx \a_{\xx,\o,s,s} \frac{Z_j^{(3)}}{Z_{j-1}} \psi^+_{\xx,\o,s}\psi^-_{\xx,\o,s}$$
with $Z^{(3)}_j /Z_{j-1} \le c_0\g^{-\th N}$. As concerns the marginal terms whose trees have a subtree
at scale $0$ of class c), the identity \pref{proU} implies that one of the $\psi$ fields of the $\a$ vertex has
to be contracted at scale $l$, while the other, by momentum conservation, can be contracted only on a finite
set of scales around the scale $j_\pp=\max \{j: |\tilde\pp|\le \g^{j_\pp}\}$. It follows that this term,
because of the compact support properties of the single scale propagators, can be connected at a larger
cluster with only two external $\psi$ field only a finite number of times; hence, the fact that these terms can
not be localized is irrelevant and the bound of the corresponding trees can be done as in the case where
all the vertices dimensions are positive, see App. \ref{appA}. Moreover, since the scale $j>l$ of the tree vertex where
one of the $\psi$ fields of the $\a$ vertex is contracted is essentially fixed to the value $j_\pp$, we can extract
from the bound the "short memory" factor $\g^{-\th(j_\pp-l)}$ ($0<\th<1$), which goes to $0$ if $l\to-\io$ at fixed $\pp$.

Let us now consider the kernels $\hat\HH^{(n+1;2m)}_{\o,s;\O}(\pp;\up';\uk^+,\uk^-)$
of the functional $\HH_{l,N}(\a,J,\h)$ defined in \pref{acca}, in the limit $L\to\io$; we are using here a notation similar to that
used for the kernels \pref{102m} (in terms of graphs, one has to substitute the $2m$ external
$\psi$ fields with external propagators linked to $\h$ fields). Let us call $\uq$ the set $(\pp;\up';\uk^+,\uk^-)$
of external momenta; the previous considerations imply that, if the momenta $\uq$ are non exceptional, see \pref{nonex},
and we take the limit $N\to\io$, the values of all the trees go to zero, except those containing a subtree
at scale $0$ of class c), which vanish in the limit $l\to-\io$, as explained above. Hence, we have proved
the following Theorem.

\begin{theorem}\lb{th3.2}
If $\bar\e$ is defined as in Theorem \ref{th2.2} and $\e_h\le \bar\e$, uniformly in $l$ and $l\le h\le 0$,
the kernels  $\hat\HH^{(n+1;2m)}_{\o,s;\O}(\pp;\up';\uk^+,\uk^-)$ are well defined for any $l$ and $N$
and, if the momenta $(\pp;\up';\uk^+,\uk^-)$ are non exceptional,
\be\lb{zero_corr} \lim_{l\to-\io} \lim_{N\to\io} \lim_{L\to\io} \hat\HH^{(n+1;2m)}_{\o,s;\O}(\pp;\up';\uk^+,\uk^-) =0\ee
so that, in the same limit, the WI \pref{grez}, \pref{char}, \pref{spin} and \pref{wis} are satisfied without the correction term.
\end{theorem}

An application of this Theorem, which is important in this paper is the following.
If we take \pref{wis} with $\h=0$ and we perform a derivative w.r.t.
$\hJ_{-\pp,\m,s}$, we obtain, in the removed cutoffs limit,
a closed expression for the Fourier transform of the density operator correlation $\r_{\kk,\o,s} =
\psi^+_{\xx,\o,s} \psi^-_{\xx,\o,s}$, that is:
\be\lb{wis1} \la\hat\r_{\pp,\o',s'} \hat\r_{-\pp,\o,s}\ra_T  =
-D_{-\o}(\pp) {\hat h(\pp)\over 4\p Z^2 c} {M^\r_{\o',\o}(\pp)+
s's M^\s_{\o',\o}(\pp)\over 2} \virg \pp\not=0\ee
which implies that
\be\lb{wis2} \la\r_{\xx,\o',s'} \r_{\yy,\o,s}\ra_T = \frac12 \lft[
G^\r_{\o',\o}(\xx-\yy) + s' s\, G^\s_{\o',\o}(\xx-\yy) \rgt] \ee
where
$$G^\g_{\o',\o}(\xx-\yy) = {1\over 4\p Z^2 c} \int{d\pp\over (2\p)^2}\;
e^{i\pp(\xx-\yy)} {p_0^2 + c^2 p^2 \over D_\o(\pp)} M^\g_{\o',\o}(\pp)
$$

\subsection{Ward Identities in the limit $L,N\to\io$ at fixed $l$.}

If we take the limit $L,N\to\io$ at fixed $l$ and we consider external momenta of order
$\g^l$, \pref{zero_corr} is of course not true, but we can say that the correlation values only depend
on the trees containing a subtree at scale $0$ of class c), which can be easily estimated by
trivial dimensional bounds. Note that this result would be true, even if there were a contribution from the
other trees; what it is important is only that the effective potential on scale $0$ is well defined in the limit $N\to\io$.
In \S\ref{ss5.1} we shall use only the following bound. Let us define:
$$\hat\HH^{(1;2)}_{\o,s}(\pp;\kk^+,\kk^-) := \lim_{\e\to 0} {\dpr^3 \HH_{l,N}\over
\dpr\ha_{\pp,\m,s} \dpr\hh^+_{\kk^+,\o,s} \dpr \hh^-_{\kk^-}}(0,0,0)$$

\begin{lemma}\lb{lm3.4}
If $\kk^+=-\kk^-=\kk$ and $\pp=\kk^+-\kk^-=2\kk$, with $|\tilde\kk|=\g^l$, then
\be\lb{H21} \big|D_\o(2\kk)^{-1} \hat\HH^{(1;2)}_{\o,s}(2\kk;\kk,-\kk)\big| \le
C \bar\e^2 \frac{\g^{-2l}}{Z_l} \ee
\end{lemma}

\bpr
Note that, if $|\tilde\kk^+|=|\tilde\kk^-|=\g^l$, the term of order $0$ in $\e$ is given by
$D_\o(2\kk)^{-1} \hat U^{(l,l)}_{l,N,\o}(\kk,-\kk)$, which vanish, as follows from its explicit
value \pref{C.6}, since $u_l(\pm \kk)=0$. The same is true for the terms of order $1$ in $\bar\e$,
which are obtained by calculating the truncated expectation on scale $j=l,l+1$ of the product
of $T_\o(\kk^+,\kk^-)= C_{\o}(\kk^+,\kk^-) \psi^+_{\kk^+,\o,s} \psi^-_{\kk^-,\o,s}$ and
the local effective interaction \pref{blue}. In fact, the two terms proportional to $\d_j$ are obtained from
the term of order $0$ multiplying it by $\d_j k \hat g^{(j)}_\o (\kk)$, while the others can be represented
in terms of Feynmann graphs, which must contain a tadpole $\hg_\o^{(j)}(0)=0$.
The factor $\g^{-2l}$ simply reflects the dimension of $D_\o(2\kk)^{-1} \hat\HH^{(1;2)}$, which is the
same as the dimension of the correlation $<\hr_{2\kk,\o,s} \hp^-_{\kk,\o,s} \hp^+_{-\kk,\o,s}>_T$,
see \pref{corr_h}. On the contrary, the factor $Z_l^{-1}$, related to the field strength renormalization,
is different from the corresponding factor of $<\hr_{2\kk,\o,s} \hp^-_{\kk,\o,s} \hp^+_{-\kk,\o,s}>_T$,
that is $Z_l^{(1)} Z_l^{-2}$, for the following two reasons:

\0a) as discussed before, the potential $\AA_0(\a,\psi)$ do not need any renormalization, hence we must put $1$
in place of $Z_l^{(1)}$;

\0b) because of the bound \pref{61a}, if $j=l+1$, or \pref{C.6a}, if $j=l$, there is a $Z_l^{-1}$ missing
in the contractions of the two $\psi$ fields of $\AA_0(\a,\psi)$.\Halmos

\section{Closed equations and vanishing of the beta function}
\lb{ss4.0}

We want to prove two relevant properties of our model:

\bit
\item[a)] If $g_{1,\perp}=0$, the assumption $\e_h\le \bar\e$,
which allows us to remove the IR cutoff, by Theorem \ref{th2.2}, is indeed satisfied.
In order to prove such highly non trivial property, we first combine the {\em Schwinger-Dyson Equation}
(SDE) with the WI, then we take the limit $L\to \io$, followed by the limit $N\to \io$,
at fixed infrared cut-off $\g^l$. We get some relations among the Schwinger functions, which, if
computed at momenta of order $\g^l$, imply that the condition $\e_h\le\bar\e$
is satisfied for $h=l$, if this is true for $h=l+1$, provide $\bar \e$ is small enough.

\item[b)] By using the previous results, we can take the limit $l\to-\io$ at fixed non exceptional
momenta in the previous relations and we get {\em exact closed equations} among the correlations;
some of them will be used in \S\ref{ss6} to calculate the two point function and some response functions
needed in the proof of Theorem \ref{th1.2}.
\eit

\subsection{Combination of Schwinger-Dyson equations and Ward Identities in the $g_{1,\perp}=0$ case}

If we put again $Z=1$, for simplicity, and we define, as in \S\ref{ss3.1},
$\tilde\chi_N(\pp) = \tilde\chi_0(2^{-1}\g^{-N-1}|\tilde\pp|)$,
where $\tilde\chi_0(t)$ is a smooth positive function of support in $[0,2]$ and equal to $1$ for $t\le 1$,
the Schwinger-Dyson equations (in the limit $\e\to 0$, see \S\ref{ss2.5c}) are generated by the identity
\be\lb{ceq} {\partial e^{\WW_{l,N}(0,\h)} \over\partial
\hh^+_{\kk,\o,s}} = \frac{\chi_{l,N}(|\tilde\kk|)}{D_\o(\kk)} \lft[
\hh^-_{\kk,\o,s}  e^{\WW_{l,N}(0,\h)}
-\sum_{\m',s'} \frac{1}{L^2}\sum_\pp \frac{\n^{\o\m'}_{s s'}(\pp)}{\t^-_N}
\tilde\chi_{N}(\pp) {\dpr^2 e^{\WW_{l,N}(J,\h)}\over \dpr\hJ_{\pp,-\m',s'}\dpr
\hh^+_{\kk+\pp,\o,s}}\Bigg|_{J=0} \rgt]\ee
where $\kk\in\DD'_L$ (hence $\kk \not=0$), $\pp\in\DD_L$ and
$\n^{\o}_{s}(\pp)$ is defined as in \pref{4.2}.

Note that $\tilde\chi_{N}(\pp)=1$  in the finite set $\{\pp\in\DD_L:|\pp|\le 2\g^{N+1}\}$, so that it
has been freely introduced to remind us that $\pp=\kk^+ -\kk^-$, where $\kk^\pm$ are the momenta
of the $\hat\psi^\pm$ variables linked to $\hJ_\pp$, whose modula are both bounded by $\g^{N+1}$,
since $\e=0$. Hence,  \pref{ceq} is a straightforward consequence of the following
property: given any $F(\psi)$ which is a power series in the
field, if $\la \cdot\ra_0$ denotes the expectation w.r.t. the free measure,
\be \la \hp^-_{\kk,\o,s}F(\ps)\ra_0 =\hg^{[l,N]}_{{\rm D},\o}(\kk) \la
{\dpr F(\ps)\over \dpr \hp^+_{\kk,\o,s}}\ra_0. \ee
so that
\be {\partial e^{\WW_{l,N}} \over\partial \hh^+_{\kk,\o,s}}(0,\h) = \la
\hp^-_{\kk,\o,s}e^{\VV(\psi,0,\h)}\ra_0 =\hg^{[l,N]}_{{\rm D},\o}(\kk) \la
{\dpr\over \dpr \hp^+_{\kk,\o,s}}e^{\VV(\psi,0,\h)}\ra_0 \ee

We now combine the SDE \pref{ceq} with the functional WI \pref{wis}, in order to
get relations among the correlations, in two different ways.

\*

\0{\bf a)} The first family of relations involves the non connected correlations; some of these
relations will be used in \S\ref{ss6} to calculate the two point function and some response functions
in the removed cutoffs limit.

First of all, we decompose the sum over $\pp$ in two parts, by using the identity
\be\lb{chiln} \tilde\chi_N(\pp)=\tilde\chi_l(\pp)+ \tilde\chi_{l,N}(\pp) \virg \tilde\chi_{l,N}(\pp) :=
\tilde\chi_N(\pp) [1-\tilde\chi_l(\pp)]\ee
where $\tilde\chi_l(\pp):=\tilde\chi_0[2^{-1}\g^{-l}|\tilde\pp|]$; by this decomposition we divide
in two parts the second term in the r.h.s. of \pref{ceq}.
Then we multiply both sides of \pref{wis} by $\exp(\WW_{l,N})$,  we take
one derivative w.r.t. $\hh^+_{\kk+\pp,\o,s}$ and we insert the
result in the part of the second term of \pref{ceq} containing
$\tilde\chi_{l,N}(\pp)$. By using the oddness of the function $\hF^{\m}_{\o,s}(\pp)$ (defined below)
and the fact that the sum over $\pp$ is a finite sum,
we see that the term that we get, if we apply the derivative w.r.t. $\hh^+_{\kk+\pp,\o,s}$ to the variable
$\hh^+_{\kk+\pp,\m,s}$ in the decomposition \pref{3.4a} of $B_{\pp,\m,s}(J,\h)$, vanishes. Hence, we get,
if $\kk\not=0$:
\be\lb{ce}\bsp &D_\o(\kk) {\partial e^{\WW_{l,N}(0,\h)} \over\partial
\hh^+_{\kk,\o,s}} = \chi_{l,N}(|\tilde\kk|)\Bigg\{
\hh^-_{\kk,\o,s}  e^{\WW_{l,N}(0,\h)}
- \frac{1}{\t_N^-} \sum_{\m,t} \frac{1}{L^4} \sum_{\pp, \qq} \tilde\chi_{l,N}(\pp)
\hF^{-\o\m}_{-\o,st}(\pp)\cdot\\
&\cdot \left[\hh^+_{\qq+\pp,\m,t} {\partial^2 e^{\WW_{l,N}(0,\h)}\over
\partial \hh^+_{\qq,\m,t} \partial\hh^+_{\kk+\pp,\o,s}}-
{\partial^2 e^{\WW_{l,N}(0,\h)}\over
\partial\hh^+_{\kk+\pp,\o,s} \partial \hh^-_{\qq+\pp,\m,t}}
\hh^-_{\qq,\m,t}\rgt]+R_{\kk,\o,s}(\h)  + \tilde R_{\kk,\o,s}(\h)\Bigg\}\esp\ee

where, if $M^\g_{\o,\o'}$ is defined as in \pref{MM} and $\HH_{l,N}$ is defined as  in \pref{acca},
$$
\hF^{\m}_{\o,s}(\pp)= \sum_{\o',s'} \n^{\o\o'}_{ss'}(\pp)
M^{s'}_{\o',\o\m}(\pp) \virg M^s_{\m,\m'}
= \frac12 (M^\r_{\m,\m'} + s M^\s_{\m,\m'})
$$
\be\lb{remT} R_{\kk,\o,s}(\h)= \frac{1}{\t^-_N} \sum_{\o',s'} \frac1{L^2} \sum_{\pp}\tilde\chi_{l,N}(\pp)
\hat F_{-\o,ss'}^{-\o\o'}(\pp) \lim_{\e\to 0} {\dpr^2 e^{\HH_{l,N}}\over \dpr
\ha_{\pp,\o',s'} \dpr \hh^+_{\kk+\pp,\o,s}}(0,0,\h) \ee
\be\lb{remp} \tilde R_{\kk,\o,s}(\h)=-\frac{1}{\t^-_N} \frac{1}{L^2} \sum_{\m',s'} \sum_\pp \tilde\chi_l(\pp)
\n^{\o\m'}_{s s'}(\pp) {\dpr^2 e^{\WW_{l,N}(J,\h)}\over \dpr\hJ_{\pp,-\m',s'}\dpr
\hh^+_{\kk+\pp,\o,s}}\Bigg|_{J=0}
\ee

We will show that

\begin{theorem}\lb{th4.1}
The correlations generated by the functionals $R_{\kk,\o,s}(\h)$ and $\tilde R_{\kk,\o,s}(\h)$
vanish in the removed cutoffs limit (that is the limit $L\to\io$, followed by the limit $N\to\io$ and, finally,
by the limit $l\to-\io$), if the external momenta are non exceptional (see \pref{nonex})
and the condition $\e_h\le\bar\e$ is satisfied for any $h$.
Hence, under these conditions, we get from \pref{ce} and \pref{3.14}
a family of {\em exact closed equations} for the
non connected correlations, generated by the functional identity
\be\lb{ce_lim}\bsp &D_\o(\kk) {\partial e^{\WW_{l,N}(0,\h)} \over\partial
\hh^+_{\kk,\o,s}} =  \hh^-_{\kk,\o,s}  e^{\WW_{l,N}(0,\h)}
-4\p c \sum_{\m,t} \int \frac{d\pp d\qq}{(2\p)^6} \hF^{-\o\m}_{-\o,st}(\pp)\cdot\\
&\cdot \left[\hh^+_{\qq+\pp,\m,t} {\partial^2 e^{\WW_{l,N}(0,\h)}\over
\partial \hh^+_{\qq,\m,t} \partial\hh^+_{\kk+\pp,\o,s}}-
{\partial^2 e^{\WW_{l,N}(0,\h)}\over
\partial\hh^+_{\kk+\pp,\o,s} \partial \hh^-_{\qq+\pp,\m,t}}
\hh^-_{\qq,\m,t}\rgt]\esp\ee
\end{theorem}

\*

\0{\bf b)} The second family of relations, which is equivalent to the first one in the
removed cutoffs limit, involves truncated correlations. We shall use one of them to prove the
vanishing of the beta function.

First of all, we choose $\kk$ so that $\chi_{l,N}(|\tilde\kk|)=1$ and we
write \pref{ceq} in the equivalent form
\be\lb{ceq1}
\bsp {D_\o(\kk)} {\partial \WW_{l,N}(0,\h) \over\partial
\hh^+_{\kk,\o,s}} &= \hh^-_{\kk,\o,s}
-\sum_{\m',s'} \frac{1}{L^2}\sum_\pp \frac{\n^{\o\m'}_{s s'}(\pp)}{\t^-_N}\tilde\chi_{N}(\pp)
{\dpr \WW_{l,N}(J,\h)\over \dpr\hJ_{\pp,-\m',s'}}\Bigg|_{J=0}
{\dpr \WW_{l,N}(J,\h)\over \dpr \hh^+_{\kk+\pp,\o,s}}-\\
-\sum_{\m',s'} &\frac{1}{L^2}\sum_\pp \frac{\n^{\o\m'}_{s s'}(\pp)}{\t^-_N}\tilde\chi_{N}(\pp)
{\dpr^2 \WW_{l,N}(J,\h)\over \dpr\hJ_{\pp,-\m',s'}\dpr\hh^+_{\kk+\pp,\o,s}}\Bigg|_{J=0}
\esp\ee
Then we make in the last term of \pref{ceq1} the decomposition \pref{chiln}, we take
on both sides of \pref{wis} one derivative w.r.t. $\hh^+_{\kk+\pp,\o,s}$ and we insert the result
in the term of \pref{ceq1} containing  $\tilde\chi_{l,N}(\pp)$. We get
\be\lb{ceq2}
\bsp &{D_\o(\kk)} {\partial \WW_{l,N}(0,\h) \over\partial
\hh^+_{\kk,\o,s}} = \hh^-_{\kk,\o,s}
-\sum_{\m',s'} \frac{1}{L^2}\sum_\pp \frac{\n^{\o\m'}_{s s'}(\pp)}{\t^-_N}\tilde\chi_{N}(\pp)
{\dpr \WW_{l,N}(J,\h)\over \dpr\hJ_{\pp,-\m',s'}}\Bigg|_{J=0}
{\dpr \WW_{l,N}(J,\h)\over \dpr \hh^+_{\kk+\pp,\o,s}}\\
&- \frac{1}{\t_N^-} \sum_{\m,t} \frac{1}{L^4} \sum_{\pp, \qq} \tilde\chi_{l,N}(\pp)
\hF^{-\o\m}_{-\o,st}(\pp)
\left[\hh^+_{\qq+\pp,\m,t} {\partial^2 \WW_{l,N}(0,\h)\over
\partial \hh^+_{\qq,\m,t} \partial\hh^+_{\kk+\pp,\o,s}}-
{\partial^2 \WW_{l,N}(0,\h)\over
\partial\hh^+_{\kk+\pp,\o,s} \partial \hh^-_{\qq+\pp,\m,t}}
\hh^-_{\qq,\m,t}\rgt]\\
&+R'_{\kk,\o,s}(\h)  + \tilde R'_{\kk,\o,s}(\h)
\esp\ee
where the correction terms $R'_{\kk,\o,s}(\h)$ and $\tilde R'_{\kk,\o,s}(\h)$ are defined as \pref{remT} and \pref{remp}, with
$\HH_{l,N}$ and $\WW_{l,N}$ in place of the corresponding exponentials.

Since the truncated correlations can be written in terms of the untruncated ones, it is a priori true that, if the external
momenta are non exceptional, the correlations generated by $R'_{\kk,\o,s}(\h)$ and $\tilde R'_{\kk,\o,s}(\h)$
go to $0$ in the removed cutoffs limit;
hence, the relations \pref{ceq2} could be used to get directly exact closed equations even for the truncated correlations.
However, in the following we are only interested to consider the limit $L,N\to\io$ at fixed $l$, for external momenta of order
$\g^l$; in such case the correction terms do not vanish. In particular,
in \S\ref{ss5.1} we shall use only the bound of the kernels we get if we take in \pref{ce} three derivatives w.r.t. the
$\h$ variables at $\h=0$. Hence we define
\be\lb{H41c} \RR^{'(3)}_{\o,s,\o',s'}(\kk_1,\kk_2,\kk_3,\kk_4) :=
\frac{\dpr^3}{\dpr\h^-_{\kk_1,\o,s} \dpr\h^+_{\kk_2,\o,s} \dpr\h^-_{\kk_3,\o',s'}}
R_{\kk_4,\o', s'}(0)\ee
In a similar way we define $\tilde \RR^{'(3)}_{\o,s,\o',s'}(\kk_1,\kk_2,\kk_3,\kk_4)$.
We will show that

\begin{lemma}\lb{lm4.3}
Given a momentum $\kk$, such that $|\tilde\kk|=\g^l$,
\bal
\lb{H41a} |D_{\o'}(\kk)|^{-1} \tilde\RR^{'(3)}_{\o,s,\o',s'}(\kk,-\kk,\kk,-\kk)|
&\le C \bar\e^2 \g^{-4l} \frac{Z_l^{(1)}}{Z_l^3}\\
\lb{H41b} |D_{\o'}(\kk)|^{-1} \RR^{'(3)}_{\o,s,\o',s'}(\kk,-\kk,\kk,-\kk)|
&\le C \bar\e^2 \g^{-4l} \frac{1}{Z_l^2}
\eal
\end{lemma}

\subsection{Proof of Theorem \ref{th4.1}}

To prove the vanishing of the (untruncated) correlations generated by $\tilde R_{\kk,\o,s}(\h)$
in the removed cutoffs limit  is an easy task. In fact these correlations can be expressed as sums of
products of truncated correlations independent of $\kk$ times the correlations generated by
$$-\frac{1}{\t^-_N} \frac{1}{L^2} \sum_{\m',s'} \sum_\pp \tilde\chi_l(\pp) \n^{\o\m'}_{s s'}(\pp)
\lft[ {\dpr^2 \WW_{l,N}(J,\h)\over \dpr\hJ_{\pp,-\m',s'}\dpr \hh^+_{\kk+\pp,\o,s}}\Bigg|_{J=0}+
{\dpr \WW_{l,N}(J,\h)\over \dpr\hJ_{\pp,-\m',s'}}\Bigg|_{J=0}
{\dpr \WW_{l,N}(J,\h)\over \dpr\hh^+_{\kk+\pp,\o,s}}\rgt]$$
All the correlations generated by the second term vanish for $l\to-\io$, by momentum conservation;
in fact $\pp$, by the condition on the external momenta, has a fixed non zero value, so that $\tilde\chi_l(\pp)=0$
for $|l|$ large enough.
On the other hand, under the hypotheses of the theorem, the kernels of the
functional $W_{l,N}(J,\h)$ with one external $J$ field (of momentum $\pp$) and $2m$ external $\h$ fields
have a bound divergent but integrable for $\pp\to 0$. This claim, which
is easy to understand on the base of rough dimensional arguments, can be made rigorous by using
the (model independent) properties of the tree expansion used to prove Theorem \ref{th2.2}.
A detailed discussion in the case $m=1$ can be found in \S2.4 of \cite{BFM007} for the Thirring model.

To get the same result for the {\em correction term} \pref{remT}, we shall proceed as in \S\ref{ss3.1},
by introducing the following functional integral:
\bea\lb{tti} e^{\TT_{\m,t,l,N}(\b,\h)} = \int\! P_Z^{[l,N]}(d\ps)
e^{\VV(\ps,0,\h) +\BB_{\m,t,0}(\ps,\b)-\BB_{\m,t,-}(\ps,\b)} \eea
where, if $\{\hb_{\kk,\o,s}, \kk\in\DD'_L\}$ are external Grassmann variables,
\bal\lb{80} \hspace{-0.5cm}\BB_{\m,t,0}(\psi,\b)&= \sum_{\o,s} \frac1{L^6}\sum_{\pp,\kk,\qq}
\hat G^{-\m}_{-\o,t}(\pp) C_{\m\o}(\qq+\pp,\qq)\; \hb_{\kk,\o,s}
\hp^-_{\kk+\pp,\o,s} \hp^+_{\qq+\pp,\m\o,ts} \hp^-_{\qq,\m\o,ts}\\
\lb{80a} \hspace{-0.5cm}\BB_{\m,t,-}(\psi,\b) &= \sum_{\o,s\atop \o',s'}
\sum_{\o,s} \frac1{L^6}\sum_{\pp,\kk,\qq} \hat G^{-\m}_{-\o,t}(\pp)
D_{-\m\o}(\pp) \n^{\m\o\o'}_{tss'}(\pp)\;\hb_{\kk,\o,s} \hp^-_{\kk+\pp,\o,s}
\hp^+_{\qq+\pp,\o',s'} \hp^-_{\qq,\o',s'}\eal
where
$$\hat G^{-\m}_{-\o,t}(\pp) :=\frac{\tilde\chi_{l,N}(\pp)) \hat F^{-\m}_{-\o,t}(\pp)}{\t_N^-}$$
Then we have:
$$
{\dpr e^{\TT_{\m,t,l,N}}\over\dpr \b_{\kk,\o,s}}(0,\h) =
\frac1{L^2} \sum_{\pp} \hat G^{-\m}_{-\o,t}(\pp)
{\dpr^2 e^{\HH_{l,N}}\over \dpr \ha_{\pp,\m\o,st}\dpr
\hh^+_{\kk+\pp,\o,s}}(0,\h)
$$
The multiscale analysis of the r.h.s. of this identity, performed in the following subsections
will allow us to complete the proof of the theorem.

\subsubsection{Analysis of the correction term \pref{remT}. Integration of the UV scales.}
\lb{ss4.2}

Let us put $K=0$; after the integration of the fields $\psi^{(N)},...\psi^{(K+1)}$, we get
\be\lb{ep3}\; e^{\TT_{\m,t,l,N}(\b,\h)}=\int\!P_Z^{[l,0]}(d\ps)\
e^{\VV^{(0)}(\ps,0,\h) +\BB_{\m,t}^{(0)}(\b,\h,\ps)}\ee
and the functional $\BB_{\m,t}^{(0)}(\b,\h,\ps)$ satisfies the identity:
\be\lb{ep3a}e^{\BB_{\m,t}^{(0)}(\b,\h,\ps)} = \int P_Z^{[1,N]} (d\z)
e^{\VV(\psi+\z,0,\h)+ \BB_{\m,t,0}(\ps+\z,\b)-\BB_{\m,t,-}(\ps+\z,\b)}\ee
As in \S\ref{ss3.2} and \S\ref{ss2.2}, we shall consider, for simplicity, only the contributions to
$\BB_{\m,t}^{(0)}(\b,\h,\ps)$  with $\h=0$; the result can be easily extended to the general case.
Hence, since we have to evaluate only the terms linear in $\b$, we shall study in detail only the terms linear in $\b$
of $\BB_{\m,t}^{(0)}(\b,0,\ps)$, whose kernels are given, for $\O=(\uo',\us')$ a generic multi-index,
by (one has to take an odd number of derivatives w.r.t. the $\psi$ variables to get a non- zero result):
\be\lb{kert} T^{(2m+1)(0)}_{\m,t;\o,s;\O}(\vv;\uu;\ux,\uy) :=
\prod_{i=1}^{m}{\dpr \over \dpr \ps^+_{\xx_i,\o'_i,s'_i}} {\dpr \over
\dpr \ps^-_{\yy_i,\o'_i,s'_i}} \times {\dpr^2 \BB_{\m,t}^{(0)}\over
\dpr\b_{\uu,\o,s} \dpr\ps^-_{\vv,\o,s}}(0,0,0)\;. \ee
A more explicit formula for the kernels $T^{(2m+1)(0)}_{\m,t;\O}$ is obtained
by using in \pref{kert} the identity:
\bal {\dpr \BB^{(0)}_{\m,t}\over\dpr \b_{\kk,\o,s}}&(0,0,\ps) =\frac1{L^4}\sum_{\pp,\qq}
\hat G^{-\m}_{-\o,t}(\pp) C_{\m\o}(\qq+\pp,\qq)
\; e^{-\VV^{(0)}(\ps,0,0)} {\dpr^3 e^{\VV^{(0)}(\ps,0,\h)}\over \dpr
\hh^+_{\qq,\m\o,ts}\dpr \hh^-_{\qq+\pp,\m\o,ts}\dpr
\hh^+_{\kk+\pp,\o,s}}\Big|_{\h=0}\nn\\
& +\sum_{\o',s'}\int{d\pp \over
(2\p)^2} \hat G^{-\m}_{-\o,t}(\pp)D_{-\m\o}(\pp)
\n^{\m\o\o'}_{tss'}(\pp) \; e^{-\VV^{(0)}(\ps,0,0)} {\dpr^2
e^{\VV^{(0)}(\ps,0,\h)}\over \dpr \hJ_{\pp,\o',s'}\dpr \hh^+_{\kk+\pp,\o,s}}\Big|_{\h=0}
\lb{prct} \eal
%
If we now use again \pref{gpc} to expand the derivatives w.r.t. the $\h$ variables in the first
line of \pref{prct}, we get, by some simple algebra, that
\be\lb{krt2}{\dpr \BB^{(0)}_{\m,t}\over\dpr \b_{\kk,\o,s}}(0,0,\ps) =
W_{\TT,1}(\psi) + W_{\TT,\#}(\psi)\ee
where $W_{\TT,1}(\psi)$, which is graphically represented in Fig. \ref{z9c}, includes
all terms such that both $\hat\psi$ variables of the interaction $\AA_0$ (that is the variables
$\hp^+_{\qq+\pp,\m\o,ts}$ and  $\hp^-_{\qq,\m\o,ts}$ of \pref{80}) are contracted, while
$W_{\TT,\#}(\psi)$ includes the other terms.
We shall call $\hT^{(2m+1)(0)}_{1,\m,t;\o,s;\O}(\kk;\kk^-;\uq^+,\uq^-)$ and
$\hT^{(2m+1)(0)}_{\#,\m,t;\o,s;\O}(\kk;\kk^-;\uq^+,\uq^-)$ the corresponding kernels
%
\begin{figure}[h!]
\begin{center}
\includegraphics[viewport=0 0 500 310, width=364.5pt, height=226pt,
clip=true]{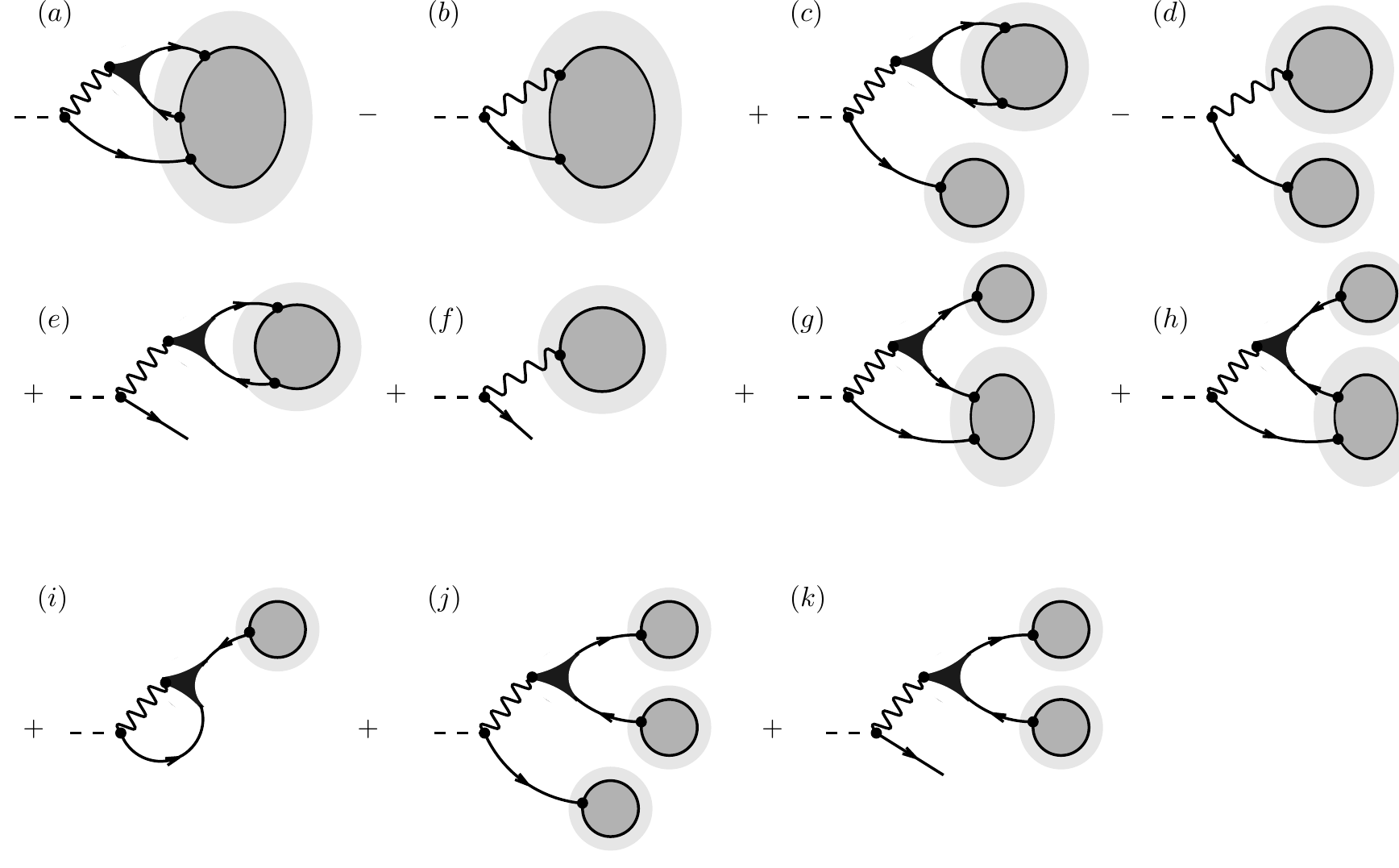}
\caption{\small Graphical representation of $W_{\TT,1}(\psi)$. The dashed line represents the external
Grassmann variable $\b$; the external full line represents a $\psi$ variable. The full triangle represents
the function $C_{\m\o}(\qq+\pp,\qq)$. The wiggling line represents the function $\hat F^{-\m}_{-\o,t}(\pp)$,
if it is linked to a triangle vertex, otherwise $\hat F^{-\m}_{-\o,t}(\pp)D_{-\m\o}(\pp)$.
\lb{z9c}}
\end{center}
\end{figure}

\begin{lemma}\lb{lm4.1}
There exist two constants  $C_1>0$  and $\th\in (0,1)$, such
that, if  $\bar g$ small enough, for  $\e=\pm1$ and $t=\pm1$,
\be\lb{hbt2}
|\hT^{(2m+1)(0)}_{1,\m,t;\o,s;\O}(\kk;\kk^-;\uq^+,\uq^-)|\le  C_0^m \bar g^2
\g^{-\th N} \ee
\end{lemma}

{\bf\0Proof.}
Let us put $\hat H_\s(\pp)=\hat G^{-\m}_{-\o,t}(\pp)D_{-\s\m\o}(\pp)$ and let us call $H_\s(\xx)$ its
Fourier transform. It is easy to see that:
\be\lb{Hbound} |\hat H_\s(\pp)| \le c_1 \bar g \virg \|H_\s\|_{L_\io} \le c_0 \bar g \ee
However, $H_\s(\xx)$  is not $L_1$ uniformly in $L$; the point is that the function $\hat H_\s(\pp)$, in the limit $L\to \io$, converges
to a function which is bounded, but has a discontinuity in $\pp=0$.
In any case, the second bound in \pref{Hbound} is sufficient to prove the bound \pref{hbt2} for the terms described, in Fig. \ref{z9c},
by graphs where the wiggling line belongs to a loop, since we can proceed in this case
as in the proof of Lemma \ref{l3.3}, by using the $L_1$ norm of
$T^{(2m+1)(0)}_{1,\m,t;\o,s;\O}$ as an upper bound of $|\hT^{(2m+1)(0)}_{1,\m,t;\o,s;\O}|$.

Let us consider first the sum $[\hT^{(a)(2m+1)(0)}_{1,\m,t;\o,s;\O}+
\hT^{(b)(2m+1)(0)}_{1,\m,t;\o,s;\O}](\kk;\kk^-;\uq^+,\uq^-)$ of the kernels associated to the
graphs $(a)$ and $(b)$ in Fig. \ref{z9c}; by using
the identity \pref{du} and the definition \pref{102m} of the kernels $\hK^{(1;2m)(0)}_{\s;\o,s;\O}$
(see also \pref{71ter}), we see that it can be written in the form:
\be \frac1{L^2} \sum_{\pp\not=0}
g^{[1,N]}(\kk+\pp) \sum_{\s=\pm} \hat G^{-\m}_{-\o,t}(\pp) D_{-\s\m\o}(\pp)
\hK^{(1;2m+2)(0)}_{\s;\o,s;\O}(\pp;0;\tilde\uq^+,\tilde\uq^-)
\ee
if we put $\uq^+ =(\tilde\uq^+, \kk+\pp)$ and $\uq^-=(\tilde\uq^-,\kk^-)$.
By using the second bound in \pref{Hbound} and \pref{hbh2}, we can bound the $L_1$ norm of the
Fourier transform of this expression by
$$c_0 \bar g \|g^{[1,N]}\|_{L_1}
\sum_\s \|K_\s^{(1;2m+2)(0)}\| \le C_1^m \bar g^2 \g^{-\th N}$$

Let us now consider the sum of the kernels associated to the graphs $(c)$ and $(d)$ in Fig. \ref{z9c}. In this case
we can not bound the l.h.s. of \pref{hbt2} by the $L_1$ norm of the
Fourier transform, but it is easy to see, by using the first bound in \pref{Hbound},
\pref{hbh2} and \pref{pc1}, that we can bound it  by
$$c_1 \bar g \|g^{[1,N]}\|_{L_1} \sum_{m_1+m_2=m\atop m_2\ge 1} \|W^{(0;2m_1+2)(0)}\|\;
\|K^{(1;2m_2)(0)}\| \le C_1^m \bar g^3 \g^{-\th N}$$
Notice that the case $m_2=0$ is excluded, because $\hat F^{-\m}_{-\o,t}(0)=0$, but, in any case,
for any fixed $\pp\not=0$, $K^{(1;0)(0)}(\pp)$ converges, for $L\to \io$, to a function which vanishes for $\pp\to 0$.

The bound of the sum of the kernels associated to the graphs $(e)$ and $(f)$ (that give
a contribution only for $m\ge 1$) is similar; we get
$$c_1 \bar g  \|K^{(1;2m)(0)}\| \le C_1^m \bar g^2\g^{-\th N}$$
The bound for the kernels associated to the remaining graphs can be done one at a time,
since there are no cancelation among them. For $(g)$, $(h)$ and $(i)$ we just use again that
the second bound in \pref{Hbound}, together with \pref{pc1}. Let us consider, for example,
the graph $(g)$; we get
\be\nn\bsp
c_0 \bar g \|g^{[1,N]}\|_{L_1} \|b_N\|_{L_1}&\sum_{j=1}^N
\|b_j\|_{L_1} \sum_{m_1+m_2=m\atop m_1\ge 1}
\|W^{(0;2m_1+2)(0)}\|\; \|W^{(0;2m_2+2)(0)}\|\;\\
&\le  C_1^m \bar g^3 \g^{-N}\esp\ee
Finally, for the graphs $(j)$ and $(0)$ we have to use the first bound in \pref{Hbound},
together to \pref{hbh2} and \pref{pc1} and the constraint
that one   of the $\ps$ field of $\AA_0$ has to be contracted on
scale $N$; we omit the details.\Halmos

\subsubsection{Analysis of the correction term \pref{remT}. Integration of the IR scales
and completion of the proof of Theorem \ref{th4.1}.}
\lb{ss4.3}

The integration of the IR scales can be done as in \S\ref{ss2.3}, by starting
the localization procedure at scale $0$, where the effective potential can be expanded in terms
of trees with one and only one $\b$-vertex, which can be divided into three classes:

\0a) Those graphically represented in Fig. \ref{z9c}, that is all those containing one $\b$ endpoint with
interaction $\BB_{\m,t,-}(\a,\psi)$ and those containing one $\b$ endpoint with interaction $\BB_{\m,t,0}(\a,\psi)$, whose $\psi$ fields of the triangle vertex are both contracted on the UV scales; their kernels, by Lemma \ref{lm4.1},
have the property that their $L^1$ norms are bounded by a dimensional factor times a $\g^{-\th N}$ factor.

\0b) Those graphically represented in Fig. \ref{z9d} (or with different  orientation of the arrows
in the lines linked to the triangle vertex), with the further condition the contracted line of the triangle vertex
has scale $N$ and gives a smooth contribution $\tilde\chi_N(\pp) u_N(\tilde\qq) D_{\o'}^{-1}$,
corresponding to the first term in \pref{3.48}.

\0c) The remaining ones, that is the interaction $\BB_{\m,t,0}(\a,\psi)$, those graphically represented
in Fig. \ref{z9d}, with the further condition the contracted line of the triangle vertex
has scale $j<N$ and gives a contribution corresponding to the second term in \pref{3.48}, and those with
both lines of the triangle vertex not contracted.

\insertplot{400}{80}{}{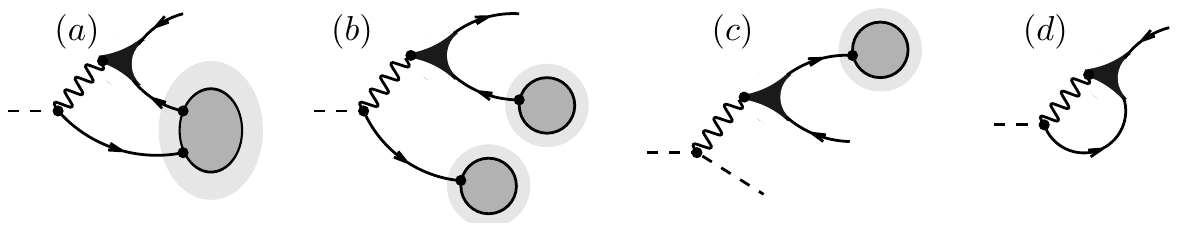}{Graphical representation of the contributions to $W_{\TT,\#}(\psi)$ with
the property that one and only one of the $\psi$ fields linked to the triangle vertex is contracted. One should
add the graphs obtained by exchanging the arrows of the lines linked to the triangle vertex.\lb{z9d}}{0}

Let us now suppose that the condition $\e_h\le \bar\e$ of Theorem \ref{th2.2} is satisfied, so that we can control the
tree expansion by localizing, besides the terms involved in the tree expansion of $W_{l,N}(J,\h)$, the
new marginal terms, that is those with one $\b$ external field and two or three external $\psi$ fields. As we shall see, it is sufficient to
localize only the terms with one external $\psi$ field, by putting (in the limit $L\to\io$)
$$\LL \int d\xx d\yy \b_{\xx,\o,s} w(\xx-\yy) \psi_{\yy,\o,s}= \hat w(0) \int d\xx \b_{\xx,\o,s}\psi_{\xx,\o,s}$$
Since $\hat w(0)=0$, by the oddness of the free propagator, this can be done without introducing any new
renormalization constant and we are left with only marginal terms. The consequence is that, given any tree
with root of scale $l-1$, all the tree vertices have positive dimension, except those belonging to the path $\CC$ which
connects the endpoint of type $\b$ with the root; these vertices may have dimension $0$.
However, if the tree has a subtree with root of scale $0$ belonging to the class a) above, its value has
a $\g^{\th N}$ factor in front of the dimensional bound.
If we write $\g^{\th N} = \g^{\th (N+l)/2} \g^{\th (N-l)/2}$, we can use the factor $\g^{\th (N-l)/2}$ to
regularize the vertices on $\CC$; hence, if we consider the tree expansion of any particular correlation function,
we can safely bound the sum over the trees of class a), and we are left with a factor $\g^{\th (N+l)/2}$ in front to it,
which goes to $0$, if $N\to \io$ at fixed $l$.
A similar argument can be done for the trees of class b), by using the bound \pref{61}.

Let us now consider the trees of class c). Because of the identity \pref{proU}, they are characterized
by the fact that one of the two lines linked to the triangle vertex in Fig. \ref{z9d} is contracted at scale $l$, while
the other is contracted at a scale $j\in [l,N-1]$ in a tree vertex that we call $v^*$, belonging to the path $\CC$
which connects the $\b$ endpoint $v_\b$, of scale $j_\b\ge j+1$, with the root. Note that $j_\b+1$ can be larger than
$j$ only if $j>l$ and the two $\psi$ fields linked to the triangle vertex  are still not contracted in the effective
potential of scale $h\in[j+1,j_\b]$. However, these vertices may have dimension $0$ only if the remaining $\psi$ field
of the $\BB_{\m,t,0}$ interaction is contracted in a propagator which enters a cluster with two external fields;
by the support properties of the free single scale propagators, this can happens at most two times; hence these vertices
do not need a regularization. As concerns the vertices between $v^*$ and the root, the regularization is ensured
by the bound \pref{61a}, which implies a ``gain factor'' $\g^{-(j-l)}$ with respect to the dimensional bound;
from this factor we can extract a factor $\g^{-(j_\b-l)}$, without destroying the summability of the tree expansion
of the correlation functions. On the other hand, if the external momenta are non exceptional, the momentum $\kk$
of the $\b$ field is different from $0$ and this implies that $j_\b$ is bounded from below, so that even the contribution
of the trees of class c) goes to $0$ as $l\to -\io$. This completes the proof of Theorem \ref{th4.1}

\subsection{Proof of Lemma \ref{lm4.3}}

First of all, note that the connectivity structure of both $\tilde\RR^{'(3)}_{\o,s,\o',s'}$ and
$\RR^{'(3)}_{\o,s,\o',s'}$ has to be the same as that of graph $a)$ of Fig. \ref{z9c} (with three $\h$
fields extracted from the halo and a $J$-vertex in place of the $\a$-vertex, for $\tilde\RR^{'(3)}_{\o,s,\o',s'}$), that is any Feymann
graph has to be connected without using the $\b$ vertex. It immediately follows that the lowest order contributions
are of the second order in $\bar\e$.

The bound for $\tilde\RR^{'(3)}_{\o,s,\o',s'}(-\kk,\kk,-\kk,\kk)$ can be proved by the same arguments
used in \cite{BM004} for proving a similar result (see Lemma A1.2 of that paper)
in the case of a model with spin $0$, local interaction and a fixed UV cutoff. Hence, we shall skip the proof,
but only note that the dimensional factors of the bound differ from those of the bound in \pref{corr_h} for the
four point correlations, which have the same formal scaling dimension, only because

\0a) the external propagator in the $\b$-vertex is substituted with $D_{\o'}(\kk)$, so that we ``loose" a $Z_l^{-1}$ factor;

\0b) we have to take into account that the internal $J$-vertex is renormalized, so that we ``gain" a factor $Z_l^{(1)}/Z_l$.

Let us finally consider $\RR^{'(3)}_{\o,s,\o',s'}(-\kk,\kk,-\kk,\kk)$. First of all note that, for a generic choice
of the external momenta, $\RR^{'(3)}_{\o,s,\o',s'}(\kk_1,\kk_2,\kk_3,\kk_4)$ is obtained from
$\RR^{(3)}_{\o,s,\o',s'}(\kk_1,\kk_2,\kk_3,\kk_4)$ by subtracting
the Feynmann graphs which are only connected through the $\b$ vertex. However, these graphs give a vanishing
contribution if $(\kk_1,\kk_2,\kk_3,\kk_4)=(-\kk,\kk,-\kk,\kk)$, thanks to the cutoff function $\tilde\chi_{l,N}(\pp)$
present in the definition \pref{remT}; in fact, for these graphs $\pp=\kk_4-\kk_3=2\kk$ and $\tilde\chi_{l,N}(\pp)=0$,
if $|\pp|\ge 2\g^l$. It follows that the tree expansion of $\RR^{'(3)}_{\o,s,\o',s'}(\kk_1,\kk_2,\kk_3,\kk_4)$
is that discussed in the proof of Theorem \ref{th4.1}; in particular it
only depends on the trees containing a subtree at scale $0$ of class c), defined in \S\ref{ss4.3}. The
contributions of these trees can be easily estimated by trivial dimensional bounds and we get
\pref{H41b}; the ``missing factor" $Z_l^{(1)}/Z_l$ with respect to the bound \pref{H41a} can be explained
as in the proof of Lemma \ref{lm3.4}.


\subsection{Vanishing of the Beta function}
\lb{ss5.1}

We want now to prove the property a), defined at the beginning of \S\ref{ss4.0},
of the model with $g_{1,\erp}=0$  (at $N,L=+\io$) and, at the same time, to analize the dependence
of the r.c.c. on a small perturbation of the ``velocity" $c$, which appears in the free propagator \pref{gth}.
Hence, we put
\be c=c_0+\d \virg c_0>0 \virg |\d|\le c_0/2\ee
The introduction of the new parameter $\d$ is essential to derive from the property a) a property of the beta function,
that we call {\em partial vanishing of the beta functions}, see (2.108), (2.159) and App. C of the companion paper
\cite{BFM012_1}. In its turn, this property is essential to control the IR integration in the Hubbard model.

Lemma \ref{t3.2} and Theorem \ref{t2.1} imply that, if $\bar g=\max\{|g_{1,\erp}|,|g_\arp|, |g_\erp|, |g_4|\}$ is small enough
and $K=0$,
\be e^{\WW_{[l,N]}(0,0)}=e^{- L^2 E_1} \int
P_Z(d\psi^{[l,1]})e^{\VV^{(1)}(\sqrt{Z}\psi^{[l,1]},0)}\label{ssss}
\ee
where $\VV^{(1)}(\psi,0)$ is given by \pref{2.17b} with $W^{(n;2m)(1)}$
verifying \pref{pc1}. Starting from \pref{ssss} we can move from
the free integration to the effective interaction the term $-\d F_\d(\sqrt{Z}\psi^{[l,1]})$,
with $F_\d(\psi)$ defined as in \pref{blue}.

The new propagator differs from \pref{gth} because, in place of $c$, we have
$c_0 + \d u_l(\kk)$, with $u_l(\kk) = 1 -\chi^{[l,1]}(|(k_0, c k)|)$. On the
other hand, for $\d$ small enough, $\chi^{[l,1]}(|(k_0, c k)|)$
and $\chi^{[l,1]}(|(k_0, c_0 k)|)$ differ only for values of $\kk$ of size $\g$ or $\g^l$ and
\be\lb{ul} u_l(\kk)=0\virg \text{if\ } \chi^{[l+1,0]}(|(k_0, c_0 k)|)>0 \ee
Hence, we can write
\be \chi^{[l,1]}(|(k_0, c k)|) = \bar \chi^{(1)}(\kk) +
\chi^{[l+1,0]}(|(k_0, c_0 k)|) + \bar \chi^{(l)}(\kk) \ee
with $\bar \chi^{(1)}(\kk)$ and $\bar \chi^{(l)}(\kk)$ smooth
functions, whose support is on values of $\kk$ of size $\g$ or
$\g^l$, respectively; moreover, if we define
\be \tilde C_l(\kk) =
\lft[\chi^{[l+1,0]}(|(k_0, c_0 k)| + \bar
\chi^{(l)}(\kk))\rgt]^{-1} \ee
then $\tilde C_l(\kk)=1$, if $1\ge |(k_0, c_0 k)|\ge \g^{l+1}$ and
$\tilde C_l(\kk)^{-1} \bar \chi^{(l)}(\kk) \le 1$.
It follows that the free measure $P(d\psi^{[l,1]})$ can be written
as $P(d\bar\psi^{(1)}) P(d\tilde\psi^{[l,0]})$, where
$\bar\psi^{(1)}$ is a field whose covariances has the same scale
properties of $\psi^{(1)}$, while
\be P(d\tilde\psi^{[l,0]}) = \NN^{-1} \exp\lft\{ - {Z\over L^2}
\sum_{\kk,\o,s} \tilde C_l(\kk) [-ik_0+\o c_0 (1+ u_l(\kk) \d)k]
\tilde\psi_{\kk,\o,s}^{+[l,0]} \tilde\psi_{\kk,\o,s}^{-[l,0]}
\rgt\} \ee
The integration of the single scale field $\bar\psi^{(1)}$ can be done without any problem.
At this point, we start the multiscale integration of the field $\tilde\psi^{[l,0]}$ as in \S\ref{ss2.3}
up to scale $l+1$, the only difference being that we have $c_0$ in place of $c$ in the single scale propagators
\pref{prop_ir}, thanks to \pref{ul}. This property is not true only in the last
step, $j=l$, but this is not a problem, since we have to study the RG flow at
fixed $j$ and $l\to-\io$ and, moreover, the contribution of the IR scale
fluctuations to the Schwinger functions at fixed space-time coordinates
vanishes as $l\to-\io$.

We prove the following lemma.

\begin{lemma}\lb{lm4.4} Let $\bar\e$ and $\e_h$ be defined as in Theorem \ref{th2.2}. Then there are
constants $\e_1$ and $c_2$, independent of $l$, such that, if $\e_0\le \e_1$, then $\e_h\le c_2\e_0\le \bar\e$,
for any $h\in [l+1,0]$.
\end{lemma}

\bpr The proof is by contradiction. Assume that there exists a $h\le 0$ such that
\be \e_{h+1}\le c_2\e_0 < \e_h\le 2 c_2\e_0\le \bar\e\label{2.3}\ee
We show that this is not possible, if $\e_1, c_2$ are suitably
chosen. Let us consider the model \pref{vv1} with $l=h$, that is
with infrared cutoff $\g^h$.

Let us consider the first equation in \pref{corr_h}, which says that, in the limit $L,N\to\io$ at fixed $l=h$,
if $|\kk|=\g^h$, $\hat S_\o(\kk)=<\hp^-_{\kk,\o,s} \hp^+_{\kk,\o,s}> = (Z_h D_\o(\kk))^{-1}[1+O(\e_h)]$. It is easy to
see that, the only term of order one in $\e_h$ contributing to $\hat S_\o(\kk)$ is given by the graph with one insertion of a $\d_l$
vertex in the renormalized free propagator. Hence, if we move from the interaction to the renormalized free measure
the term $-\d_l F_\d(\sqrt{Z_l}\psi^{(l)})$, before the integration of the field $\psi^{(l)}$, we get
\be \hat S_\o(\kk) \= \la\hat\psi_{\kk,\o,s}^-
\hat\psi^+_{\kk,\o,s}\ra_h={1\over Z_h
D_{h,\o}(\kk)}[1+W_2^{(h)}(\kk)]\label{gg1} \ee
where $\la\cdot\ra_l$ denotes the expectation with propagator \pref{gth} ($c=c_0+\d$) and
\be\lb{kajkj} D_{\o,h}(\kk)=-i k_0+\o k( c_0+\d_h) \virg |W_2^{(h)}(\kk)| \le C \e_h^2\ee
In the same manner we get, by using the second equation in \pref{corr_h}, that
\be \la\hat\r_{2\kk,\o,s} \hat\psi^-_{\kk,\o,s}
\hat\psi^+_{-\kk,\o,s}\ra_h = - \frac{Z^{(1)}_h }{Z_h^2 D_{\o,h}(\kk)^2}
 [1+W_{2,1}^{(l)}(\kk)]\label{gg2} \ee
where $|W_{2,1}^{(l)}(\kk)| \le C \e_h^2$.
Moreover, by using the WI \pref{grez}, we can write an equation relating \pref{gg1} with \pref{gg2}. Let us
put in \pref{grez} $\m=\o$ and let us take two derivatives with respect to $\hh^-_{\kk,\o,s}$ (the first) and
$\hh^+_{\kk+\pp,\o,s}$ at $J=\h=0$.
\be\label{ce1} D_\o(\pp) \la\hat\r_{\pp,\o,s} \hat\psi^-_{\kk+\pp,\o,s}
\hat\psi^+_{\kk,\o,s}\ra_h= \hat S_\o(\kk) - \hat S_\o(\kk+\pp) + D_{-\o}(\pp) R(\pp,\kk)
+\hat\HH^{(1;2)}_{\o,s}(\pp;\kk+\pp,\kk)\ee
where $D_{-\o}(\pp) R(\pp,\kk)$ is the contribution of the second term in the first line of \pref{grez}
and $\hat\HH^{(1;2)}_{\o,s}(\pp;\kk+\pp,\kk)$ is defined as in Lemma \ref{lm3.4}. Note that $R_h(\pp,\kk)$
is of the second order in $\e_h$, since it is a linear combination, with coefficients of order $\e_h$ of the correlations
$\la\hat\r_{\pp,\o,s} \hat\psi^-_{\kk+\pp,\o',s'} \hat\psi^+_{\kk,\o',s'}\ra_h$ with $(\o',s')\not=(\o,s)$, which vanish at order $0$.
It follows, by using again the second equation in \pref{corr_h}, that, if $\kk=-\bar\kk$, with $|\kk|=\g^h$, and
$\pp=2\bar\kk$, then
\be\lb{gg3} |R(2\bar\kk,\bar\kk)| \le C \g^{-2h} \frac{Z^{(1)}_h}{Z_h^2} \e_h^2\ee
Noreover, by Lemma \ref{lm3.4},
\be\lb{gg4} |D_{\o}(2\bar\kk) \hat\HH^{(1;2)}_{\o,s}(2\bar\kk;\bar\kk,-\bar\kk) \le
C \g^{-2h} \frac{1}{Z_h} \e_h^2\ee

If we insert \pref{gg1}, \pref{gg2}, \pref{gg3}, \pref{gg4} in \pref{ce1}, we easily get
$$\frac{Z^{(1)}_h}{Z_h} \big[-i k_0 + \o k(c_0+\d)\big]\; \big[1 + O(\e_h^2) \big] =
\big[-i k_0 + \o k(c_0+\d_h)\big]\; \big[1+ O(\e_h^2) \big]$$
which implies that
\be\lb{gg5} \frac{Z^{(1)}_h}{Z_h} = 1 + O(\e_h^2) \virg \d_h = \d+ O(\e_h^2) \ee

\*

Let us now consider the equation we get, if we take three derivatives w.r.t. the field $\h$ at $\h=0$
on both sides of the WI \pref{ceq2}, written in the limit $N,L\to\io$. Let us define
$$ \hat S_4(\uk,\o,s,\o',s') := <\hp^+_{\kk_1,\o,s} \hp^-_{\kk_2,\o, s} \hp^+_{\kk_3,\o',s'} \hp^-_{\kk_4,\o', s'}>_T $$
with $\uk = (\kk_1, \kk_2, \kk_3, \kk_4)$, $\kk_1-\kk_2+\kk_3-\kk_4=0$, and
$$ \hat S_{1,2}(\pp,\kk,\o,s,\o',s') = <\hr_{\pp,\o,s} \hp^-_{\kk-\pp,\o', s'} \hp^+_{\kk,\o',s'}>_T$$
Hence, if we consider \pref{ceq2} with $\kk=\kk_4$ and $(\o',s')$ in place of $(\o,s)$ and we take the derivative w.r.t. $\hh^-_{\kk_3,\o',s'}$
followed by the derivatives w.r.t. $\hh^+_{\kk_2,\o,s}$ and $\hh^-_{\kk_1,\o,s}$, we get, by using \pref{4.2}, \pref{3.14},
\pref{defh} (with $K=0$) and the definition \pref{H41c}, if we put $\O:=(\o,s,\o',s')$,
\be\lb{WI4}
\bsp
D_\o(\kk_4) &S_4(\uk,\O)  = -\sum_{\m,t} \hat h(\kk_1-\kk_2) \Big[\d_{\m\o',1} \d_{ts',-1} g_{\erp} +\\
+\d_{\m\o',1} &\d_{ts',1} g_{\arp} + \d_{\m\o',-1} \d_{ts',-1} g_4 \Big] S_{1,2}(\kk_1-\kk_2,\kk_2,-\m,t,\o,s) S_2(\kk_3)-\\
-\sqrt{4\p c} &\int \frac{d\pp}{(2\p)^2} \Big[\hat F^{-\o\o'}_{-\o',s s'}(\pp) S_4(\kk_1,\kk_2-\pp,\kk_3,\kk_4+\pp,\O)-\\
-\hat F^{-1}_{-\o',+1}(\pp) &S_4(\kk_1,\kk_2,\kk_3+\pp,\kk_4+\pp,\O)
-\hat F^{-\o\o'}_{-\o',s s'}(\pp) S_4(\kk_1+\pp,\kk_2,\kk_3,\kk_4+\pp,\O) \Big] +\\
&+R^{'(3)}_{\O}(\kk_1,\kk_2,\kk_3,\kk_4) + \tilde R^{'(3)}_{\O}(\kk_1,\kk_2,\kk_3,\kk_4)
\esp\ee
Let us now put $(\kk_1,\kk_2,\kk_3,\kk_4) = (\kk,-\kk,\kk,-\kk)$, with $|\tilde\kk|=\g^h$. Under this condition,
the terms in the third and fourth line of \pref{WI4}, can be bounded by $C\e_h^2 \g^{-3h} Z_h^{-2}$; this can be proved
as in the proof of the analogous bound (4.21) in the paper \cite{BM004}. A similar bound is valid for
$R^{'(3)}_{\O}(\kk,-\kk,\kk,-\kk)$ and $\tilde R^{'(3)}_{\O}(\kk,-\kk,\kk,-\kk)$, by Lemma \ref{lm4.3}. Hence,
if we extract in both sides of \pref{WI4} the terms of order one in $\e_h$ and use Theorem \ref{th3.3}
and the first identity in \pref{gg5}, we easily get
the identity
\be\lb{gg6} [g_{\a,h} + O(\e_h^2)] {\g^{-4 h}\over Z_h^2} = g_{\a,0} [1+O(\e_h)] {\g^{-4 h}\over Z_h^2} +
O(\e_h^2) {\g^{-4 h}\over Z_h^2} \virg \a=\erp,\arp,4\ee
Note that, if we call $\uv_j$ the set of r.c.c. of scale $j$, its value is independent of the IR cutoff scale up to $j=h$,
see the remark after \pref{LB}. Hence, \pref{gg6} is valid also if we interpret $\e_h$
as that calculated in the model with IR cutoff $l<h$. Hence,
the second identity in \pref{gg5} and \pref{gg6} imply that, if the hypothesis \pref{2.3} is satisfied, there exists
a constant $c_3$, independent of $c_2$ and $\e_1$, such that
\be\lb{2.4} \e_h-\e_0 \le c_3 \e_h^2\ee
By the hypothesis \pref{2.3}, $\e_h\le 2c_2\e_0$; hence, the bound \pref{2.4} implies that
$\e_h \le \e_0 (1+ 4c_3 c_2^2\e_0)$. On the other hand, the second member of this inequality
can not be larger than $c_2\e_0$, if, for instance, $c_2=2$ and $\e_0\le \e_1:=1/(16 c_3)$.
\qed

\*

Lemma \ref{lm4.4} implies that the r.c.c.'s are well defined and uniformly bounded for any
$h$ and that, as a consequence, also the ren.c.'s, the effective potentials and the correlation functions
are well defined. We can then write, for any $h\le 0$, equations of the form
\be\lb{bf} \uv_{h-1} = \uv_{h} + \uB^{(h)}\lft(\uv_h, \ldots, \uv_0, \uv\rgt)
\virg \uv:= (g_{\arp}, g_{\erp}, g_{4},\d)\ee
\be \frac{Z_{h-1}^{(1)}}{Z_{h-1}} = \frac{Z_{h}^{(1)}}{Z_{h}} \Big[ 1 +
\tilde\uB^{(h)}\lft(\uv_h, \ldots, \uv_0, \uv\rgt) \Big]\ee
Let us call $\ub^{(j)}(\uv')$ and $\tilde\ub^{(j)}(\uv')$ the functions which are obtained from
$\uB^{(h)}$ and $\tilde\uB^{(h)}$,
by subtracting the contribution of the trees containing endpoints of scale grater
than $0$ and by putting everywhere a fixed value $\uv'$, with $|\uv'| \le \bar\e$ in place of $\uv_j$,
for all $j\le 0$, in the remaining trees contribution (which do not depend explicitly on the parameters
$\uv$ of the interaction). It is easy to see, by using the {\em short memory property}
of the tree expansion and its analyticity in the r.c.c.'s, that $\ub^{(j)}(\uv')$ and $\tilde\ub^{(j)}(\uv')$
converge, as $h\to-\io$, to analytic functions $\ub^{(-\io)}(\uv')$ and $\tilde\ub^{(-\io)}(\uv')$,
such that $|\ub^{(j)}(\uv')-\ub^{(-\io)}(\uv')|$ and $|\tilde\ub^{(j)}(\uv')-\tilde\ub^{(-\io)}(\uv')|$
are bounded by $C |\uv'|^2 \g^{\th j}$, with $\th\in(0,1)$. On the other hand, it is not too hard to
prove that, as a consequence of the bound \pref{2.4}, $\ub^{(-\io)}(\uv') = \tilde\ub^{(-\io)}(\uv')=0$;
see pag. 156 of \cite{BGPS994}. Hence, we get the very important property:
\be\lb{van} \lft|\ub^{(j)}(\uv') \rgt| \le C |\uv'|^2 \g^{\th j}\virg
\lft|\tilde \ub^{(j)}(g_\arp, g_\erp, g_4, \d) \rgt|\le C |\uv'|^2 \g^{\th j}\ee
that we call the {\em asymptotic vanishing of the beta function}. This property has been used
to control the flow of the r.c.c.'s also in the companion paper \cite{BFM012_1},
not only in the case $g_{1,\perp}=0$, but also in the spin-symmetric and repulsive
case, that is $g_\arp=g_{\perp}-g_{1,\perp}$ and $g_{1,\perp}>0$.

As one can easily guess, the bound \pref{van} allows us to prove that the r.c.c.'s, the correlation functions at fixed
space coordinates and their Fourier transforms at non exceptional external momenta, converge as
$l\to-\io$ to functions, which are analytic in $\uv$ around the origin.
A detailed discussion of this point, which only depends on the structure of the tree
expansion, can be found, in the case of the Thirring model, in \cite{BFM007}.


\section{Closed equations in the limit of removed cut-offs}\lb{ss6}

In order to prove Theorem \ref{th1.2}, we need to calculate the explicit expression of the two-points function and of
the truncated correlations of some quadratic operators in the space coordinates. We shall do that by taking
the inverse Fourier transform of the closed equations obtained in the previous sections for the Fourier transforms of the correlations.

There is in principle a problem in this procedure, because one could be afraid that the removed cutoffs limit does not
commute with the Fourier transform because of the presence of delta functions in the calculations. However, we have shown
in a paper on the Thirring model(see \S A.3 of \cite{BFM009_1}) that this is not the case and the arguments given there are model independent.

\subsection{The two-points function}

If we perform in \pref{ce_lim} one derivative w.r.t. $\hh^-_{\o,s}$ at $\h=0$ and we put
$\la\psi^-_{\xx,\o,s}\psi^+_{\yy,\o',s'}\ra = \d_{\o,\o'}\d_{s,s'} S_{\o}(\xx-\yy)$, we get, in the space coordinates:
\be\lb{eqx} \lft(\dpr_\o S_{\o}\rgt)(\xx) -F^{-}_{-\o,+}(\xx) S_{\o}(\xx) = \frac1Z \d(\xx) \ee
with $\dpr_\o=\dpr_{x_0}+i\o c\dpr_{x_1}$; note that we have added the dependence on $Z$, which
was put equal to $1$ in \pref{ce_lim}.

The solution of \pref{eqx} is:
\be\lb{Som+} S_{\o}(\xx)= e^{\D^{-}_{+}(\xx|0)}g_\o(\xx) \virg
g_\o(\xx) = \frac1{2\p Z} \frac1{cx_0+i\o x}\;, \ee
having defined $\D^\e_{s}$ such that $\dpr_{\o}^\xx
\D^\e_{s}(\xx|\zz)= F^\e_{-\o,s}(\xx)$:
\be\lb{DD} \D^\e_{s}(\xx|\zz) = \int\!{d\kk\over (2\p)^2} \
{e^{-i\kk\xx} - e^{-i\kk\zz} \over D_\o(\kk)}
\hF^\e_{-\o,s}(-\kk)=\D^{\e}_\r(\xx|\zz)+s\D^{\e}_\s(\xx|\zz)\;.
\ee
for
$$
\hat \D_\r^\e(\pp)=g_\r \hat h(-\pp) {M^\r_{-\o,-\o\e}(-\pp)\over
D_\o(\pp)}+ {g_4\over 2} \hat h(-\pp) {M^\r_{\o,-\o\e}(-\pp)\over
D_\o(\pp)}
$$
$$
\hat \D_\s^\e(\pp)=g_\s \hat h(-\pp) {M^\s_{-\o,-\o\e}(-\pp)\over
D_\o(\pp)}- {g_4\over 2} \hat h(-\pp) {M^\s_{\o,-\o\e}(-\pp)\over
D_\o(\pp)}
$$
In order to evaluate the asymptotic behavior of
$\D^\e_{s}(\xx|0)$, we need to study functions of the type
\be\lb{II} I_{\o,\e}(\xx) = \int{d^2\pp\over (2\p)^2}\ a(\pp)
{e^{-i\pp\cdot\xx}-1\over (p_0+i\o c p_1)[v_+(\pp) p_0-i\e\o
v_-(\pp) c p_1]} \ee
where $a(\pp)$ and $v_s(\pp)>0$ are even smooth functions of fast
decrease. It is easy to show that
\be\lb{II1} I_{\o,\e}(\xx) = {a(0)\over v_+(0)} \tilde
I_{\o,\e}(\xx) + A + O(1/|\xx|) \ee
where $A$ is a {\it real} constant and, if $v=v_-(0)/v_+(0)$,
$$
\tilde I_{\o,\e}(\xx) = \int_{-1}^{+1} {d p_1\over (2\p c)}
\int_{-\io}^{+\io} {d p_0\over (2\p)} {e^{-i(p_o x_0 + p_1
x_1/c)}-1\over (p_0+i\o p_1)(p_0- i\e\o v p_1)}
$$
One can see that, if $v>0$, $v\not=1$ and $\xx\not=0$,
$$
\tilde I_{\o,\e}(\xx) = {1\over 2\p c(1+\e v)} \left[ F(x_0,\o
x_1/c) + \e F(v x_0, -\e\o x_1/c) \right]
$$
where
$$
F(x_0,x_1) = \int_0^1 {dp_1\over p_1} \left[ e^{-p_1(|x_0|+i {\rm
sgn}(x_0) x_1)} -1\right]= \ln|z| +i{\rm Arg}\big({\rm
sgn}(x_0)z\big)+B+{\rm O}(1/z)
$$
where $z=x_0+ix_1$, $B$ is a real constant and $|{\rm Arg}(z)|\le
\p$. Since
$$
{\rm Arg}\big({\rm sgn}(x_0)z\big)= {\rm Arg}(z)-\th(x_0){\rm
sgn}(x_1)\p
$$
the function $F(\xx)$ (considered only for $|\xx|>1$) is
discontinuous at $x_0=0$, while $\tilde I_{\o,\e}(\xx)$ is
continuous. We can then write
\be\lb{II2} \tilde I_{\o,\e}(\xx) = -{1\over 2\p c(1+\e v)} \left[
\log (x_0+i\o x_1/c) + \e \log (v x_0 -i\e \o x_1/c) \right] + C
+O(1/|\xx|) \ee
where $C$ is again a real constant. By using \pref{MM}, \pref{II},
\pref{II1} and \pref{II2}, one can easily check that
\be\lb{delta} \bsp \D^\e_\g(\xx|0)= &-{H^\e_{\g,\e}\over 4\p c}
\ln\lft(v^2_\g x^2_0+ (x_1/c)^2\rgt)
-{H^\e_{\g,-}+H^\e_{\g,+}\over 4\p c} \ln{x_0+i\o x_1/c\over
v_\g x_0+i\o x_1/c}\\
&+ C^\e_\g +O(1/|\xx|) \esp \ee
for
$$
H^+_{\g,\e}={2g_\g u_{\g,\e}+g_{4,\g} w_{\g,\e}\over v_{\g,+}+\e
v_{\g,-}}={\e g_\g\over v_{\g,+}v_{\g,-}}
$$
$$
H^-_{\g,\e}={2g_\g w_{\g,\e}+ g_{4,\g} u_{\g,\e}\over v_{\g,+}-\e
v_{\g,-}}=-{4\p\e\over 2\n_{\g,+}\n_{\g,-}} \lft[1-{(v_{\g,-}-\e
v_{\g,+})^2\over 4}\rgt]
$$
where $C^\pm_\g$ are real constants and
$v_\g=v_{\g,+}(0)/v_{\g,-}(0)$ (and $g_{4,\r}=g_4$ while
$g_{4,\s}=-g_4$).

By using \pref{MM1} and \pref{MM2},
\be {H^+_{\g,+}\over 4\p c}= {\n_{\g}\over
v_{\g,+}v_{\g,-}}={\z_\g\over 2} \qquad
{H^+_{\g,-}+H^+_{\g,+}\over 4\p c}=0 \ee
\be {H^-_{\g,-}\over 4\p c}= {1-\frac14 (v_{\g,+} +
v_{\g,-})^2\over 2 v_{\g,+}v_{\g,-}}={\h_\g\over 2} \qquad
{H^-_{\g,-}+H^-_{\g,+}\over 4\p c}=-{1\over 2} \ee
Note that this expression is continuous in $v_\g=1$, as one
expects, and that, at least at small coupling, $\h_\g \ge 0$.

By using \pref{Som+} and \pref{DD}, we finally get
\be\lb{4.23} S_\o(\xx) = \frac1{2\p Z} { (c^2 v_\r^2 x_0^2 +
x_1^2)^{-\h_\r/2} (c^2v_\s^2 x_0^2 +x_1^2)^{-\h_\s/2} \over (c
v_\r x_0+i\o x_1)^{1/2} (c v_\s x_0+i\o x_1)^{1/2}} e^{C
+O(1/|\xx|)} \ee
where $C$ is a real constant $O(g)$ and $z^{1/2} = |z|^{1/2} e^{i
Arg(z)/2}$. Note that the leading term is well defined and
continuous at any $\xx\not=0$.

Note also that, if $g_4=0$, $v_\r=v_\s=1$ and $\h_\r=\h_\s\=\h/2$,
so that
\be S_\o(\xx) = \frac1{2\p Z} { (c^2 x_0^2 + x_1^2)^{-\h/2} \over
c x_0+i\o x_1} e^{C +O(1/|\xx|)} \ee
If we also put $g_\erp=0$ and $g_\arp=\l$, we get for $\h$ the
value found for the regularized Thirring model, that is
$\h=2\t^2/(1-\t^2)$, with $\t=\l/(4\p c)$; see eq. (4.21) of
\cite{BFM009}.

\subsection{The four point functions and the densities correlations}\lb{ss4.4}

We want to calculate the truncated correlations $\la O^{(t)}_\xx O^{(t)}_\yy\ra^T$ of the local quadratic operators
$O^{(t)}_\xx$, $t=(1,\a)$ or $(2,\a)$, defined as the analogous operators of the
Hubbard model in \S 2.4 of the previous paper \cite{BFM012_1}, that is
\be\nn
O^{(1,C)}_\xx = \sum_{\o,s} \psi^+_{\xx,\o,s} \psi^-_{\xx,\o,s}\;,
O^{(1,S_i)}_\xx = \sum_{\o,s,s'} \psi^+_{\xx,\o,s}\s^{(i)}_{s,s'}\psi^-_{\xx,\o,s'}\;,
O^{(1,SC)}_\xx = \sum_{\e,\o,s} s\, e^{2i \e\o p_F x} \psi^\e_{\xx,\o,s}\psi^\e_{\xx,\o,-s}
\ee
\bal
&O^{(2,C)}_\xx = \sum_{\o,s} e^{2i \o p_F x} \psi^+_{\xx,\o,s} \psi^-_{\xx,-\o,s}\;,
&O^{(2,S_i)}_\xx = \sum_{\o,s,s'} e^{2i \o p_F x} \psi^+_{\xx,\o,s}\s^{(i)}_{s,s'}\psi^-_{\xx,-\o,s'}\nn\\
&O^{(2,SC)}_\xx = \sum_{\e,\o,s} s\,\psi^\e_{\xx,\o,s} \psi^\e_{\xx,-\o,-s} \virg
&O^{(2,TC_i)}_\xx = \sum_{\e,\o,s,s'} e^{-i\e \o p_F} \psi^\e_{\xx,\o,s}
\tilde\s^{(i)}_{s,s'} \psi^\e_{\xx,-\o,s'}\nn
\eal
Note that $p_F$ has no special meaning in the effective model, but it
is left there since we want to compare the correlations in the two models, in the proof of Theorem \ref{th1.2}.

Our UV regularization implies that $\la O^{(t)}_\xx\ra=0$ for any
$t$; hence we can make the calculation very simply, by using the
explicit expressions of the (untruncated)  four points functions which follow
from the closed equation \pref{ce} and then evaluating them so
that the two coordinates corresponding to each $O^{(t)}$ operator
coincide, if this is meaningful. This works for all values of $t$,
except $(1,C)$ and $(1,S_3)$, where there is a singularity,
related to the fact that the operators $\r_{\xx,\o,s} =
\psi^+_{\xx,\o,s} \psi^-_{\xx,\o,s}$ are not well defined in the
limit $N\to\io$, because of the singularity of the free propagator
at $\xx=0$. However, in these cases we can use directly the WI
\pref{wis1} to calculate correctly, in the limit $N\to \io$, the
correlations of $O_\xx^{(1,C)}$ and $O_\xx^{(1,S_3)}$, by using
\pref{wis2}, \pref{MM} and the equations \pref{II}, \pref{II1},
\pref{II2}. We get, for $|\xx|>1$,
\be\nn \bsp G^\g_{\o,\o}(\xx) &\simeq {1-v_\g^2\over 8\p^2 c^2
Z^2} \lft[{u_{\g,+}\over v_{\g,+}-v_{\g,-}} { 1\over (v_\g x_0+i\o
x_1/c)^2}- {u_{\g,-}\over
v_{\g,+}+v_{\g,-}} { 1\over (v_\g x_0-i\o x_1/c)^2}\rgt]\\
G^\g_{-\o,\o}(\xx) &\simeq {1-v_\g^2\over 8\p^2 c^2 Z^2}
\lft[{w_{\g,+}\over v_{\g,+}-v_{\g,-}} { 1\over (v_\g x_0+i\o
x_1/c)^2}- {w_{\g,-}\over v_{\g,+}+v_{\g,-}} { 1\over (v_\g
x_0-i\o x_1/c)^2}\rgt] \esp \ee
the corrections being of order $1/|\xx|^3$. This implies that, for
$|\xx|>1$,
\be \la O^{(1,C)}_{\bf 0} O^{(1,C)}_\xx \ra^T = \frac{v_\r^2
(1-\n_4+2\n_\r) + (1+\n_4-2\n_\r)}{2\p Z^2 c^2 v_{\r,+} v_{\r,-}}
{v^2_\r x^2_0-x^2/c^2\over (v^2_\r x^2_0+x^2/c^2)^2} +O(1/|\xx|^3)
\ee
while $\la O^{(1,S_3)}_{\bf 0} O^{(1,S_3)}_\xx \ra^T$ is obtained
from this expression, by replacing $\n_4$ with $-\n_4$ and $\n_\r$
with $\n_\s$ (hence also $v_\r$, $v_{\r,+}$ and $v_{\r,-}$ with
$v_\s$, $v_{\s,+}$ and $v_{\s,-}$).

%
%

In order to calculate the other correlations, we first note that
the only four points functions different from zero are those
defined by the equation
$$
G^{\o_1,\o_2}_{s_1,s_2}(\xx,\yy,\uu,\vv) = \la
\psi^-_{\xx,\o_1,s_1} \psi^-_{\yy,\o_2,s_2} \psi^+_{\uu,\o_2,s_2}
\psi^+_{\vv,\o_1,s_1}\ra
$$
By \pref{ce}, $G^{\o_1,\o_2}_{s_1,s_2}(\xx,\yy,\uu,\vv)$ is the
solution of the equation:
\be\bsp \lft(\dpr_{\o_1}^\xx
G^{\o_1,\o_2}_{s_1,s_2}\rgt)(\xx,\yy,\uu,\vv) = \d(\xx-\vv)
S_{\o_2}(\yy-\uu)-\d_{\o_1,\o_2}\d_{s_1,s_2} \d(\xx-\uu)
S_{\o_1}(\yy-\vv)+ \\
\Big[ -F^{-\o_1\o_2}_{-\o_1,s_1s_2}(\xx-\yy)
+F^{-\o_1\o_2}_{-\o_1,s_1s_2}(\xx-\uu)
+F^{-}_{-\o_1,+}(\xx-\vv)\Big]
G^{\o_1,\o_2}_{s_1,s_2}(\xx,\yy,\uu,\vv) \esp \ee
For the two-points correlation of $O^{(2,\a)}_\xx$ we are
interested in the case $\o_1=-\o_2=\o$. For
$G^\o_s(\xx,\yy,\uu,\vv) =G^{\o,-\o}_{s',ss'}(\xx,\yy,\uu,\vv)$ we
find
\be\lb{eqG} G^\o_s(\xx,\yy,\uu,\vv) =
e^{-\Big[\D^+_s(\xx-\yy|\vv-\yy) - \D^+_s(\xx-\uu,\vv-\uu)\Big]}
S_{\o}(\xx-\vv) S_{-\o}(\yy-\uu)\;. \ee

Therefore, for $\a=C,S_3$ we set $\xx=\uu$, $\yy=\vv$ and $s=+$,
while for $\a=S_1.S_2$ we set $s=-$; for $TC_1,TC_3$ we set
$\uu=\vv$, $\xx=\yy$ and $s=+$; while for $TC_2,SC$ we set $s=-$.

For the two-points correlation of $O^{(1,\a)}_\xx$,
$\a\not=C,S_3$, we are interested in the case $\o_1=\o_2=\o$. If
$\bar G^\o_s(\xx,\yy,\uu,\vv)
=G^{\o,\o}_{s',ss'}(\xx,\yy,\uu,\vv)$ we find
\be\lb{eqGbar} \bsp \bar G^\o_s(\xx,&\yy,\uu,\vv) = e^{-\Big[
\D^-_s(\xx-\yy|\vv-\yy) - \D^-_s(\xx-\uu,\vv-\uu)\Big]}
S_{\o}(\xx-\vv) S_{\o}(\yy-\uu)\\ &-\d_{s,+} e^{-\Big[
\D^-_+(\xx-\yy|\uu-\yy) - \D^-_+(\xx-\vv,\uu-\vv)\Big]}
S_{\o}(\xx-\uu) S_{\o}(\yy-\vv)\;. \esp \ee
For $\a=SC$ we set $\xx=\yy$, $\uu=\vv$ and $s=-$; for
$\a=S_1,S_2$ we set $\xx=\uu$, $\yy=\vv$ and $s=-$; for $\a=S_3,C$
we set $\xx=\uu$, $\yy=\vv$ and $s=+$.

Therefore, it is easy to see, by using \pref{DD} and \pref{4.23},
that, for $|\xx|>1$,
\bal &\la O^{(2,\a)}_{\bf 0} O^{(2,\a)}_\xx \ra^T = \frac{1}{\p^2
Z^2 c^2} {\cos(2p_F x)^{m_\a}\over (v^2_\r
x^2_0+x^2/c^2)^{x_{\r,t}}} {1\over (v^2_\s
x^2_0+x^2/c^2)^{x_{\s,t}}} +O(1/|\xx|^3)\virg \forall\a\nn\\
&\la O^{(1,SC)}_{\bf 0} O^{(1,SC)}_\xx \ra^T = -\frac{1}{\p^2 Z^2
c^2} {\cos(2p_F x)\over (v^2_\r x^2_0+x^2/c^2)^{2\h_\r}} {v^2_\r
x^2_0-x^2/c^2\over
(v^2_\r x^2_0+x^2/c^2)^2} +O(1/|\xx|^3) \lb{4.28}\\
&\la O^{(1,\a)}_{\bf 0} O^{(1,\a)}_\xx \ra^T = \frac{1}{\p^2 Z^2
c^2} {1\over (v^2_\s x^2_0+x^2/c^2)^{2\h_\s}} {v^2_\s
x^2_0-x^2/c^2\over (v^2_\s x^2_0+x^2/c^2)^2}+O(1/|\xx|^3),\;
\a=S_1,S_2\nn \eal
where $m_\a=1$, if $\a=C,S_i$, while $m_\a=0$, if $\a=SC, TC_i$,
and
\be x_{\g,t} =
\begin{cases}
\h_\g-\z_\g+1/2 & t=(2,C), (2,S_3)\\
\h_\g-s(\g)\z_\g+1/2 & t=(2,S_1), (2,S_2)\\
\h_\g+\z_\g+1/2 & t=(2,TC_1), (2,TC_3)\\
\h_\g+s(\g)\z_\g+1/2 & t=(2,SC), (2,TC_2)
\end{cases}
\ee

Let us now consider the special case $g_\s=0$ (i.e.
$\h_\s=\z_\s=0$), which we use as a effective model for the
Hubbard model. In this case, the equations \pref{4.28} imply that
$\la O^{(t)}_{\bf 0} O^{(t)}_\xx \ra$ decays, for $|\xx|\to\io$,
as $|\xx|^{-2 X_t}$, with
\be\lb{4.30} 2X_t =
\begin{cases}
2+ 2\h_\r - 2\z_\r & t=(2,C), (2,S_i)\\
2+ 2\h_\r + 2\z_\r & t=(2,SC), (2,TC_i)\\
2+ 4\h_\r & t=(1,SC)\\
2 & t=(1,C), (1,S_i)
\end{cases}
\ee

Note that
\be \h_\r=-{1\over 2}+ {4-v_{\r+}^2- v_{\r-}^2\over 4 v_{\r+}
v_{\r-}}\virg \x_\r={2 \n_\r\over v_{\r+} v_{\r-}} \ee
Let us now define $K=2 X_{2,C}-1$ and $\tilde K=2 X_{2,SC}-1$. By
using \pref{MM2}, we see that
\be\lb{KK} K=\frac{(1-2\n_\r)^2 - \n_4^2}{v_{\r+} v_{\r-}} =
\sqrt{(1-\n_4)-2\n_\r\over (1-\n_4)+2\n_\r}
\sqrt{(1+\n_4)-2\n_\r\over (1+\n_4)+2\n_\r}\ee
$$\tilde K=\frac{(1+2\n_\r)^2 - \n_4^2}{v_{\r+} v_{\r-}}=K^{-1}$$
$$4\h_\r = K+\tilde K -2$$
These equations imply that all the critical indices $X_t$ and the
parameter $\h_\r$ can be expressed in terms of the single
parameter $K$, only depending on $g_2/c$ and $g_4/c$. In the
following section we will show that the coupling of the model
\pref{vv1} can be chosen so that its exponents coincide with the
Hubbard ones; then, by some simple algebra, one can check the
validity of the scaling relations \pref{1.8}.

\section{Proof of Theorem 1.1} \lb{ss2.5d}
\subsection{The extended scaling relations: proof of \pref{1.8}}

In this section we finally establish contact with the
Renormalization Group analysis of the Hubbard model done in the
companion paper \cite{BFM012_1}, and we prove Theorem 1.1.

Let us call $\tilde v_h=(\tilde g_{2,h},\tilde g_{4,h},\tilde
\d_h)$, $h\le 0$, the running coupling constants in the effective
model with ultraviolet cutoff $\g^N$ and parameters
\be\lb{5.1} g_{1,\perp}=0\virg g_{\arp}=g_{\erp}=\tilde g_{2,N}
\virg g_4=\tilde g_{4,N} \virg \d =\tilde \d_N \ee
so that, in particular, $c=c_0(1 +\tilde \d_N)$, and put $\tilde
v_N=(\tilde g_{2,N}, \tilde g_{4,N}, \tilde \d_N)$. We call
$v_h=(g_{2,h}, g_{4,h}, \d_h)$, $h\le 0$, the analogous constants
in the Hubbard model, and $\vec v_h=(v_h,g_{1,h},\n_h)$. The
analysis of the RG flow given above implies that, for $h\le 0$,
\bal \lb{asasa} \tilde v_{h-1} &=\tilde v_h + \beta^{(0,h)}(\tilde
v_h,..,\tilde
v_0) + \tilde r^{(h)}(\tilde v_h,..,\tilde v_0,\tilde v_N)\\
\lb{asasa2} v_{h-1} &= v_h + \beta^{(0,h)}(v_h,..,v_0) +
r^{(h)}(\vec v_h,..,\vec v_0,\l) \eal
where $\beta^{(0,h)}(\tilde v_h,..,\tilde v_0)$ is the beta
function of the effective model with parameters \pref{5.1},
modified so that the endpoints have scale $\le 0$. Note that
$\beta^{(0,h)}(v_h,..,v_0)$ is the function
$\beta^{(h)}(v_h,..,v_0)$ modified so that, in its tree expansion,
no trees containing endpoints of type $g_1$ appear and the space
integrals are done in terms of continuous variables, instead of
lattice variables (the difference is given by exponentially
vanishing terms). The crucial bound \pref{van} and the short
memory property imply that $|\tilde r^{(h)}(\tilde v_h,..,\tilde
v_N)|\le C [\max_{k\ge h} |\tilde v_k|]^2\; \g^{\th h}$, and a
similar bound is verified from $r^{(h)}(\vec v_h,..,\vec v_0,\l)$.

\begin{lemma}\lb{lm5.1}
Given the Hubbard model with coupling $\l$ such that $g_{1,0}\in
D_{\e,\d}$, it is possible to choose $\tilde v_N$ as analytic
function of $\l$, so that
\be\lb{5.4} \tilde g_{2}=2\l \left[\hat v(0)-{1\over 2}\hat
v(2p_F) \right] + O(\l^{3/2})\virg \tilde g_{4}=2\l\hat
v(0)+O(\l^2)\virg \tilde\d=O(\l) \ee
and, if $\tilde v_h$ are the r.c.c.  of the effective model with
parameters satisfying \pref{5.1}, while $v_h$ are the r.c.c.  of
the Hubbard model, then, $\forall h\le 0$,
\be\lb{144} |v_h-\tilde v_h|\le C {|g_{1,0}|\over 1+(a/2)
|g_{1,0}| \,|h|} \ee
Moreover, the r.c.c.  $\tilde v_h$ have a well definite limit as
$N\to +\io$ and this limit still satisfies \pref{144}.
\end{lemma}

\0{\bf Proof} - We have seen in the previous sections that the
flows \pref{asasa} and \pref{asasa2} have well defined limits
$\tilde v_{-\io}$ and $v_{-\io}$, as $h\to -\io$, if the initial
values are small enough and $g_{1,0}\in D_{\e,\d}$. Moreover, the
proof of this property for the flow \pref{asasa} implies that
$\tilde v_{-\io}$ is a smooth invertible function of $\tilde v_N$,
such that $\tilde v_{-\io}=\tilde v_N+O(\tilde v_N^2)$; let us
call $\tilde v_N(\tilde v_{-\io} )= \tilde v_{-\io} + O(\tilde
v_{-\io}^2)$ its inverse. It is also clear that $\tilde v_N(\tilde
v_{-\io} )$ has a well defined limit as $N\to\io$, that we shall
call $\tilde v(\l)$, and that this is true also for the r.c.c.
$\tilde v_h$, $h\le 0$.

The previous remarks imply that it is possible to choose $\tilde
v_N$, satisfying \pref{5.4}, so that
\be\lb{147}\tilde v_{-\io}-v_{-\io}=0\ee
In order to prove \pref{144}, we note that, because of the bound
\pref{van} and the short memory property, in the effective model
with couplings satisfying \pref{5.4},
\be |\tilde v_h - \tilde v_{-\io}| \le C \l^2 \g^{\th h}\ee
On the other hand
\be |v_h - v_{-\io}| \le C \sum_{j=-\io}^h
[|g_{1,j}|^2+\l\g^{{\th\over 2}j}] \le C_1 {|g_{1,0}|\over 1+(a/2)
|g_{1,0}| \,|h|} \ee
These two bounds immediately imply \pref{144}.\Halmos

Let us now note that the critical indices of the effective model
can be calculated in terms of $\tilde v_{-\io}$ by the same
procedure used for the Hubbard model and that we get an equation
{\it with the same function $\b_t^{(0,j)}$}. Hence, the above
lemma allows us to conclude that the critical indices in the
Hubbard model and in the effective model coincide, provided that
the value of $\tilde v= \lim_{N\to\io} \tilde v_N$ is chosen
properly. It follows that all the indices are given by the
equations \pref{4.30}, with
\be\lb{5.9} \bsp \n_\r &= {\tilde g_2(\l)\over 4\pi c}= \l {\hat
v(0)-\hat v(2p_F)/2\over 2\pi
\sin \bar p_F} + O(\l^{3/2})\\
\n_4 &= {\tilde g_4(\l)\over 4\pi c} = \l {\hat v(0)\over 2\pi
\sin \bar p_F} +O(\l^2)\esp \ee
where \pref{5.4} has been used, together with $c=\sin \bar p_F +
O(\l)$. Moreover, \pref{5.9} and \pref{KK} imply that $K= 1- 2\l
[\hat v(0)-\hat v(2p_F)/2]/(\pi \sin \bar p_F)+ O(\l^{3/2})$, in
agreement with \pref{g1}.

\subsection{Susceptibility and Drude weight: proof of \pref{den}} \lb{ss2.5e}

The effective model is not invariant under a gauge transformation
with the phase depending both on $\o$ and $s$, if $g_{1,\perp}>0$;
however, it is still invariant under a gauge transformation with
the phase only depending on $\o$. This is true, in particular, if
the interaction is spin symmetric, that is if
$g_\arp=g_{\erp}-g_{1,\perp}$. Since also the Hubbard model is
spin symmetric, it is natural to see if one can use this
``restricted " gauge invariance to get some useful information on
the asymptotic behavior of the Hubbard model, by comparing it with
the effective model with $g_{1,\perp}>0$.

Let us put $g_{\erp}\equiv \bar g_2$, $g_{1\perp}\equiv \bar g_1$
and $g_\arp=\bar g_2- \bar g_1$. We want to show that we can
choose the parameters of the effective model $\bar g_1$, $\bar
g_2$, $\bar g_4$, $\bar\d$, so that the running coupling constants
are asymptotically close to those of the Hubbard model. This
result is stronger of the similar one contained in Lemma
\ref{lm5.1}, since now all the running couplings are involved, and
this implies also that the values of $\bar g_2$, $\bar g_4$ and
$\bar\d$ are {\it different} with respect to the analogous
constants defined in Lemma \ref{lm5.1}. The main consequence of
these considerations is that we can use the restricted WI of this
new effective model to get non trivial information on some Hubbard
model correlation functions, not plagued by logarithmic
corrections.

Let $\vec l_h=(\bar g_{1,h}, \bar g_{2,h}, \bar g_{4,h},
\bar\d_h)$, $h\le 0$, be the running coupling constants appearing
in the integration of the infrared part of the effective model.
The smoothness properties of the integration procedure imply that,
in the UV limit, $\vec l_0$ is a smooth invertible function of the
interaction parameters $\vec l= (\bar g_{1}, \bar g_{2}, \bar
g_{4}, \bar\d)$, whose inverse we shall call $\vec l(\vec l_0)$;
hence we can fix the effective model by giving the value of $\vec
l_0$ and by putting $\vec l= \vec l(\vec l_0)$. In a similar way
we call $\vec g_h=(g_{1,h}, g_{2,h}, g_{4,h}, \d_h)$, $h \le 0$,
the running couplings of the Hubbard model with coupling $\l$.

We now define $\vec x_h=\vec l_h-\vec g_h$, $h\le 0$. By using the
decomposition for $\vec g_h$ and the similar one for $\vec l_h$,
we can write
\be \vec x_{h-1}=\vec x_h+ [\vec\b^{(1)}_h(\vec g_h,..,\vec
g_0)-\vec\b^{(1)}_h(\vec l_h,..,\vec l_0)] +\vec\b^{(2)}_h(\vec
g_h,\n_h...\vec g_0,\n_0,\l)+ \vec\b^{(3)}_h(\vec l_h,..,\vec
l_0,\vec l) \ee
%
In the usual way, one can see that
\be |\vec\b^{(1)}_h(\vec g_h,..,\vec g_0)-\vec\b^{(1)}_h(\vec
l_h,..,\vec l_0)|\le C \left[ |\l|+\sup_{k\ge h}|\vec l_k| \right]
\sum_{k=h}^0 \g^{-\th (k-h)}|\vec x_k| \ee
and that $|\vec\b^{(2)}_h|\le C|\l|\g^{\th h}$,
$|\vec\b^{(3)}_{h}|\le C [\sup_{k\ge h}|\bar l_k|]^2$. Note that
the different power in the coupling of these two bounds is due to
the terms linear in $\l$ in the beta function for $\d_h$, which
are present in the Hubbard model, while similar terms are absent
in the effective model.

We want to show that, given $\l$ positive and small enough, it is
possible to choose  $\vec l_0$, hence $\vec x_0$, so that $\vec
x_{-\io}=0$; we shall do that by a simple fixed point argument.
Note that $\vec x_{-\io}=0$ if and only if
\be\lb{5.3} \vec x_{\bar h}=-\sum_{h=-\io}^{\bar h} \{
[\vec\b^{(1)}_h(\vec g_h,..,\vec g_0)-\vec\b^{(1)}_h(\vec
l_h,..,\vec l_0)] +\vec\b^{(2)}_h+ \vec\b^{(3)}_h\}\ee
We consider the Banach space $\MM_\th$, $\th<1$, of sequences
$\vec x=\{\vec x_h\}_{h\le 0}$ with norm\\
$\|\vec x\|=\sup_{k\le 0}|\vec x_k|\g^{-(\th/2)k}$
and the operator ${\bf T}:\MM_\th\to \MM_\th$, defined as the
r.h.s. of \pref{5.3}. Given $\x>0$, let $\BB_\x = \{\vec x\in
\MM_\th: \|\vec x\| \le \x \l\}$; if $\l$ is small enough, say
$\l\le \e_0$ and $\x\l \le \e_0$, and $\vec l_h=\vec g_h + \vec
x_h$, the functions $\vec\b^{(1)}_h(\vec l_h,..,\vec l_0)$ and
$\vec\b^{(3)}_h(\vec l_h,..,\vec l_0,\vec l)$ are well defined and
satisfy the bounds above, even if $\vec x$ is not the flow of the
effective model corresponding to $\vec l_0$. Hence, we have:
\be \g^{-(\th/2)h} |{\bf T}(\vec x)_h|\le c_0 \l(\x \l+1)
\sum_{k=-\io}^{h-1}\g^{{\th\over2} k}\le c_1 \l (1+\x \l)\ee
so that $\BB_\x$ is invariant if $\x= 2 c_1$ and $\l\le \e_1=\min
\{\e_0, \e_0/(2c_1), 1/(2c_1)\}$. Moreover
\be\nn\bsp {\bf T}(\vec x)_h-{\bf T}&(\vec x')_h =
\sum_{h=-\io}^{\bar h} \big\{ [\vec\b^{(1)}_h( \{\vec g_k + \vec
x_k\}_{k\ge h}) -
\vec\b^{(1)}_h(\{\vec g_k + \vec x'_k\}_{k\ge h})]\\
&+ [\vec\b^{(3)}_h(\{\vec g_k + \vec x'_k\}_{k\ge h})  -
\vec\b^{(3)}_h(\{\vec g_k + \vec x_k\}_{k\ge h})] \big\}\esp\ee
and $|{\bf T}(\vec x)_h-{\bf T}(\vec x')_h|\le c_2\l \|x-x'\|$,
thanks to the fact that all the terms in the r.h.s. of this
equation are of the second order in the running couplings. It
follows that, if $c_2\l<1$, ${\bf T}$ is a contraction in
$\BB_\x$, so that \pref{5.3} has a unique solution $\vec x^{(0)}$
in this set; moreover, if we put $\vec l_h = \vec g_h + \vec
x^{(0)}_h$, $\{\vec l_h\}_{h\le 0}$ is the flow of the effective
model corresponding to a value of $\vec l$ such that
\be\lb{5.5} |\vec g_h-\vec l_h|\le C|\l|\g^{{\th \over 2}h}\ee
Finally, this solution is such that $\vec l$ is equal to $\vec
g_0$ at the first order; hence we get
\be\lb{7.8} \bar g_1= 2\l\hat v(2p_F)+O(\l^2)\virg \bar g_2=2\l
\hat v(0) + O(\l^2)\virg \bar g_4=2\l \hat v(0)+O(\l^2)\virg
\bar\d=O(\l) \ee

Thanks to the bound \pref{5.5}, this choice of $\vec l$, allows us
to say that there are constants $Z=1+O(\l^2)$, $Z_3=1+O(\l)$ and
$\tilde Z_3=v_F+O(\l)$ such that, if $\k \le 1$ and $|\pp|\le \k$,
\be\lb{fff2}\bsp \hat\O_C(\pp) &=Z_3^2 \la\hat\r_\pp
\hat\r_{-\pp}\ra^{(g)} +A_c + o(\pp)\\
\hat D(\pp) &= -\tilde Z_3^2\la\hat j_\pp \hat j_{-\pp}\ra^{(g)}
+A_j +o(\pp) \esp\ee
where $\la\cdot\ra^{(g)}$ denotes the expectation in the effective
model satisfying \pref{5.5}, $A_c$ and $A_j$ are suitable $O(1)$
constants and
\be \r_\xx=\sum_{\o,s}\psi^+_{\xx,\o\,s}\psi_{\xx,\o\,s}\quad\quad
j_\xx=\sum_{\o,s}\o\psi^+_{\xx,\o\,s}\psi_{\xx,\o\,s}\;.\ee
Moreover, if we put $\pp_F^\o=(\o p_F,0)$ and we suppose that
$0<\k\le |\pp|,|\kk'|,|\kk'-\pp|\le 2\k$, $0<\th<1$, then
\bal \hat G^{2,1}_\r(\kk'+ \pp_F^\o, \kk'+\pp+\pp_F^\o) &=Z_3
\la\hat\r_\pp
\psi_{\kk',\o}\psi^+_{\kk'+\pp,\o}\ra^{(g)}[1+O(\k^\th)]\nn\\
\hat G^{2,1}_j(\kk'+ \pp_F^\o, \kk'+\pp+\pp_F^\o) &=\tilde Z_3
\la\hat
j_\pp \psi_{\kk',\o}\psi^+_{\kk'+\pp,\o}\ra^{(g)}[1+O(\k^\th)]\lb{h10a} \\
\hat S_2(\kk'+\pp_F^\o) &=
\la\psi^-_{\kk',\o,\s}\psi^+_{\kk,\o,\s}
\ra^{(g)}[1+O(\k^\th)]\;.\nn \eal
where $G^{2,1}_\r(\xx)$ and $G^{2,1}_j(\xx)$ are defined after
\pref{eqm}, while the functions $\la\hat\r_\pp
\psi_{\kk',\o}\psi^+_{\kk'+\pp,\o}\ra^{(g)}$ and $\la\hat j_\pp
\psi_{\kk',\o}\psi^+_{\kk'+\pp,\o}\ra^{(g)}$ coincide with the
functions \pref{vertr} and \pref{vertj}, respectively, with
$c=v_F(1+\bar\d)$. As already mentioned, if $g_1>0$, the effective
model is still invariant under a spin-independent phase
transformation; hence the WI \pref{grez} is satisfied, if we sum
both sides over $s$ and we substitute $\n_s^\m(\pp)$ with
\be\lb{4.2b} \bar\n^{\m}_{s}(\pp) = \lft\{\d_{\o,1} [\d_{s,-1}\bar
g_2 + \d_{s,1} (\bar g_2- \bar g_1)]+ \d_{\o,-1}\d_{s,-1}\bar
g_4\rgt\} \frac{\hat h(\pp)}{4\p \bar c}\virg \bar c=v_F(1+\bar\d)
\ee
Therefore we get the a WI similar to \pref{chara}, that is
\bal\label{ref4} &-ip_0 [1-\bar\n_4(\pp)-2\bar \n_\r(\pp)]
\la\hat\r_\pp \psi_{\kk',\o}\psi^+_{\kk'+\pp,\o}\ra^{(g)}+ \bar cp
[1+\bar\n_4(\pp)- 2\bar \n_\r(\pp)]  \la\hat j_\pp
\psi_{\kk',\o}\psi^+_{\kk'+\pp,\o}\ra^{(g)}
\notag\\
&=\frac1Z \lft[\la\psi^-_{\kk,\o,\s}\psi^-_{\kk,\o,\s}\ra^{(g)}
-\la\psi^-_{\kk+\pp,\o,\s}\psi^-_{\kk+\pp,\o,\s}\ra^{(g)}\rgt]
\eal
where
\be\lb{7.13} \bar\n_4(\pp) =\bar g_4{\hat h(\pp)\over 4\p \bar
c}\virg \bar \n_\r(\pp)= {\bar g_2-\bar g_1/2\over 4\p \bar c}\hat
h(\pp) \ee
By replacing \pref{h10a} in \pref{ref4}, and comparing with
\pref{ref1} we get, if $\bar\n_4(0)\equiv \bar\n_4$, $\bar
\n_\r(0)=\bar \n_\r$
\be\lb{5.13} {Z_3\over Z}=(1-\bar\n_4-2\bar \n_\r)\virg {\tilde
Z_3\over Z}=\bar c(1+\bar\n_4-2\bar \n_\r) \ee Similarly we get:
\be \bsp D_\o(\pp) \la\r^{(c)}_{\pp,\o}
\r^{(c)}_{-\pp,\o'}\ra^{(g)} -\bar\n_4(\pp)
D_{-\o}(\pp) \la\r^{(c)}_{\pp,\o} \r^{(c)}_{-\pp,\o'}\ra^{(g)}\\
-2\bar\n_\r(\pp) D_{-\o}(\pp) \la\r^{(c)}_{\pp,-\o}
\r^{(c)}_{-\pp,\o'}\ra^{(g)} = -\d_{\o,\o'} {D_{-\o}(\pp)\over
2\pi Z^2 \bar c} \esp \ee
Hence, by some simple algebra, we get:
\bal \la\r^{(c)}_{\pp,\o} \r^{(c)}_{-\pp,\o}\ra^{(g)} &= {1\over
2\pi Z^2 \bar c\bar v^2_{\r,+}} { D_{-\o}(\pp) [D_{-\o}(\pp) -
\bar\n_4 D_\o(\pp)]\over p_0^2 +
\bar c^2 \bar v_\r^2 p^2} + O(\pp)\\
\la\r^{(c)}_{\pp,\o} \r^{(c)}_{-\pp,-\o}\ra^{(g)} &= {1\over 2\pi
Z^2 \bar c\bar v^2_{\r,+}} { 2\bar\n_\r D_{\o}(\pp) D_{-\o}(\pp)
\over p_0^2 + \bar c^2 \bar v_\r^2 p^2} + O(\pp) \eal
where
\be \bar v_\r= \bar v_{\r,-}/ v_{\r,+} \virg \bar v_{\r,\m}^2 =
(1-\m\bar\n_4)^2- 4\bar\n_\r^2 \ee
Therefore the charge and current density correlations are given
by:
\be\label{xc} \bsp \la\r^{(c)}_{\pp} \r^{(c)}_{-\pp}\ra^{(g)} &=
{1\over \pi Z^2 \bar c\bar v^2_{\r,+}} { -p_0^2 (1- \bar\n_4 +
2\bar\n_\r) + \bar c^2 p^2 (1 +\bar\n_4 -2\bar\n_\r) \over p_0^2 +
\bar c^2 \bar v_\r^2 p^2} +
O(\pp)\\
\la j^{(c)}_{\pp} j^{(c)}_{-\pp}\ra^{(g)} &= {1\over \pi Z^2 \bar
c\bar v^2_{\r,+}} { -p_0^2 (1- \bar\n_4 - 2\bar\n_\r) + \bar c^2
p^2 (1 +\bar\n_4 +2\bar\n_\r) \over p_0^2 + \bar c^2 \bar v_\r^2
p^2} + O(\pp) \esp \ee

Note the above equations are true also for $g_1<0$, provided that
an infrared cut-off is present and for coupling small enough
(vanishing removing the cut-off).

From the WI \pref{ref1} we see that \be \hat\O_C(0,p_0)=0\virg
\hat D(p,0)=0 \ee
and this fixes the values of the constants $A_c$ and $A_j$ in
\pref{fff2}, so that
\be \bsp \hat \O_C(\pp) &= {Z_3^2\over \pi Z^2 \bar c\bar
v^2_{\r,+}} [ (1+\bar\n_4- 2\bar\n_\r)+\bar
v_\r^2(1-\bar\n_4+2\bar\n_\r)] {\bar c^2 p^2\over
p_0^2+\bar v_\r^2 \bar c^2 p^2}+o(\pp)\\
\hat D(\pp) &= {\tilde Z_3^2\over \pi Z^2 \bar c\bar v^2_{\r,+}
v_\r^2} [ (1+\bar\n_4+ 2\bar\n_\r)+\bar
v_\r^2(1-\bar\n_4-2\bar\n_\r)] {p_0^2\over p_0^2+\bar v_\r^2 \bar
c^2 p^2}+o(\pp)\esp \ee
If we insert \pref{5.13} in the previous equations, we get, for
the susceptibility \pref{kk} and the Drude weight \pref{cc}, the
values
\be \bsp \k &={(1-\bar\n_4-2\bar \n_\r)^2 \over \pi  \bar c\bar
v^2_{\r,+}\bar v_\r^2}[ (1+\bar\n_4- 2\bar\n_\r)+\bar
v_\r^2(1-\bar\n_4+2\bar\n_\r)] =
\frac{\bar K}{\p \bar c\bar v_\r}\\
D &= {\bar c(1+\bar\n_4-2\bar \n_\r)^2\over \pi\bar v^2_{\r,+}
\bar v_\r^2} [ (1+\bar\n_4+ 2\bar\n_\r)+\bar
v_\r^2(1-\bar\n_4-2\bar\n_\r)]= \frac{\bar K \bar c\bar
v_\r}{\p}\esp \ee
where
\be\lb{bKK} \bar K=\frac{(1-2\bar\n_\r)^2 - \bar\n_4^2}{\bar
v_{\r+} \bar v_{\r-}} = \sqrt{(1-\bar\n_4)- 2 \bar\n_\r\over (1-
\bar\n_4) + 2\bar\n_\r} \sqrt{(1+ \bar\n_4)-2\n_\r\over (1+
\bar\n_4)+2\bar\n_\r} \ee
so that
\be {\k\over D}= {1\over \bar c^2\bar v_\r^2} \ee
and this completes the proof of Theorem \ref{th1.2}. \vskip.3cm

{\bf Remark 1}  We are unable to see if $\bar c\bar
v_\r=v_F(1+\bar\d)\bar v_\r$ coincides with the velocity $c v_\r =
v_F(1+\tilde\d) v_\r$ appearing in the two-point function
asymptotic behavior \pref{4.23}, with $v_\r$ given (see
\pref{MM2}) by
\be v_\r= {(1+\n_4)^2-4\n_\r^2\over (1-\n_4)^2-4\n_\r^2}\ee
with $\n_\r$ and $\n_4$ defined as in \pref{5.9}. In fact, it is
easy to see that $\tilde\d$ is equal to $\bar\d$ at the first
order and this is true also for $\bar v_\r$ and $v_\r$ by
\pref{7.8}, \pref{7.13} and \pref{5.9}; however, our arguments are
not able to exclude that the two velocities are different.
Moreover, \pref{bKK} and \pref{KK} imply that $\bar K=K$ at first
order, but they also could be different. Note that the equality of
$\bar K$ and $K$, would imply that $\k=K/v$, with $v=\bar c\bar
v_\r$ being the charge velocity, a relation proposed in
\cite{Ha080} which, together with \pref{1.8} and \pref{hh1}, would
allow for the exact determination of the exponents in terms of the
susceptibility and the Drude weight. \vskip.3cm {\bf Remark 2}
Note that the Ward Identities \pref{xc} are true also for
$g_{1,\perp}<0$ and $l=-\io$, provided that $L$ is finite.


\appendix

\section{Proof of Theorem 2.1}\lb{appA}
%
%
By using \pref{B.13} and the definition \pref{lcz} of $\LL$, we get in the following way a
simple expression for $\VV^{(0)}(\t,\a,\psi^{[l,k]},J)$, for any $\t\in\TT_{N,k,n_g,n_J}$.

We associate with any vertex $v$ of the tree a subset $P_v$ of $I_v$, the
{\it external $\psi$ fields} of $v$, and the set $\xx_v$ of all space-time points
associated with one of the end-points following $v$; moreover, we shall
denote $\xx^J_v\subset \xx_v$ the set of all
space time points associated with the special endpoints following $v$.
The subsets $P_v$ must satisfy various
constraints. First of all, $|P_v|\ge 2$, if $v>v_0$; moreover, if $v$ is not
an endpoint and $v_1,\ldots,v_{s_v}$ are the $s_v\ge 1$ vertices immediately
following it, then $P_v \subseteq \cup_i P_{v_i}$; if $v$ is an endpoint,
$P_v=I_v$. If $v$ is not an endpoint, we shall denote by $Q_{v_i}$ the
intersection of $P_v$ and $P_{v_i}$; this definition implies that $P_v=\cup_i
Q_{v_i}$. The union ${\cal I}_v$ of the subsets $P_{v_i}\setminus Q_{v_i}$
is, by definition, the set of the {\it internal fields} of $v$, and is not
empty if $s_v>1$. Given $\t\in\TT_{N,k,n_g,n_J}$, there are many possible choices
of the subsets $P_v$, $v\in\t$, compatible with all the constraints. We shall
denote ${\cal P}_\t$ the family of all these choices and ${\bf P}$ the
elements of ${\cal P}_\t$. Then we can write:
\be \VV^{(0)}(\t,\a,\psi^{[l,k]},J)=\sum_{\bP\in\PP_\t}\VV^{(0)}(\t,\a,\bP)
\ee
%
%
\be \VV^{(0)}(\t,\a,\bP)=\int d\xx_{v_0}  \tilde\psi^{[l,k]} (P_{v_0})
\tilde J(S_{v_0}) K_{\t,\a,\bP}^{(k+1)}(\xx_{v_0})\lb{3.37}\ee
where $S_v$ denotes the set of special endpoints following $v$,
$\tilde\psi^{[l,k]} (P_v):=\prod_{f\in P_v}
\psi^{[l,k] \e(f)}_{\xx_(f),s(f)}$, $\tilde J(S_v)= \prod_{v\in S_v} J_{\xx_v,\Th_v}$ and
$K_{\t,\a,\bP}^{(k+1)}(\xx_{v_0},\yy_{v_0})$ is defined inductively
(recall that $h_{v_0} =k+1$) by the equation, valid for any
$v\in\t$ which is not an endpoint,
\be K_{\t,\a,\bP}^{(h_v)}(\xx_v)={1\over s_v !} \prod_{i=1}^{s_v}
[K^{(h_v+1)}_{v_i}(\xx_{v_i})] \EE^T_{h_v}[
\tilde\psi^{(h_v)}(P_{v_1}\bs Q_{v_1}),\ldots,
\tilde\psi^{(h_v)}(P_{v_{s_v}}\bs Q_{v_{s_v}})] \label{3.38}\ee
If $v$ is an endpoint, $K^{(h_v)}_{v}(\xx_v)$ is equal to one of
the kernels in $\LL\VV^{(h_v)}$, otherwise $K^{(h_v)}_{v}=\RR K^{(h_v)}_{\t_i,\bP_i}$, where
$\t_1,\ldots,\t_{s_v}$ are the subtrees of $\t$ with root $v$,
$\bP_i=\{P_v,v\in\t_i\}$. Hence, if we use the {\em Brydges-Battle-Federbush} identity (see \cite{Le087}) to
expand the truncated expectation in \pref{3.38}, the kernel in \pref{3.37} can be rewritten as
\be\label{3.80} \bsp
& K_{\t,\a,\bP}^{(k+1)}(\xx_{v_0})= \sum_{T\in {\bf T}}
\Big[ \prod_{i=1}^n K^{h_i}_{v^*_i}(\xx_{v^*_i})\Big]\\
&\Big\{\prod_{v\,\hbox{\rm not e.p.}}{1\over s_v!} \int
dP_{T_v}(\tt_v)\cdot\; \det G^{h_v,T_v}(\tt_v) \Big[\prod_{l\in
T_v} g^{(h_v)}_{\o_l}(\xx_l-\yy_l)]\Big]\Big\} \esp\ee
where ``e.p.''  is an abbreviation of ``endpoint'', $v^*_i$ are the end-points,
$K^{h_i}_{v^*_i}(\xx_{v^*_i})$ is $W^{(0,4),h_i}$, $W^{(0,2),h_i}$ or $W^{(1,2),h_i}$ ,
${\bf T}$ is the set of the tree graphs on $\xx_{v_0}$, obtained by putting together an
anchored tree graph $T_v$ for each non trivial vertex $v$ and
adding the lines connecting the space-time points belonging to the sets
$\xx_{v_i^*}$; moreover, $dP_{T_v}({\bf t_v})$ is a probability measure with support on
the set of ${\bf t_v =\{t_{v,i,i'}\in [0,1], 1\le i,i' \le s_v\}}$
such that $t_{v,i,i'}={\bf u}_i\cdot{\bf u}_{i'}$ for some family of vectors
${\bf u}_i\in \RRR^s$ of unit norm; finally, if $L_v= |{\cal I}_v|/2 -s_v+1$,
$G^{h_v,T_v}(\tt_v)$ is a $L_v \times L_v$ is a matrix with elements of the form
\be G^{h_v,T_v}_{v,ij,i'j'}=t_{v,i,i'}
g_{\o_{ij,i'j'}}^{(h_v)}(\xx_{ij}-\yy_{i'j'})\label{jhoi}\ee
with $g_\o^{(h)}(\xx)$ admitting  a Gram representation:
$g_\o^{(h)}(\xx-\yy)=\frac1Z \int d\zz\, A_h^*(\xx-\zz)\cdot B_h(\yy-\zz)$,
with
\be\lb{B.4} \bsp A_h(\xx)&={1\over L^2}\sum_{\kk}
\sqrt{f_h(\tilde\kk)}\frac{e^{-i\kk\xx}}{ k_0^2+c^2 k^2}\\
B_h(\xx)&={1\over L^2}\sum_{\kk}
\sqrt{f_h(\tilde\kk)}\,e^{-i\kk\xx}(ik_0+\o c k)\esp\ee
and
\be ||A_h||^2=\int d\zz |A_h(\zz)|^2\le C\g^{-2h}\;,\quad\quad
||B_h||^2\le C \g^{4h}\;,\lb{B.5xx}\ee
for a suitable constant $C$. Therefore the Gram--Hadamard
inequality, combined with the dimensional bound on
$g^{(h)}_\o(\xx)$, implies that
\be |{\rm det} G^{h_v,T_v}({\bf t}_v)| \le
C^{\sum_{i=1}^{s_v}|P_{v_i}|-|P_v|-2(s_v-1)}\cdot\; \g^{{h_v}
\left(\sum_{i=1}^{s_v}|P_{v_i}|-|P_v|-2(s_v-1)\right)}\;.\lb{2.54}\ee
By the decay properties of $g^{(h)}_\o(\xx)$, it also follows that
\be \prod_{v\ {\rm not}\ {\rm e.p.}} {1\over s_v!}\int \prod_{l\in
T_v} d(\xx_l-\yy_l)\, ||g^{(h_v)}_{\o_l}(\xx_l-\yy_l)||\le C^{n+m}
\prod_{v\ {\rm not}\ {\rm e.p.}} {1\over s_v!}
\g^{-h_v(s_v-1)}\;.\lb{2.55b}\ee
Therefore, proceeding as in the proof of Lemma 2.2 in the
companion paper \cite{BFM012_1}, we get the bound
\be \|W^{(n;2m)(h)}\|\le \sum_{n\ge d_{n,m}}  C^{n+m} \e_0^n
\sum_{\t\in {\cal T}_{N,h,n,m}\atop \a\in A_\t}
\sum_{\bP\in{\cal P}_\t\atop |P_{v_0}|=2m}
\g^{- D_{v_0} h} \Big[\prod_{v\ {\rm not}\
{\rm e.p.}\atop v>v_0} \g^{-D_v}\Big]\;,
\lb{B.18aa}\ee
where
\be D_v=-2+{|P_v|\over 2}+n^J_v \ee
where $n^J_v$ is the number of $J$-endpoints of the subtree with root $v$.
The definition of the $\RR$ operation (which is applied to all vertices, except
the endpoints and $v_0$) implies that the tree value vanish, if there is even one vertex,
except the endpoints and $v_0$, with $D_v\le 0$. Hence, we can suppose that in
\pref{B.18aa}, $D_v>0$ in all vertices of the product and the bound \pref{pc1}
follows from \pref{B.18aa} by the combinatorial arguments used in
the proof of Lemma 2.2 of companion paper \cite{BFM012_1}.

As concerns the proof of the uniform convergence as $N\to\io$, which only depends
on the structure of the tree expansion and on the corresponding bounds, we again
refer to Lemma 2.2 of \cite{BFM012_1}. \qed

\section{Proof of \pref{prc1} and \pref{prc2}}\lb{appB}
Let us denote by $\VV(\ps, J, \h)$ the expression inside the braces
in the r.h.s. side of \pref{vv1} and let us define $\VV^{(0)}(\psi,J,\h)$
so that
\be\lb{ep2}\; e^{\VV^{(0)}(\ps,J,\h)}=\int\!P_{1,N}(d\z)\
e^{\VV(\ps+\z, J, \h)} \ee
where $P_{1,N}(d\z)\equiv P(d\psi^{[1,N]})$. Note that $\VV^{(N)}(\ps,J,\h) = \VV(\ps,J,\h)$
and that the functional $\bar V^{(0)}(\ps, J)$ used in \pref{prc1} and  \pref{prc2} is equal to
$\VV^{(0)}(\ps, J, 0)$. The properties of the Grassmannian integral imply that, for $\e=\pm $,
\bal\lb{wck} \int\!P_{k+1,N}(d\z)\ \z_{\xx,\o,s}^{-\e}
F(\z)=\e\int\!d\uu\ g^{[1,N]}_\o(\xx-\uu) \int\!P_{k+1,N}(d\z)\
{\dpr F(\z)\over \dpr \z_{\uu,\o,s}^\e} \eal
and therefore
\be\lb{wt2} \bsp &{\dpr e^{\VV^{(0)}}\over\dpr
\h^\e_{\xx,\o,s}}(\ps,J,\h) =\int\!P_{k+1,N}(d\z)\
\lft(\ps_{\xx,\o,s}^{-\e}+\z_{\xx,\o,s}^{-\e}\rgt)
e^{\VV(\ps+\z,J,\h)} \\ &= \ps_{\xx,\o,s}^{-\e}e^{\VV^{(0)}(\ps,J,\h)}
+\e\int\! d\uu\ g^{[1,N]}_\o(\xx-\uu) {\dpr e^{\VV^{(0)}}\over
\dpr \ps_{\uu,\o,s}^\e}(\ps,J,\h)\esp \ee
which is \pref{gpc}. Besides, using the explicit
expression of $\VV(\ps,J,\h)$, obtained by adding to \pref{ep0} the terms
linear in $\h$, by explicit computation of the derivatives on both sides of \pref{ep2} we find
the following two identities; if $\Th=(\o,s,t_0)$,
\be\lb{aprc1} \bsp &\frac{\dpr e^{\VV^{(0)}}}{\dpr \ps^+_{\xx,\o,s}}(\ps,J,0)
=\sum_{t_0}J_{\xx,\Th}{\dpr e^{\VV^{(0)}}\over\dpr
\h^+_{\xx,\o,t_0}}(\ps,J,0) \\
&+\sum_{t_0,\Th'}\int\!d\ww\ h^{L,K}_{\Th,\Th'}(\xx-\ww) {\dpr^2e^{\VV^{(0)}}\over\dpr J_{\ww,\Th'}
\dpr \h^+_{\xx,\o,t_0}}(\ps,J,0)\esp\ee
\be\lb{aprc2}\frac{\dpr e^{\VV^{(0)}}}{\dpr J_{\xx,\Th}}(\ps,J,0)
=-{\dpr^2e^{\VV^{(0)}}\over
\dpr\h^+_{\xx,\o,t_0}\dpr\h^-_{\xx,\o,s}}(\ps,J,0)
\ee

Finally,  plug \pref{wt2} for $\e=+$ into \pref{aprc1} and
get \pref{prc1}. Also from \pref{wt2} and \pref{aprc2} we obtain:
\bea {\dpr e^{\VV^{(0)}}\over \dpr J_{\xx,\Th}}(\ps,J,\h) =-
\ps^-_{\xx,\o,t_0} {\dpr e^{\VV^{(0)}}\over
\dpr\h^-_{\xx,\o,s}}(\ps,J,\h) \hskip5em \cr\cr -\int\! d\uu\
g^{[1,N]}_\o(\xx-\uu) {\dpr ^2 e^{\VV^{(0)}}\over \dpr
\ps_{\uu,\o,t_0}^+\dpr\h^-_{\xx,\o,s}}(\ps,J,\h) \eea
which gives \pref{prc2} by applying \pref{wt2} again and using that
$g_\o^{[1,N]}(0)=0$.

\section{Some properties of the functions $\hU^{(i,j)}_{l,N,\o}(\kk^+,\kk^-)$}\lb{appC}

%
%

By using \pref{defU} and \pref{proU}, we easily see that, if $i,j>l$,
\be\lb{C.1} \hspace{-0.15cm}\lim_{\e\to 0} \hU^{(i,j)}_{l,N,\o}(\kk^+,\kk^-) = \frac1{Z_{i-1}Z_{j-1}}
\frac{\d_{j,N} u_N(\tilde\kk^-) \tilde f_i(|\tilde\kk^+|) D_\o(\kk^-) -
\d_{i,N} u_N(\tilde\kk^+) \tilde f_j(|\tilde\kk^-|) D_\o(\kk^+)}{D_\o(\kk^+) D_\o(\kk^-)}
\ee
where $U_N(\kk)$ is a smooth function such that $u_N(\kk)=0$ for $|\kk| \le \g^N$
and $u_N(\kk)=1-f_N(|\kk|)$ for $|\kk| \ge \g^N$; note that
$u_N(\kk)$ is a smooth function and that $u_N(\kk)=1-\chi_{[-\io,N]}(|\kk|)$.
If we put $\pp=\kk^+ -\kk^-$ and we use that $D_\o(\pp) = D_\o(\kk^+) - D_\o(\kk^-)$, we see
that $\hU^{(i,j)}_{l,N,\o}(\qq+\pp,\qq)$ satisfies the identity \pref{du}, if we put
\bal \hat\SS^{(i,j)}_{l,N,\o,\o}(\qq+\pp,\qq) &=  -\tilde\chi_N(\pp)
\frac{\d_{i,N} u_N(\tilde\qq+\tilde\pp) \tilde f_j(|\tilde\qq|)}{D_\o(\qq) D_\o(\qq+\pp)}\lb{C2}\\
\hat\SS^{(i,j)}_{l,N,-\o,\o}(\qq+\pp,\qq) &= \tilde\chi_N(\pp) \frac{\d_{j,N} u_N(\tilde\qq)
\tilde f_i(|\tilde\qq+\tilde\pp|)-
\d_{i,N} u_N(\tilde\qq+\tilde\pp) \tilde f_j(|\tilde\qq|)}{D_{-\o}(\pp) D_\o(\qq+\pp)}\lb{C3}
\eal
The proof of the bound \pref{61} for $i=N$ and $j>l$ easily follows from the smoothness and support properties of the functions $f_i(|\tilde\kk|)$, $u_N(\kk)$ and $\tilde\chi_N(\pp)$.
Note that this definition of the functions $\hat\SS^{(i,j)}_{l,N,\o',\o}$ is useful only in the case $j>N-2$, in
order to exploit the fact that $\hU^{(i,j)}_{l,N,\o}(\kk,\kk)=0$;
see \S4.2 of \cite{BM002} for more details in the case $N=0$. If $i=N$ and $l< j\le N-2$, we could write
the simpler decomposition
\be\lb{C.1a} \tilde\chi_N(\pp) \lim_{\e\to 0} \hU^{(N,j)}_{l,N,\o}(\kk^+,\kk^-) = -\frac12 \sum_{\o'=\pm \o}
D_{\o'}(\pp) \frac1{Z_{j-1}} \tilde\chi_N(\pp)\frac{\tilde f_j(|\tilde\kk^-|) u_N(\tilde\kk^+)}{D_\o(\kk^-)
D_{\o'}(\pp)}\ee
which also satisifies the bound \pref{61}.

If $i=N$ and $j=l$, one has to add to the r.h.s. of \pref{C.1a} the term
$$\frac{\tilde\chi_N(\pp)}{\tilde Z_{l-1}(\kk^-)} \frac{f_N(|\tilde\kk^+|) u_l(\tilde\kk^-)}{D_\o(\kk^+)}
=\frac12 \sum_{\o'=\pm \o}
D_{\o'}(\pp) \frac1{\tilde Z_{l-1}(\kk^-)} \tilde\chi_N(\pp)\frac{f_N(|\tilde\kk^+|) u_l(\tilde\kk^-)}{D_\o(\kk^+)
D_{\o'}(\pp)}$$
with $\tilde Z_{l-1}(\kk)$ defined as in \pref{prop_ir}; then we get the bound \pref{61} even for $j=l$,
since $\tilde Z_{l-1}(\kk)\ge 1$.

In order to prove the bound \pref{61a}, note that, if $N>j> l$,
\be
[1-\tilde\chi_l(\pp)] \lim_{\e\to 0} \hU^{(j,l)}_{l,N,\o}(\kk^+,\kk^-) =\frac12 \sum_{\o'=\pm \o}
D_{\o'}(\pp) \frac{[1-\tilde\chi_l(\pp)] }{Z_{j-1} \tilde Z_{l-1}(\kk^-)}
\frac{\tilde f_j(|\tilde\kk^+|) u_l(\tilde\kk^-)}{D_\o(\kk^+) D_{\o'}(\pp)}
\ee
where $u_l(\kk)$ is a smooth function
such that $u_l(\kk)=0$ for $|\kk|\ge \g^l$ and $u_l(\kk)=1-f_l(|\kk|)$ for $|\kk|\le \g^l$;
the identity is valid also for $j=l$, with $\tilde Z_{l-1}(\kk^-)$ in place of $Z_{j-1}$.
Then the bound \pref{61a} follows from the remarks that $\tilde Z_{l-1}(\kk)\ge 1$ and that
$[1-\tilde\chi_l(\pp)]\tilde f_j(|\tilde\kk^+|) u_l(\tilde\kk^-)\not=0$ only if $j>l$ and, in
such case $|D_\o(\pp)^{-1}| \le c\g^{-j}$.

We are also interested in the bound of
\be\lb{C.6}
\frac{\hU^{(l,l)}_{l,N,\o}(\kk^+,\kk^-)}{D_{\o'}(\pp)} =
 \frac1{\tilde Z_{l-1}(\kk^+) \tilde Z_{l-1}(\kk^-)} \lft[
\frac{\tilde f_l(|\tilde\kk^+|) u_l(\tilde\kk^-)}{D_\o(\kk^+) D_{\o'}(\pp)} -
\frac{\tilde f_l(|\tilde\kk^-|) u_l(\tilde\kk^+)}{D_\o(\kk^-) D_{\o'}(\pp)} \rgt]
\ee
By using the remark after \pref{prop_ir}, we see that its Fourier transform $\bar S_{\o,\o'}^{l,l}(\zz;\xx,\yy)$
satisifies, for any positive integer $r$, the bound
\be\lb{C.6a}
|\bar S_{\o,\o'}^{l,l}(\zz;\xx,\yy)| \le \frac1{Z_{l-1}} b_{r,l}(\xx-\zz) b_{r,l}(\xx-\zz)
\ee
with $b_{r,l}(\xx)$ defined as in \pref{61}.

Finally, we want to prove the bounds \pref{3.14}. To begin with, note that the r.h.s. of \pref{C2} and \pref{C3}
do not depend on $l$, if $i,j\ge K+1$, and that $\sum_{i=K+1}^N f_i(|\tilde\qq|) = \chi_{[-\io,N]}(|\tilde\qq|)$,
if $|\tilde\pp|\le 1$ and $N$ is large enough. Hence, if $|\tilde\pp|\le 1$,
\bal \sum_{i,j=K+1}^N {1\over L^2}\sum_{\qq\in\DD'_L} \hat\SS^{(i,j)}_{l,N,\o,\o}(\qq+\pp,\qq) &=
- {1\over L^2}\sum_{\qq\in\DD'_L}\frac{u_N(\tilde\qq+\tilde\pp) \chi_{[-\io,N]}(|\tilde\qq|)}{D_\o(\qq) D_\o(\qq+\pp)}\lb{C2a}\\
\sum_{i,j=K+1}^N {1\over L^2}\sum_{\qq\in\DD'_L} \hat\SS^{(i,j)}_{l,N,-\o,\o}(\qq+\pp,\qq) &=
{1\over L^2}\sum_{\qq\in\DD'_L}\frac{\chi_{[-\io,N]}(|\tilde\qq+\tilde\pp|)-
\chi_{[-\io,N]}(|\tilde\qq|)}{D_{-\o}(\pp) D_\o(\qq+\pp)}\lb{C3a}
\eal
where we also used the fact that $u_N(\kk^-) \chi_{[-\io,N]}(\kk^+) - u_N(\kk^+) \chi_{[-\io,N]}(\kk^-)=
\chi_{[-\io,N]}(\kk^+) - \chi_{[-\io,N]}(\kk^-)$. Since $|\tilde\qq|$ is of order $\g^N$, it is easy to see that
we can substitute $L^{-2}\sum_{\qq\in\DD'_L}$ with $(2\p)^{-2}\int  d\qq$ in the r.h.s. of \pref{C2a}, for any small
$\pp$, and in the r.h.s. of \pref{C3a}, for $\pp\not=0$ and small, by making an error of order $\g^{-N}/L$.
On the other hand, by using the fact that under the change of variables $\tilde\qq=(cq,q_0) \rightarrow (q_0,-cq)$,
which leaves invariant $|\tilde\qq|$, $D_\o(\qq) \rightarrow i\o D_\o(\qq)$, we see that
\be\nn \int \frac{d\qq}{(2\p)^2} \frac{u_N(\tilde\qq) \chi_{[-\io,N]}(|\tilde\qq|)}{D_\o(\qq)^2}=0\ee
This proves \pref{3.14} for $\t_N^+$. As concerns $\t_N^-$, an easy calculation shows that
\be\nn \bsp &\lim_{\pp\to 0} \int \frac{d\qq}{(2\p)^2} \frac{\chi_{[-\io,N]}(|\tilde\qq+\tilde\pp|)-
\chi_{[-\io,N]}(|\tilde\qq|)}{D_{-\o}(\pp) D_\o(\qq+\pp)}=\\
&-\frac1{2c}\int\!\frac{d\qq}{(2\p)^2}\; \frac{\c'(|\qq|)}{|\qq|}
=-\frac{1}{4\p c} \int_0^\io\! dt\;\c'(t) = -\frac{1}{4\p c}
\lft[\c(\io)-\c(0)\rgt]=\frac{1}{4\p c}\esp\ee
which proves \pref{3.14} for $\t_N^-$.

\section{Approximate Spin-Charge Separation} \lb{ss2.5dhy}

\begin{theorem}\lb{th1.3}
Under the same condition of the previous Theorem, the Fourier
transform of the 2-point Schwinger function is given by \be \hat
S_2(\kk+{\bf p}_F^\o)=Z(\kk)\hat S_{M,\o}(\kk)[1+R(\kk)]\virg {\bf
p}_F^\o = (\o p_F,0)\ee
where
\be |R(\kk)|\le C {\l^2\over 1+ a |\l \log |\kk||}\virg a\ge 0\;,
\ee
\be Z(\kk)=L(|\kk|^{-1})^{\z_z}[1+R'(\kk)]\virg |R'(\kk)|\le C|\l|
\ee
$L(t)$, $t\ge 1$, is the function defined in Theorem 1.1 of the
companion paper \cite{BFM012_1} and $\hat S_{M,\o}(\kk)$ is a function
whose Fourier transform is of the form
\be\lb{hh} S_{M,\o}(\xx)= \frac1{2\p v_F} { [v_\r^2 x_0^2
+(x_1/v_F)^2]^{-\h_\r/2}\over (v_\r x_0+i\o x_1/v_F)^{1/2} (v_\s
x_0+i\o x_1/v_F)^{1/2}} e^{C +O(1/|\xx|)} \ee
with $v_{\r,\s}=1 +O(\l)$, $\h_\r=O(\l^2)$,
$v_\r-v_\s=c_v\l+O(\l^2)$, with $c_v\not=0$.
\end{theorem}

Similar expressions are true also for the density correlations
(the explicit formulae are in \S 5 and are not reported here for
brevity). The above theorem says that the two point function can
be written, up to a logarithmic correction, as the 2-point
function of the Mattis model \cite{Ma064}, a model which shows an {\it
anomalous dimension} and the phenomenon of {\it spin-charge
separation}. This last property means that there is, in the Mattis
model, an exact decoupling of the hamiltonian as sum of two
independent hamiltonians describing the spin and charge degrees of
freedom. A manifestation of spin charge separation is that the
2-point function is factorized in the product of two functions,
similar to Schwinger functions of particles with different
velocities. In this sense, the above theorem says that the
spin-charge separation occurs approximately also in the Hubbard
model, but is valid only at large distances and up to logarithmic
corrections.

If $\kk\not=0$, the Fourier transform $\hat S_2(\kk+\o \pp_F)$ of
the two-point Schwinger function $S_2(\xx)$ in the Hubbard model
an be written as a tree expansion, in a way similar to eq. (2.64)
of \cite{BFM007}, whom we shall refer to for the notation:
\be \hat S_2(\kk+\pp^\o_F) = \sum_{n=0}^\io \sum_{j_0=-\io}^{0}
\sum_{\t\in\TT_{j_0,n,2,0}} \sum_{\bP\in \PP \atop |P_{v_0}|=2}
\hG^{2}_{\t,\o}(\kk)\ee
where $\pp_F^\o=(\o p_F,0)$, $\o=\pm$. Here $\hG^{2}_{\t,\o}(\kk)$
represents the contribution of a single tree $\t$ with $n$
endpoints and root of scale $j_0$; if $|\kk|\in [\g^{h_\kk},
\g^{h_\kk+1})$, it obeys the bound:
\be |\hG_{\t,\o}^{2}(\kk)| \le C \g^{-(h_\kk-j_0)}
{\g^{-h_\kk}\over Z_{h_\kk}} \sum_{n=0}^\io \sum_{j_0=-\io}^{0}
\sum_{\t\in\TT_{j_0,n,\kk}} \sum_{\bP\in \PP \atop |P_{v_0}|=2}
(C|\l|)^n \prod_{\rm v\ not\ e.p.} \g^{-d_v}{Z_{h_v}\over
Z_{h_v-1}} \;, \ee
where $\TT_{j_0,n,\kk}$ denotes the family of trees whose special
vertices (those associated with the external lines) have scale
$h_\kk$ or $h_{\kk+1}$. Moreover, $d_v>0$, except for the vertices
belonging to the path connecting the root with $v^*$, the higher
vertex (of scale $h^*$) preceding both the two special endpoints,
where $d_v$ can be equal to $0$. These vertices can be regularized
by using a factor $\g^{-(h^*-j_0)}$, extracted from the factor
$\g^{-(h_\kk-j_0)}$, so that we can safely perform the sum over
all the trees with a fixed value of $h^*$ and we get
\be\lb{mmm} |\hat S_2(\kk+\pp^\o_F)| \le C {\g^{-h_\kk}\over
Z_{h_\kk}} \sum_{h^*=-\io}^{h_\kk} \g^{-(h_\kk - h^*)} \le C
{\g^{-h_\kk}\over Z_{h_\kk}} \ee
A similar bounds can be obtained for the effective model with
$g_{1\erp}=0$ and couplings chosen as in Lemma \ref{lm5.1}. We
shall call $\hat S_\o^M(\kk)$ and $\tilde Z_h$ the two-point
function Fourier transform and the renormalization constants,
respectively, in this model.

Let us put
\be \hat S_2(\kk+\pp^\o_F)={1\over Z_{h_\kk}} \bar G^2_\o(\kk)
\virg \hat S_\o^M(\kk) = {1\over \tilde Z_{h_\kk}} \bar
G^{2,M}_\o(\kk) \ee
We can write
\be \hat S_2(\kk+\pp^\o_F) = {\tilde Z_{h_\kk}\over Z_{h_\kk}}
{\bar G^2_\o(\kk)\over \tilde Z_{h_\kk}} = {\tilde Z_{h_\kk}\over
Z_{h_\kk}} \hat S_\o^M(\kk)+ {1\over Z_{h_\kk}} [\bar
G^2_\o(\kk)-\bar G^{2,M}_\o(\kk)] \ee
Note now that $\bar G^2_\o(\kk)$ differs from $\bar
G^{2,M}_\o(\kk)$ for three reasons:

\begin{itemize}

\item[1)] the propagators are different, which produces a
difference  exponentially small thanks to the bounds (2.102) and (2.103) of \cite{BFM012_1} and the short
memory property;

\item[2)] the r.c.c.  $v_h$ and $\tilde v_h$ are different, which
produces a  difference of order $\tilde g_{1,h_\kk}$, thanks to \pref{144} and the
short memory property;

\item[3)] in the tree expansion of $\bar G^2_\o(\kk)$ and of the
ratios
    $Z_j/Z_{j-1}$ there are trees with endpoints of type $g_1$, not present
    in the tree expansion of $\bar G^{2,M}_\o(\kk)$ and $\tilde Z_j/\tilde
    Z_{j-1}$; this fact produces again a difference of order $\tilde
    g_{1,h_\kk}$.

\end{itemize}

\0 These remarks, together with the fact that there is no tree
with only one endpoint in the tree expansion, implies that
\be\lb{bs1} \left|{1\over Z_{h_\kk}} [\bar G^2_\o(\kk)-\bar
G^{2,M}_\o(\kk)] \right| \le C|\l \tilde g_{1,h_\kk}|
{\g^{-h_\kk}\over Z_{h_\kk}} \ee

For similar reason, we have
\be\lb{bs2} {\tilde Z_{h_\kk}\over Z_{h_\kk}} = \prod_{j=h_\kk}^0
{\tilde Z_{h_j}\over Z_{h_{j-1}}} {\tilde Z_{h_0}\over Z_{h_0}} =
[1+O(\l^2)] e^{O(\l) \sum_{j=h_\kk}^0 \tilde g_{1,j}} = [1+O(\l)]
L(|\kk|^{-1})^{O(\l)} \ee
%


\vskip.3cm {\it Acknowledgements} G.B. and V.M. acknowledge the
financial support of MIUR, PRIN 2008. V.M. gratefully acknowledges
also the financial support from the ERC Starting Grant
CoMBoS-239694.

\end{document}